\documentclass[aps,prb,twocolumn,superscriptaddress,showpacs,apsrev]{revtex4-1}
\usepackage[T1]{fontenc}
\usepackage{lmodern}
\usepackage[latin1]{inputenc}
\usepackage[swedish,english]{babel}
\usepackage{amsmath}
\usepackage{amsfonts}
\usepackage{graphicx} 
\usepackage{comment}
\usepackage{xfrac}
\usepackage{color}
\usepackage{multirow}
\usepackage{bm}
\usepackage{hyperref}
\usepackage{braket}
\usepackage{caption}
\usepackage{subcaption}

\hypersetup{
	colorlinks = true,
	pdftitle = {},
	pdfauthor = {},
	pdfkeywords = {},
	linkcolor = blue,
	citecolor = blue,
	filecolor = black,
	urlcolor = magenta
}

\def\mean#1{\left< #1 \right>}
\newcommand{\dd}{\mathrm{d}}
\newcommand{\ee}{\mathrm{e}}
\newcommand{\img}{\mathrm{i}}

\graphicspath{{./figures/}}

\begin{document}

\title{Magnetic Effects in the Paraxial Regime of Elastic Electron Scattering}
 
\author{Alexander Edstr\"om}
\affiliation{Department of Physics and Astronomy, Uppsala University, Box 516, 75121 Uppsala, Sweden}
\author{Axel Lubk}
\affiliation{Triebenberg Laboratory, Technische Universit\"at Dresden, Germany}
\author{J\'{a}n Rusz}
\affiliation{Department of Physics and Astronomy, Uppsala University, Box 516, 75121 Uppsala, Sweden}

\date{\today}

\begin{abstract}
Based on a recent claim [Phys. Rev. Lett. 116, 127203 (2016)] that electron vortex can be used to image magnetism at the nanoscale in elastic scattering experiments, using transmission electron microscopy, a comprehensive computational study is performed to study magnetic effects in the paraxial regime of elastic electron scattering in magnetic solids. Magnetic interactions from electron vortex beams, spin polarized electron beams and beams with phase aberrations are considered, as they pass through ferromagnetic FePt or antiferromagnetic LaMnAsO. The magnetic signals are obtained by comparing the intensity over a disk in the diffraction plane for beams with opposite angular momentum or aberrations. The strongest magnetic signals are obtained from vortex beams with large orbital angular momentum, where relative magnetic signals above $10^{-3}$ are indicated for $10\hbar$ orbital angular momentum, meaning that relative signals of one percent could be expected with the even larger orbital angular momenta, which have been produced in experimental setups. All results indicate that beams with low acceleration voltage and small convergence angles yield stronger magnetic signals, which is unfortunately problematic for the possibility of high spatial resolution imaging. Nevertheless, under atomic resolution conditions, relative magnetic signals in the order of $10^{-4}$ are demonstrated, corresponding to an increase with one order of magnitude compared to previous work.  
\end{abstract}

\maketitle

\section{Introduction}

Nanoengineering of magnetic materials allows for miniaturization of magnetic technology, design of nanostructures with new or improved properties as well as exploration of novel aspects of fundamental phenomena. A complete understanding of magnetic systems at the length scales relevant in modern technologies requires readily available characterization methods capable of reaching high spatial resolution, down to the atomic distances. The use of circularly polarized X-rays, available in synchrotron facilities, allows for element specific imaging of magnetism with resolution down to approximately 10 nm in so called X-ray magnetic circular dichroism\cite{thole, carra, stohr1999470} (XMCD) experiments. The discovery of an electron equivalence to XMCD, electron magnetic chiral dichroism\cite{nature} (EMCD) opened up new possibilities to observe magnetism in the transmission electron microscope (TEM), potentially allowing spatial resolution well below the \AA ngstr\"om regime\cite{PhysRevLett.102.096101} in scanning TEM (STEM) mode. EMCD, however, suffered from low signal-noise ratio (SNR) but gained renewed attention with the discovery of electron vortex beams\cite{Uchida2010, Verbeeck2010, McMorran2011}, i.e., electron beams with well defined orbital angular momentum (OAM), which also facilitate an EMCD signal in electron energy loss (EELS) experiments\cite{Verbeeck2010}. Later it was shown that vortex beam EMCD can only be observed at atomic resolution\cite{Rusz2013} and it remains technologically challenging to perform such experiments with convincing and reproducible experimental results, although new breakthroughs might be expected with further improvements in atomic size vortex beam generation\cite{Beche2016}. 

In EMCD experiments the magnetic signal appears in a part of the EELS spectrum where merely a small fraction of the scattered electrons are found, making the signal weak. Magnetic effects should, however, also appear in the elastic scattering regime and albeit previous suggestions that such effects are very weak\cite{Fujiwara, Rother2009}, it was recently shown in computational work\cite{edstrom16} that electron vortex beams carrying tens of quanta of angular momentum should yield a magnetic signal in elastic scattering, which is deemed feasible to detect with modern technology. This appears promising considering that vortex beams with as much as hundreds of $\hbar$ of OAM have been produced\cite{McMorran2011, Grillo2015}. The magnetic signal is observed as a difference in the intensity distribution of electrons in the diffraction plane for opposite OAM beams, and will in the forthcoming be termed OAM magnetic signal. Applying the same analysis to the signal of oppositely spin polarized beams, which is referred to as spin magnetic signal in the following, is of great interest for the emerging development of spin polarized TEM technology\cite{tanaka}. Further understanding of these phenomena and how to best detect such magnetic signals in experiments requires a more comprehensive study as will be presented in this work, where a discussion of the relevant theory and computational methods is first provided in Sec.~\ref{TheoryMethods}. In particular, the paraxial Pauli equation with relativistically corrected kinetic energy used is derived in Sec.~\ref{ParPauliEq} while its solution by a multislice approach is discussed in Sec.~\ref{PauliMS}. The description of the magnetism in a magnetic solid, which is required and here obtained from first-principles electronic structure theory calculations, is provided in Sec.~\ref{MagFields}. A comprehensive study of the OAM and spin magnetic signals and how they depend on various beam parameters, including acceleration voltage, convergence angle and angular momentum, are then provided for the ferromagnetic compound FePt in Secs.~\ref{FePtzmag}-\ref{FePtxmag}. The case of anti-ferromagnet LaMnAsO is also considered in Sec.~\ref{LaMnAsOresults}. The previous work\cite{edstrom16} suggested that an experimentally measurable magnetic signal was only obtainable with large OAM beams, which unfortunately hinders the possibility of high resolution STEM imaging because large OAM also results in large beam dimensions. One important question to be addressed is therefore whether more beneficial conditions can be found, e.g., by varying the beam parameters, in order to allow for atomic resolution magnetic measurements with the suggested technique. 

In electron vortex beams the OAM manifests itself as a phase winding such that $\psi_l (\mathbf{r}) \sim \ee^{i l \phi} $, resulting in an OAM of $l\hbar$, where $l$ is an integer. Recently, it was suggested\cite{emcdc34, PhysRevB.93.104420} and subsequently experimentally corroborated\cite{Idrobo2016}, that this is merely one of many types of phase distributions which can result in a magnetic interaction in inelastic scattering EMCD experiments. Alternatives correspond to beam aberrations that can be controlled in modern aberration corrected (S)TEMs, thereby opening new paths towards high resolution imaging of magnetism. The possibility of magnetic signals based on aberrated beams also in elastic scattering is thus explored in Sec.~\ref{aberrations}. For further insight into whether the discussed effects are realistic to observe in experiments a final Section~\ref{noise} presents a discussion regarding possible noise and errors. 

\section{Theory and Methodology}\label{TheoryMethods}

A description of the elastic scattering of fast electrons with wave vector $k$ travelling along the $z$-direction in an electrostatic potential $V(\mathbf{r})$, is often based on the paraxial Schr\"odinger equation\cite{kirkland}
\begin{equation}
	\frac{\partial}{\partial z} \psi(\mathbf{r}) = i  \left( \frac{1}{2k}\nabla_{xy}^2   +\frac{meV(\mathbf{r})}{\hbar^2 k} \right) \psi(\mathbf{r}),
\label{paraxialSeq}
\end{equation}
for the envelope wave function $\psi(\mathbf{r})$ related to the complete wave function $\psi_\text{f}(\mathbf{r}) = \psi(\mathbf{r})\ee^{ikz}$. Here $m=\gamma m_0$ is the relativistically corrected mass, $-e$ is the electron charge, $\hbar$ the reduced Planck's constant and $\nabla_{xy}^2$ is the two-dimensional Laplacian. In the paraxial regime, 
\begin{equation}
	\left| \frac{\partial^2\psi}{\partial z^2} \right| \ll \left| k \frac{\partial\psi}{\partial z} \right|, 
\label{paraxialAppr}
\end{equation} 
Eq.~\ref{paraxialSeq} provides a well established and accurate description of elastic scattering processes\cite{kirkland}. As it is a first order equation in $z$ it can be solved, for example, through multislice algorithms, where the solution is computed slice by slice from the knowledge of the initial wave function at $z=0$. The input required about the system of interest is the electrostatic potential, $V(\mathbf{r})$, which can be obtained, e.g., via tabulated values\cite{kirkland} or from calculations based on electronic structure theory\cite{Meyer2011}. However, Eq.~\ref{paraxialSeq} neglects magnetism as it does not consider the spin and orbital angular momentum of the electron beam nor the magnetic fields in the scatterer. With the development of electron vortex beams and spin polarized electron microscopes the effects of magnetism in electron scattering processes are of increasing relevance. Therefore, in the coming Sections, \ref{ParPauliEq}-\ref{MagFields}, a paraxial equation, which takes into consideration magnetic effects, will be derived from a relativistically corrected Pauli equation. The relativistically corrected form of the Pauli equation can be obtained from a squared form of the Dirac equation by neglecting certain effects, such as spin-orbit coupling. A multislice solution to this equation is then presented and a description of the magnetic fields in a solid will be discussed. In Sec.~\ref{constantB} a brief discussion is given regarding effects expected in a constant $\mathbf{B}$-field. 

\subsection{Paraxial Pauli Equation}\label{ParPauliEq}

The time-independent Pauli equation reads
\begin{equation}
	\left[ \frac{1}{2m}  \left[ \boldsymbol{\sigma} \cdot \left( \hat{\mathbf{p}} + e \mathbf{A}(\mathbf{r}) \right) \right]^2  -eV(\mathbf{r}) \right] \mathbf{\Psi}_\text{f}(\mathbf{r}) = E\mathbf{\Psi}_\text{f}(\mathbf{r})
\label{Peq}
\end{equation}
where $\hat{\mathbf{p}}=-i\hbar\nabla$ is the momentum operator, $\mathbf{A}$ is the vector potential and $\boldsymbol{\sigma} = (\sigma_x, \sigma_y, \sigma_z)$ contains the Pauli matrices 
\begin{equation}
	\sigma_x =  \begin{pmatrix} 0 & 1 \\ 1 & 0 \end{pmatrix} , \quad \sigma_y =  \begin{pmatrix} 0 & -i \\ i & 0 \end{pmatrix} , \quad \sigma_z =  \begin{pmatrix} 1 & 0 \\ 0 & -1 \end{pmatrix}. 
\label{Pmat}
\end{equation}
We are considering an elastic process with energy 
\begin{equation}
	E = \frac{h^2}{2m\lambda^2} = \frac{\hbar^2k^2}{2m}, 
\end{equation}
where $\mathbf{k}$ is the wave vector of the incoming electron and $\lambda = \frac{2\pi}{|\mathbf{k}|} = \frac{2\pi}{k}$ its wavelength. Both mass and wavelength are relativistically corrected according to Fujiwara\cite{Fujiwara} so $m=\gamma m_0$ and $\lambda = \frac{hc}{\sqrt{(m_0c^2 + T)^2 - m_0^2c^4}}$, with $h$ and $c$ being Planck's constant and the speed of light. $T$ is the kinetic energy which is typically expressed in terms of the acceleration voltage $V_\text{acc}$, i.e., $T=eV_\text{acc}$.
\begin{equation}
	\mathbf{\Psi}_\text{f}(\mathbf{r}) = \begin{pmatrix} \psi_{\text{f}\uparrow}(\mathbf{r}) \\ \psi_{\text{f}\downarrow}(\mathbf{r}) \end{pmatrix}
\end{equation}
is the two component wave function with a spin up ($\uparrow$) and a spin down ($\downarrow$) part. For fast incoming electrons with wave vector $(0,0,k)$, it is suitable to use the ansatz
\begin{equation}
	\mathbf{\Psi}_\text{f}(\mathbf{r}) = \ee^{\img k z} \mathbf{\Psi}(\mathbf{r}) = \ee^{\img k z} \begin{pmatrix} \psi_\uparrow(\mathbf{r}) \\ \psi_{\downarrow}(\mathbf{r}) \end{pmatrix}
\end{equation} 
so that $\psi_{\uparrow\downarrow}(\mathbf{r})$ are slowly varying with $z$, $k$ is large and 
\begin{equation}
	\left| \frac{\partial^2\psi_{\uparrow\downarrow}}{\partial z^2} \right| \ll \left| k \frac{\partial\psi_{\uparrow\downarrow}}{\partial z} \right| .
\label{fast}
\end{equation} 

In Coulomb gauge, $\nabla \cdot \mathbf{A} = 0$ (which is used throughout this work), the momentum part, $ \left[ \boldsymbol{\sigma} \cdot \left( \hat{\mathbf{p}} + e \mathbf{A}(\mathbf{r}) \right) \right]^2$, of Eq.~\ref{Peq} is equivalent to  
\begin{equation}
	\left( -\hbar^2\nabla^2 - 2i{\hbar}e\mathbf{A}\cdot\nabla + e\hbar\boldsymbol{\sigma}\cdot\mathbf{B} + e^2A^2 \right) \mathbf{\Psi}_\text{f},
\end{equation}
where the magnetic flux density $\mathbf{B}=\nabla\times\mathbf{A}$ has been introduced. In the following the term proportional to $A^2$ is neglected as it is small compared to all other terms\cite{Rother2009}. From the Dirac equation\cite{Strange} a term proportional to $V^2$, related to the $A^2$ term, would also appear and both would in principle be straight forward to include here but would only provide minor quantitative corrections. Furthermore, these terms are diagonal in spin space and do not couple to orbital angular momentum whereby they should not be important for the effects which are in focus of this work. The gradient operator yields
\begin{equation}
	\nabla \psi_{\text{f}\uparrow\downarrow} = \text{e}^{ i k z} \left( \frac{\partial }{\partial x} , \frac{\partial}{\partial y} , \frac{\partial}{\partial z} + i k \right) \psi_{\uparrow\downarrow}
\label{nablas}
\end{equation}
and similarly the Laplacian equates to
\begin{equation}
	\nabla^2 \psi_{\text{f}\uparrow\downarrow} = \text{e}^{ i k z} \left( \nabla_{xy}^2 + \frac{\partial^2}{\partial z^2} + 2 i k \frac{\partial}{\partial z} - k^2 \right) \psi_{\uparrow\downarrow}.
\label{laplace}
\end{equation}
Neglecting the term containing the second derivative with respect to $z$ according to  Eq.~\ref{fast} and rearranging, Eq.~\ref{Peq} becomes
\begin{widetext}\begin{equation}
	\frac{\partial}{\partial z} \begin{pmatrix} \psi_\uparrow(\mathbf{r}) \\ \psi_{\downarrow}(\mathbf{r}) \end{pmatrix} = \frac{i m}{\hbar}\left( \hbar k + e A_z \right)^{-1} \left\{ \frac{\hbar^2}{2m}\nabla_{xy}^2   + \frac{ie\hbar}{m}\mathbf{A}_{xy}\cdot\nabla_{xy} - \frac{\hbar k e A_z}{m} - \frac{e\hbar}{2m}\boldsymbol{\sigma}\cdot\mathbf{B} + eV \right\}  \begin{pmatrix} \psi_\uparrow(\mathbf{r}) \\ \psi_{\downarrow}(\mathbf{r}) \end{pmatrix} \equiv \hat{H} \begin{pmatrix} \psi_\uparrow(\mathbf{r}) \\ \psi_{\downarrow}(\mathbf{r}) \end{pmatrix},
\label{paraxialPeq}
\end{equation}\end{widetext}
which, upon setting $\mathbf{A} = \mathbf{B} = 0$, reduces to the paraxial Schr\"odinger equation in Eq.~\ref{paraxialSeq} for each of the spin components.

Eq.~\ref{paraxialPeq} is a matrix equation, where each term on the right hand side is diagonal except the $\boldsymbol{\sigma}\cdot\mathbf{B}$ term, which includes off-diagonal contributions proportional to $B_x \pm i B_y$. Spin flip effects are therefore only caused by the $x$- and $y$-components of the $B$-field whereby such effects are expected to be stronger for magnetizations parallel to the $xy$-plane as has been suggested before\cite{GrilloKarimi} and will be further discussed in Sec.~\ref{FePtxmag}

\subsection{Multislice Solution}\label{PauliMS}

In order to numerically integrate Eq.~\ref{paraxialPeq}, formally written as 
\begin{equation}
\frac{\partial \psi}{\partial z} = \hat{H} \psi,
\label{firstorderPsiEq}
\end{equation}
a multislice method\cite{kirkland} is applied. There are a number of different such methods, including the conventional method\cite{Cowley1957}, invoking Fourier transforms or the real space version\cite{VanDyck_RSMS} where the propagator is computed via a series expansion of the exponential function. Here the real space version will be used as it has been reported to efficiently yield high numerical precision\cite{Cai2009} and is easy to generalize from the Paraxial Schr\"odinger equation to the Paraxial Pauli equation presented in Eq.~\ref{paraxialPeq}. That the real space version of the multislice algorithm does not require periodicity in the $xy$-plane is also important as it will be necessary to include non-periodic vector potentials as discussed further in Sec.~\ref{MagFields} and Appendix~\ref{AppA}. The first step in this approach is to note that the formal solution to Eq.~\ref{firstorderPsiEq} is 
\begin{equation}
\mathbf{\psi}(x,y,z+\Delta z) = \hat{Z}\{ \ee^{\int_z^{z+\Delta z} \hat{H}(x,y,z') \dd z' } \}  \mathbf{\psi}(\mathbf{r})  \\
\label{multislicesol}
\end{equation}
where $\hat{Z}$ is Dyson's path ordering operator for the variable $z$ needed when the $\hat{H}(x,y,z')$ operators do not commute for different $z'$. With $\hat{h} = \hat{Z}\frac{1}{\Delta z} \int_z^{z+\Delta z} \hat{H}(x,y,z') \dd z' $, the exponential can be expanded: 
\begin{equation}
\mathbf{\psi}(x,y,z+\Delta z) =  \sum_{n=1}^\infty \frac{\Delta z^n}{n!}{\hat{h}}^n(\mathbf{r}) \mathbf{\psi}(\mathbf{r}).
\label{multisliceser}
\end{equation}
For thin enough $\Delta z$ the series in Eq.~\ref{multisliceser} will converge with a small number terms and can be truncated with suitable numerical accuracy and furthermore $\hat{h} \approx \hat{H}$.

\subsection{Interactions with a Constant B-Field}\label{constantB}

The spin and orbital angular momentum operators are
\begin{equation}
	\hat{\mathbf{S}} = \frac{\hbar}{2}\boldsymbol{\sigma} \quad \text{and} \quad \hat{\mathbf{L}} = \mathbf{r} \times \hat{\mathbf{p}} = -i\hbar \mathbf{r} \times \nabla 
\end{equation}
and the relevant part of the Hamiltonian discussed in Sec.~\ref{ParPauliEq}, which describes the magnetic interactions between the electron beam and the sample is 
\begin{equation}
	\hat{H}_\text{mag} = \frac{ie\hbar}{m}\mathbf{A}_{xy}\cdot\nabla_{xy} - \frac{e}{m}\hat{\mathbf{S}}\cdot\mathbf{B}.
\end{equation}
If the magnetic field is a constant along the $z$-direction, then as detailed in Sec.~\ref{MagFields}, in the Coulomb gauge 
\begin{equation}
	\mathbf{A} = \frac{1}{2} \mathbf{B} \times \mathbf{r}, 
\end{equation}
yielding 
\begin{align}
	\hat{H}_\text{mag} &= -\frac{e}{m}\mathbf{A}\cdot\hat{\mathbf{p}} - \frac{e}{m}\hat{\mathbf{S}}\cdot\mathbf{B} = \nonumber \\  &= -\frac{e}{2m}\left( \left( \mathbf{B} \times \mathbf{r} \right) \cdot \hat{\mathbf{p}} + 2 \hat{\mathbf{S}}\cdot\mathbf{B} \right) = \nonumber \\ &= -\frac{e}{2m}\left( \left( \mathbf{r} \times \hat{\mathbf{p}} \right) \cdot \mathbf{B} + 2 \hat{\mathbf{S}}\cdot\mathbf{B} \right) = \nonumber \\
	&= -\frac{e}{2m}\left( \hat{\mathbf{L}}  + 2 \hat{\mathbf{S}}\right) \cdot \mathbf{B}.
\label{Lplus2S}
\end{align}
This expression elucidates how not just the spin but also orbital angular momentum of the beam will couple to the magnetic field with a strength proportional to the angular momentum. It is also clear that a similar magnitude of magnetic effect is expected from spin and orbital angular momentum. However, the spin of the electron is fixed while orbital angular momentum of an electron vortex beam can be deliberately increased by beam shaping techniques. and, furthermore, varies significantly as the beam scatters through a crystal\cite{Loffler2012, Rusz2013}. In a magnetic material this term will be significant if the beam carries large angular momentum, which is possible with electron vortex beams\cite{McMorran2011, Saitoh2012, Grillo2015}. According to the discussion in Sec.~\ref{MagFields} the magnetic fields in a solid will be described as the sum of a periodically varying part and a uniform part related to the saturation magnetization. For a beam with size significantly larger than a unit cell, the interaction with the uniform part is expected to dominate and a proportionality is expected between angular momentum and magnetic interaction, while in the atomic resolution limit the behaviour should be different. This is also in agreement with the results of recent numerical simulations\cite{edstrom16}.

Eq.~\ref{Lplus2S} also permits a discussion of the precession of a spin vector in a magnetic field, which is relevant in Sec.~\ref{FePtxmag}, where a situation with FePt magnetized in the $x$-direction is considered. The Hamiltonian for a spin in a constant magnetic field in the $x$-direction, $\mathbf{B}=B\hat{x}$ is
\begin{equation}
\hat{H}_\mathbf{B}=\mathbf{B}\cdot\boldsymbol{\mu} = B \mu_\text{B} \sigma_x
\end{equation}
and as Eq.~\ref{paraxialPeq} has the same structure as a 2D time-dependent Pauli equation, the $z$-evolution of the expectation value of the Pauli spin vector is
\begin{align}\label{spininField}
\frac{\dd }{\dd z} \mean{\boldsymbol{\sigma}(z)} & = \img\frac{m}{\hbar^2 k} \mean{ \left[ H, \boldsymbol{\sigma} \right] } \\ \nonumber 
 & = \frac{2B\mu_\text{B}m}{\hbar^2 k} \left( 0, -\mean{\mathbf{\sigma}_z(z)}, \mean{\mathbf{\sigma}_y(z)} \right),
\end{align}
i.e., a spin originally parallel to the magnetic field will remain stationary while a spin pointing in another direction will rotate about the magnetization direction.

\subsection{$\mathbf{A}$ and $\mathbf{B}$ in a Magnetic Solid}\label{MagFields}

In order to use Eq.~\ref{paraxialPeq} for simulating magnetic scattering of electrons a realistic description of the magnetic vector potential $\mathbf{A}$ and the corresponding flux density $\mathbf{B}$ is needed. In this work, crystalline magnetic solids will be considered but as brought up in the Appendix~\ref{AppA}, even in a periodic system, $\mathbf{A}$ is in general non-periodic regardless of gauge choice. Only when the volume average of $\mathbf{B}$ vanishes, as it does in an antiferromagnet, the vector potential can be made periodic. This leads to a natural decomposition of the vector potential in a periodic ($\mathbf{A}_\text{p}$) and a non-periodic ($\mathbf{A}_\text{np}$) part, i.e., $\mathbf{A} = \mathbf{A}_\text{p} + \mathbf{A}_\text{np}$ which corresponds to a decomposition of the magnetic flux density $\mathbf{B}$ into a periodic but spatially non-uniform part with volume average zero ($\mathbf{B}_\text{p}$) and a uniform part ($\mathbf{B}_\text{avg}$), which is the volume average of the field, with the total field being $\mathbf{B} = \mathbf{B}_\text{p} + \mathbf{B}_\text{avg}$. These fields can now be related according to 
\begin{equation}
	\mathbf{B}_\text{avg} = \nabla \times \mathbf{A}_\text{np}
	\label{Bavg_Anp}
\end{equation} 
and
\begin{equation}
	\mathbf{B}_\text{p} = \nabla \times \mathbf{A}_\text{p}.
	\label{Bp_Ap}
\end{equation}
Furthermore, in Coulomb gauge Eq.~\ref{Bavg_Anp} is easily inverted to obtain 
\begin{equation}\label{AofConstB}
	\mathbf{A}_\text{np} = \frac{1}{2} \mathbf{B}_\text{avg} \times \mathbf{r}.
\end{equation} 
Since $\mathbf{B}_\text{avg}$ is the volume average of the magnetic flux density, in a magnetic material with no externally applied fields it is simply $\mathbf{B}_\text{avg} = \mu_0 \mathbf{M}$, where $\mathbf{M}$ is the magnetization of the material. How to obtain a microscopic description of $\mathbf{B}_\text{p}$ and $\mathbf{A}_\text{p}$ from first principles electronic structure theory is discussed in the following. The same procedure was applied in recent preceding work\cite{edstrom16}.

From electronic structure theory, e.g., using density functional theory (DFT), one can obtain the magnetization density as a vector field from the spin-resolved density matrix $\rho(\mathbf{r})$ according to
\begin{align}\label{magdensity}
  \mathbf{m}(\mathbf{r}) &= \mu_\text{B} \langle \boldsymbol{\sigma} \rangle = \mu_\text{B} \mathrm{Tr} [ \rho(\mathbf{r}) \boldsymbol{\sigma} ] = \\ \nonumber
  &= \mu_\text{B} \left( 2 \text{Re}\left( \psi_\uparrow^*\psi_\downarrow\right) , -2\text{Im}\left(  \psi_\downarrow^*\psi_\uparrow \right) , \rho_\text{spin} \right)
\end{align}
where $\rho_\text{spin} = \left| \psi_\uparrow \right|^2 - \left| \psi_\downarrow \right|^2$ is the spin density projected on the spin quantization axis, here chosen to be the $z$-axis. Via a Gordon decomposition\cite{Strange, Eschrig} it is possible to calculate the spin current density
\begin{equation}\label{spincurrent}
  \mathbf{j}_S(\mathbf{r}) = \nabla \times \mathbf{m}(\mathbf{r})  
\end{equation}
and in the further considerations the orbital current density is neglected as we focus on ferromagnetic transition metals with magnetism dominated by the spin, so the index $S$ is dropped in the notation of current density. From the current density, the periodic part of the magnetic vector potential $\mathbf{A}_\text{p}$ is given by the Poisson equation, since Maxwell's equations tell us that 
\begin{equation}
	\nabla \times \mathbf{B} = \mu_0 \mathbf{j}
\end{equation}
but 
\begin{equation}
	\nabla \times \mathbf{B} = \nabla \times \left( \nabla \times \mathbf{A} \right) = \nabla (\nabla \cdot \mathbf{A}) - \Delta \mathbf{A} = - \Delta \mathbf{A},
\end{equation}
in Coulomb gauge ($\nabla \cdot \mathbf{A} = 0$), so
\begin{equation} \label{Poisson}
  \Delta \mathbf{A}(\mathbf{r}) = -\mu_0 \mathbf{j}(\mathbf{r}).
\end{equation}
Clearly, $\Delta \mathbf{A}_\text{np} = 0$ (as the second derivatives of Eq.~\ref{AofConstB} are all zero) whereby Eq.~\ref{Poisson} reads 
\begin{equation} \label{Poisson_p}
  \Delta \mathbf{A}_\text{p}(\mathbf{r}) = -\mu_0 \mathbf{j}(\mathbf{r}).
\end{equation}
and $\mathbf{A}_\text{p}$ can be obtained by solving the Poisson equation with periodic boundary conditions. A unique solution additionally requires knowledge of the value of the vector potential at some point or its volume average, which can be set to an arbitrary value by the remaining gauge freedom\cite{Jackson}. Finally, $\mathbf{B}_\text{p}$ is easily calculated from Eq.~\ref{Bp_Ap}. This procedure uniquely determines $\mathbf{A}$ and $\mathbf{B}$, given the density matrix. Furthermore, the fields have been constructed so that $\mathbf{B}$ fulfills Maxwell's equations with physical boundary conditions (periodicity with a volume average corresponding to $\mu_0 \mathbf{M}$) and so that $\mathbf{A}$ fulfills its defining equation and Coloumb gauge. 

Finally, it is noted that some simplifications occur in the case of collinear magnetism with $\hat{\mathbf{z}}$ defining the spin quantization axis. The magnetization density in Eq.~\ref{magdensity} then simplifies to 
\begin{equation}
	\mathbf{m}(\mathbf{r}) = \mu_\text{B} \rho_\text{spin} \hat{\mathbf{z}}, 
\end{equation}
so the current density in Eq.~\ref{spincurrent}, and hence also $\mathbf{A}$, has non-zero $x$- and $y$-components only. Therefore the Poisson equation needs to be solved only in two dimensions, independently for each value of $z$, which simplifies the numerical work. The result of this methodology applied to ferromagnet FePt and antiferromagnet LaMnAsO  will now be presented. 

\subsubsection{FePt}

FePt in the tetragonal L1$_0$ structure is a ferromagnetic material with lattice parameters $a=2.71~\text{\AA}$ and $c=3.72~\text{\AA}$\cite{Gilbert2013} and a Curie temperature in the vicinity of $700~\text{K}$\cite{Gilbert2013}, which has gained much attention, for example, due to its large magnetocrystalline anisotropy\cite{Gilbert2013, Burkert2005a}. 

\begin{figure}[hbt!]
	\centering
	\includegraphics[width=0.23\textwidth]{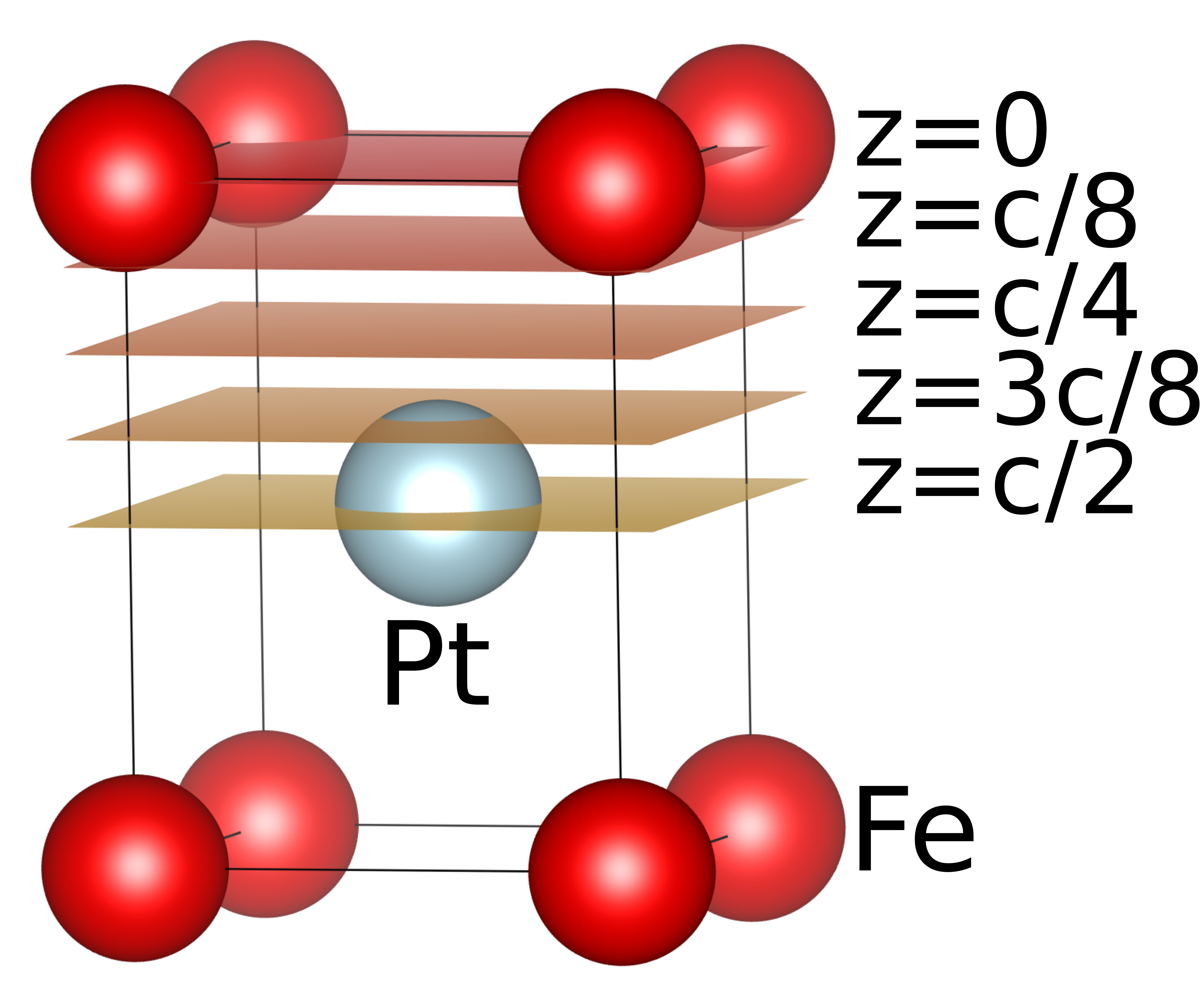}
	\includegraphics[width=0.23\textwidth]{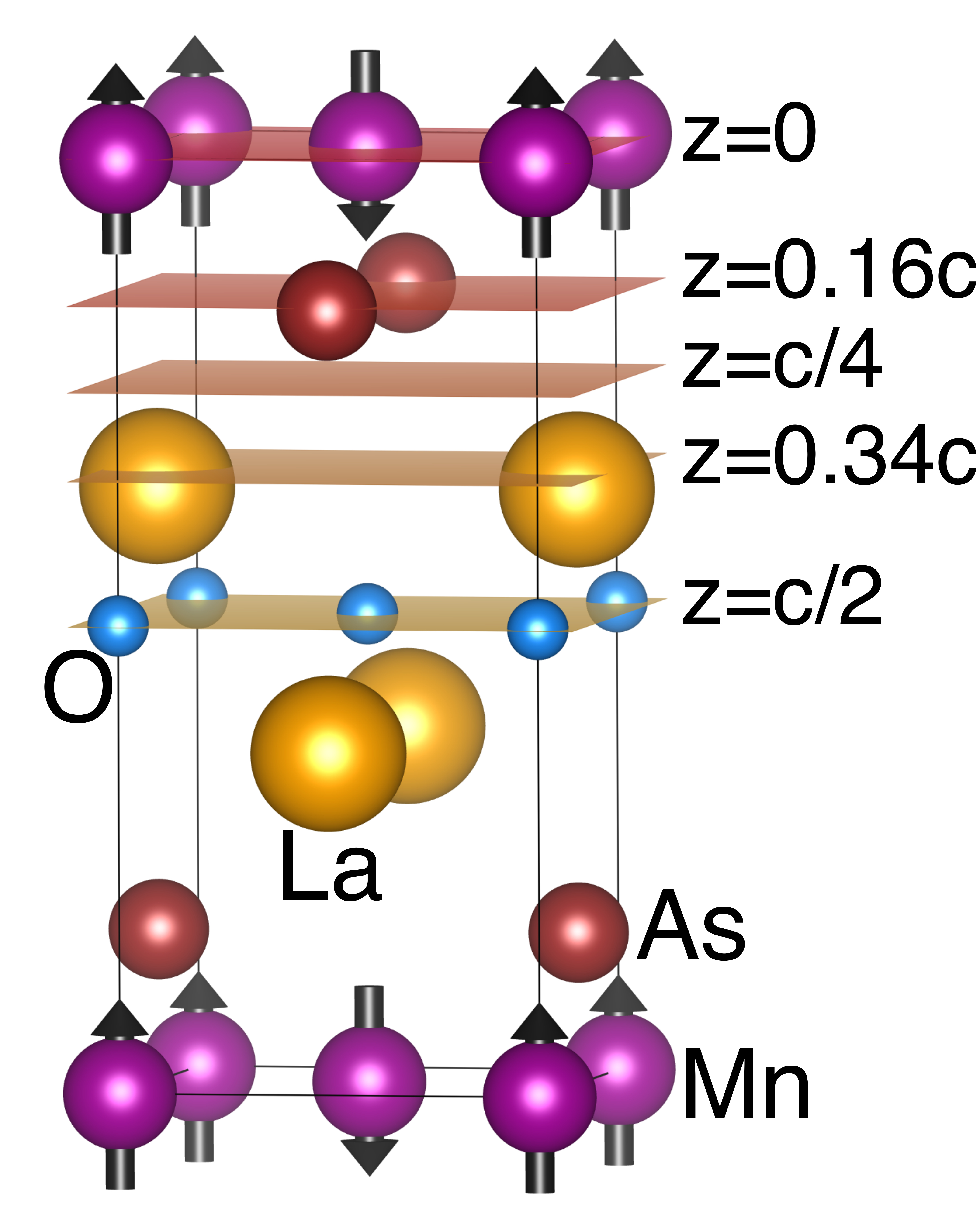}
	\caption{One $a \times a \times c$ sized unit cell of a) FePt in the L1$_0$ structure with planes at $z=0$, $z=\frac{c}{8}$, $z=\frac{c}{4}$, $z=\frac{3c}{8}$ and $z=\frac{c}{2}$ and b) LaMnAsO in the tetragonal crystal structure of space group p4/nmm and planes at $z=0$, $z=0.16c$, $z=\frac{c}{4}$, $z=0.34c$ and $z=\frac{c}{2}$.}
	\label{fig.crystalstructs}
\end{figure}
\begin{figure}[hbt!]
	\centering 
	\includegraphics[width=0.46\textwidth]{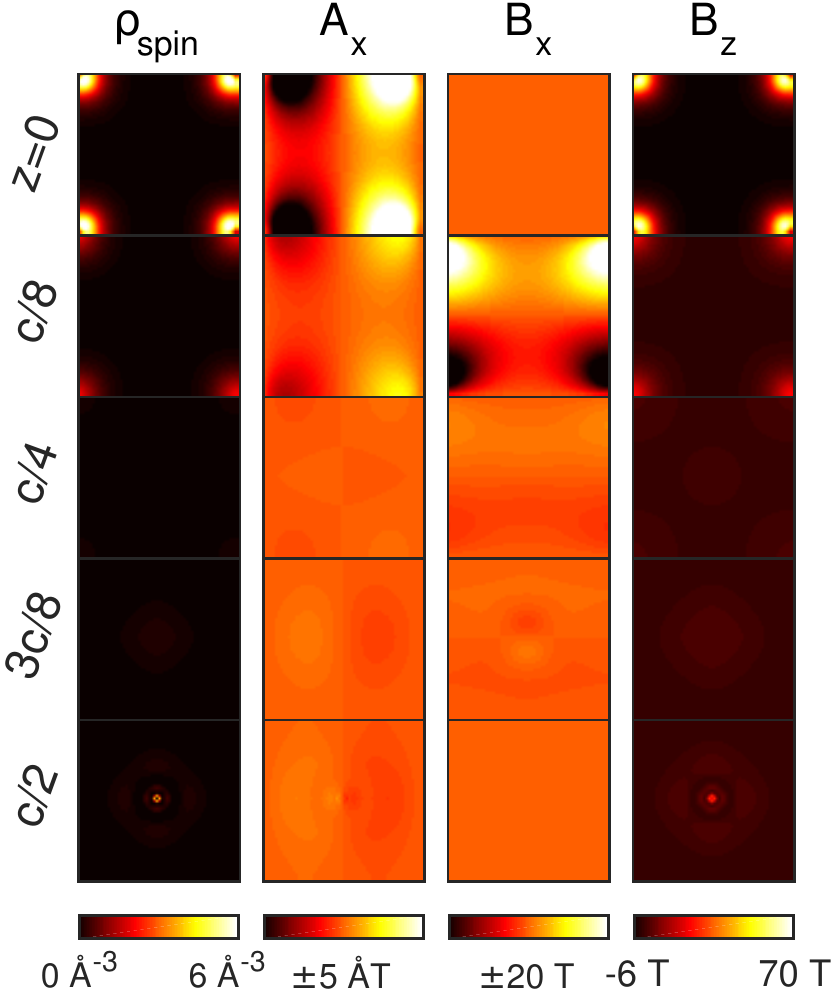}
	\caption{Spin density, $x$-component of the $\mathbf{A}_\text{p}$-field, $x$-component of $\mathbf{B}_\text{p}$-field and $z$-component of $\mathbf{B}_\text{p}$ in a unit cell of L1$_0$ FePt from the procedure described in the text.}
	\label{fig.FePt-fields}
\end{figure}

  
Fig.~\ref{fig.crystalstructs}a) shows one unit cell of FePt with planes at $z=0$, $z=\frac{c}{8}$, $z=\frac{c}{4}$, $z=\frac{3c}{8}$ and $z=\frac{c}{2}$. The spin density of this FePt structure was computed in a collinearly spin polarized DFT calculation using the full-potential linearized augmented plane wave (FP-LAPW)\cite{Blaha2001} method in the generalized gradient approximation (GGA)\cite{Perdew1996}. Calculations were performed with the experimental values of the lattice parameters. Such calculations produce a magnetic moment of $2.93\mu_\text{B}$ on the Fe atom and a smaller induced moment of $0.37\mu_\text{B}$ on the Pt atom, which corresponds to a saturation magnetization of $\mu_0 M = 1.38~\text{T}$, in good agreement with the experimentally reported value of $1.36~\text{T}$\cite{Seki2006}. In the left column of Fig.~\ref{fig.FePt-fields} the spin density is shown in the planes indicated in Fig.~\ref{fig.crystalstructs}a). The remaining columns in Fig.~\ref{fig.FePt-fields} show the $x$-component of $\mathbf{A}_\text{p}$, the $x$-component of $\mathbf{B}_\text{p}$ and finally the $z$-component of $\mathbf{B}_\text{p}$, respectively. The $z$-component of $\mathbf{A}_\text{p}$ is zero because collinear magnetism is considered and as a result of the crystal symmetry, the $y$-components of $\mathbf{A}_\text{p}$ and $\mathbf{B}_\text{p}$ are the same as the $x$-components but rotated by $90^\circ$ about the $z$-axis.
The shape of the $z$-component of $\mathbf{B}_\text{p}$ is very similar to the spin density, but even though only collinear magnetism with a magnetisation density along the $z$-direction is taken into account, $\mathbf{B}_\text{p}$ has non-zero $x$- and $y$-components in contrast to $\mathbf{m}(\mathbf{r})$. Together with the constant saturation field $\mathbf{B}_\text{avg} = \mu_0 M \hat{\mathbf{z}}$, and the non-periodic part $\mathbf{A}_\text{np} = \mu_0 \mathbf{M} \times \mathbf{r} = \mu_0 M \left( -y, x, 0 \right)$, the fields in Fig.~\ref{fig.FePt-fields} are used as input in Sec.~\ref{FePtzmag} and Sec.~\ref{FePtxmag}.

\subsubsection{LaMnAsO}

In Sec.~\ref{LaMnAsOresults} the antiferromagnetic compound LaMnAsO in tetragonal crystal structure (space group p4/nmm) with lattice parameters $a=4.114~\text{\AA}$ and $c=9.030~\text{\AA}$\cite{Emery2011}, illustrated in Fig.~\ref{fig.crystalstructs}b), is studied. This material has two antiferromagnetically coupled Mn atoms with crystallographic positions $\left( 0,0,0 \right)$ and $\left( \frac{1}{2},\frac{1}{2},0 \right)$, i.e. the different columns (in $z$-direction) of Mn atoms have antiparallel spins while the different planes have parallel spins, whereby magnetic STEM imaging over the $xy$-plane should be able to distinguish the different Mn columns. The additional atoms are As at  $\left( \frac{1}{2},0,0.1684 \right)$ and $\left( 0,\frac{1}{2},1-0.1684 \right)$, La at  $\left( 0,\frac{1}{2},0.3674 \right)$ and $\left( \frac{1}{2},0,1-0.3674 \right)$ and O at $\left( 0,0,\frac{1}{2} \right)$ and $\left( \frac{1}{2},\frac{1}{2},\frac{1}{2} \right)$. The Neel temperature of the compound has been experimentally reported as $317~\text{K}$\cite{Emery2010}, although more recent work suggests that this value is due to an impurity whereas the correct Neel temperature of LaMnAsO should be 360 K\cite{PhysRevB.93.054404}. The magnetic moment of Mn at $2~\text{K}$ is $3.54\mu_\text{B}$\cite{Emery2011}. A FP-LAPW calculation in the GGA, with experimental lattice parameters but computationally relaxed internal atomic positions yields a magnetic moment on Mn of $3.51\mu_\text{B}$, in good agreement with the experimental value\cite{LMAOnote}. 

\begin{figure}[hbt!]
	\centering
	\includegraphics[width=0.46\textwidth]{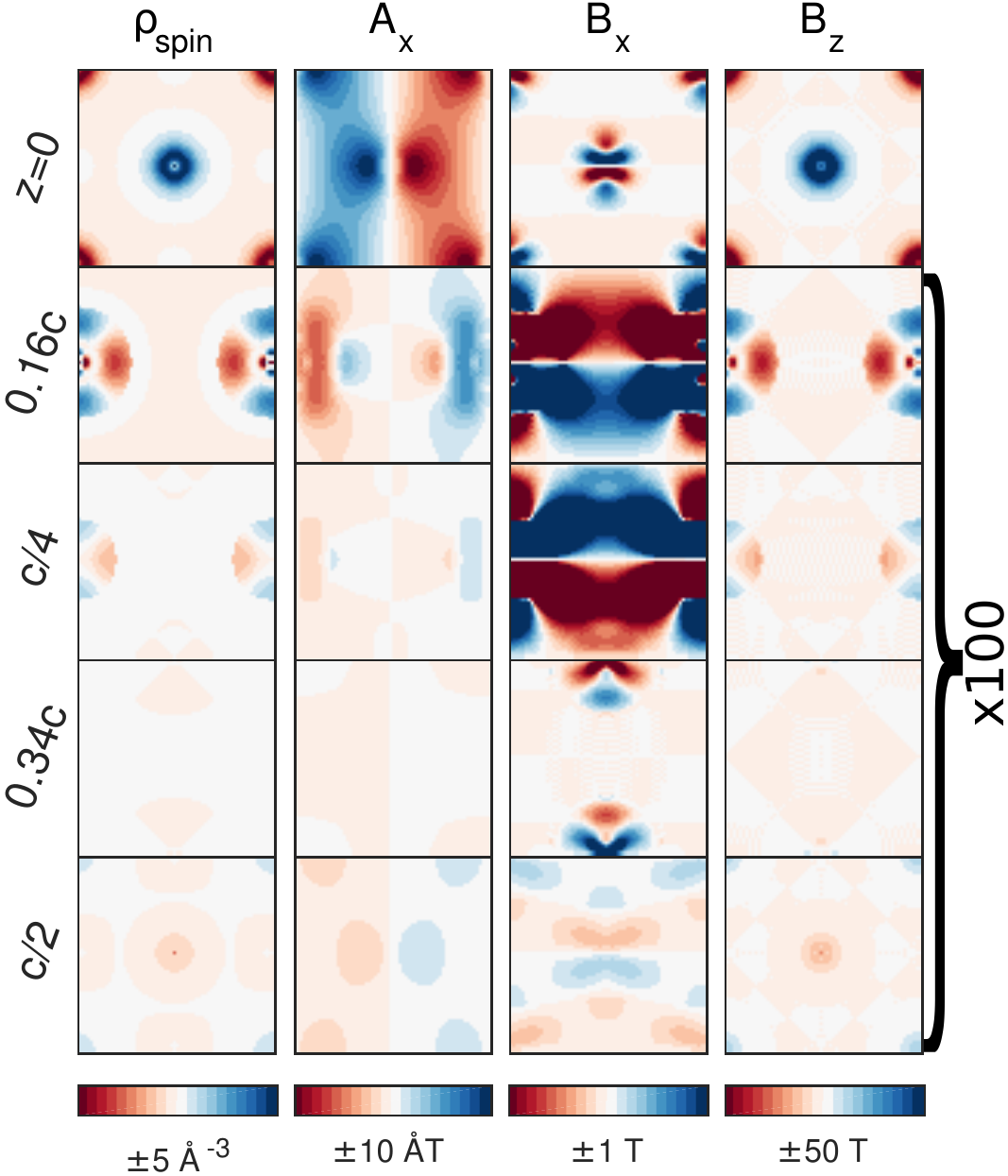}
	\caption{Spin density, $x$-component of the $\mathbf{A}_\text{p}$-field, $x$-component of $\mathbf{B}_\text{p}$-field and $z$-component of $\mathbf{B}_\text{p}$ in a unit cell of LaMnAsO from the procedure described in the text, in the planes where $z=0$, $z=0.16c$, $z=\frac{c}{4}$, $z=0.34c$ and $z=\frac{c}{2}$. The numbers in the planes not containing Mn are multiplied by 100 for visibility. }
	\label{fig.LMAO-fields}
\end{figure}

Fig.~\ref{fig.LMAO-fields} presents to same type of data as in Fig.~\ref{fig.FePt-fields} but for LaMnAsO, with the first row showing the spin density as obtained from the FP-LAPW GGA calculation in the planes where $z=0$, $z=0.16c$, $z=\frac{c}{4}$, $z=0.34c$ and $z=\frac{c}{2}$ as illustrated in Fig.~\ref{fig.crystalstructs}b).
Observations which can be made in Fig.~\ref{fig.LMAO-fields} are similar as in Fig.~\ref{fig.FePt-fields}. For example, the shape of the $\mathbf{B}$-field is very similar to the spin density again. Most of the spin density is located in the plane of the Mn atoms and it can be seen that also the small induced spin moments on the O atoms are antiparallel to that on the Mn atoms in the same atomic columns. Furthermore, it can be seen that even though both La and As have magnetic moments which are identically zero, there is a non-zero distribution of spin density around the atoms, however, as it integrates to zero over a sphere centered at the atoms, there is a net moment of zero. Nevertheless, this spin density yields a non-zero magnetic field around these atoms, which is, however, 2-3 orders of magnitude weaker than the field around the Mn atoms.

\section{Results}\label{results}

In this section, results of numerical simulations based on the methodology introduced in the preceding parts of the text, will be presented. First, results of calculations for FePt with magnetization parallel ($z$-direction) or perpendicular ($x$-direction) to the propagation direction are presented in Secs.~\ref{FePtzmag}-\ref{FePtxmag}, respectively, and then results for LaMnAsO are shown in Sec.~\ref{LaMnAsOresults}. The results from calculations with aberrated electron beams are presented in Sec.~\ref{aberrations}. The calculations were performed with the $\mathbf{A}(\mathbf{r})$ and $\mathbf{B}(\mathbf{r})$ input as presented in the Sec.~\ref{MagFields}, while the electrostatic potential, $V(\mathbf{r})$, was taken from tabulated data\cite{kirkland}. All electron vortex beams are generated as disks in reciprocal space, according to
\begin{equation}
\psi_l (k_\perp, \phi_k) \sim \ee^{\img l \phi_k} \Theta(q_\text{max} - k_\perp), 
\label{eq.vortex}
\end{equation}
where $k_\perp = \sqrt{k_x^2 + k_y^2}$ and $\phi_k$ are cylindrical coordinates and $q_\text{max}$ the beam size in reciprocal space, which can be related to convergence angle, $\alpha$, and wavelength, $\lambda$, according to $q_\text{max} = \frac{\alpha}{\lambda}$.

\subsection{FePt with Magnetization Parallel to the Propagation Direction}\label{FePtzmag}

Multislice simulations were performed for an FePt system with $60\times60~\text{u.c.}^2 = 16.26\times16.26~\text{nm}^2$ in the $xy$-directions and thicknesses up to $t=200~\text{u.c.} = 74.4~\text{nm}$ with various combinations of acceleration voltage, $V_\text{acc}$, convergence angle, $\alpha$ and OAM, $l$, listed in Table~\ref{table1}, where also the corresponding wavelengths, $\lambda$, of the electron beams are given. Each unit cell was discretized on a $64\times64\times32$ grid. Parameter combinations with $\alpha=6$ and $l \geq 20$ were excluded as they result in large beam size requiring larger supercell sizes. 
\begin{table}
\caption{\label{table1}Table containing the parameter values of the acceleration voltage $V_\text{acc}$ and its corresponding wavelength, $\lambda$, convergence angle, $\alpha$ and OAM, $l$ for which calculations were performed for the FePt system.}
\begin{ruledtabular}
\begin{tabular}{lr}
Parameter & Values\\
\hline
$V_\text{acc}$ (kV) & 60 100 200 300 1000 \\
$\lambda$ (pm) & 4.87 3.70 2.51 1.97 0.87\\
$\alpha$ (mrad)& 6 15 30 60 \\
$l$ & 0 1 2 4 5 10 20 30 \\
\end{tabular}
\end{ruledtabular}
\end{table}
The calculations for non-zero $l$ were performed with a beam initially spin polarized with spin up electrons in the propagation ($z$) direction, since the OAM magnetic interaction should be very similar regardless of spin\cite{edstrom16}. The $l=0$ calculations were performed both with spin up and spin down beams in order to look at spin effects. In Sec.~\ref{FePtxmag} we investigate the case where the magnetization direction is perpendicular to the beam spin quantization axis instead of parallel. Calculations were performed for beam position $(x,y) = (0,0)$, which is on a column of Fe atoms. For beams with spatial extent significantly beyond atomic distances, the beam position is not important\cite{edstrom16}. For smaller beams the position is important and will be explored in the context of atomic resolution STEM imaging later in this section as well as in Sec.~\ref{LaMnAsOresults}.  

\begin{figure}[hbt!]
	\centering
	\begin{subfigure}[b]{0.23\textwidth}
        \includegraphics[width=\textwidth]{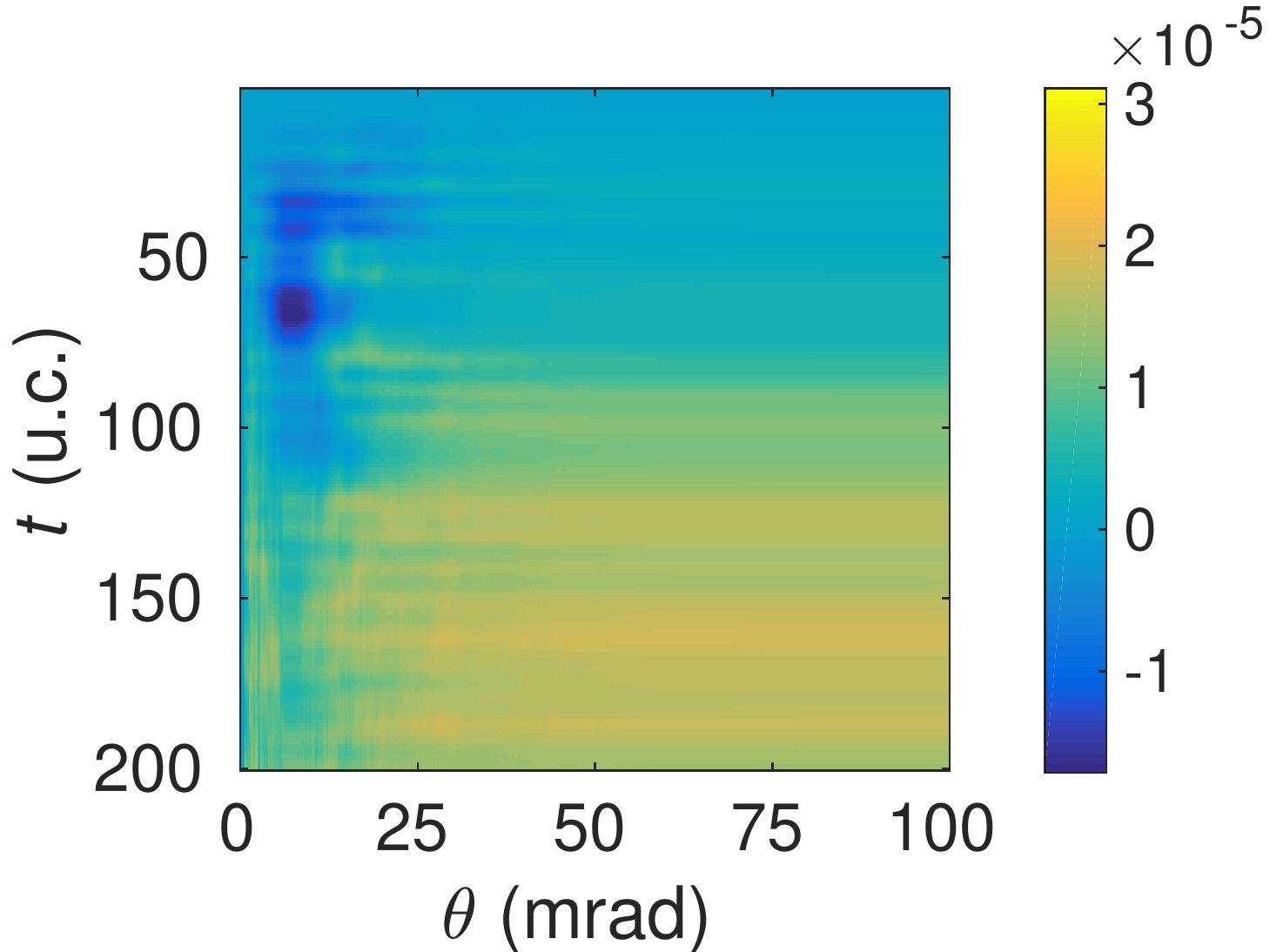}
        \caption{$V_\text{acc} = 100~\text{kV}$, $\alpha = 6~\text{mrad}$, $l=\pm1$.}
    \end{subfigure}
    \begin{subfigure}[b]{0.23\textwidth}
        \includegraphics[width=\textwidth]{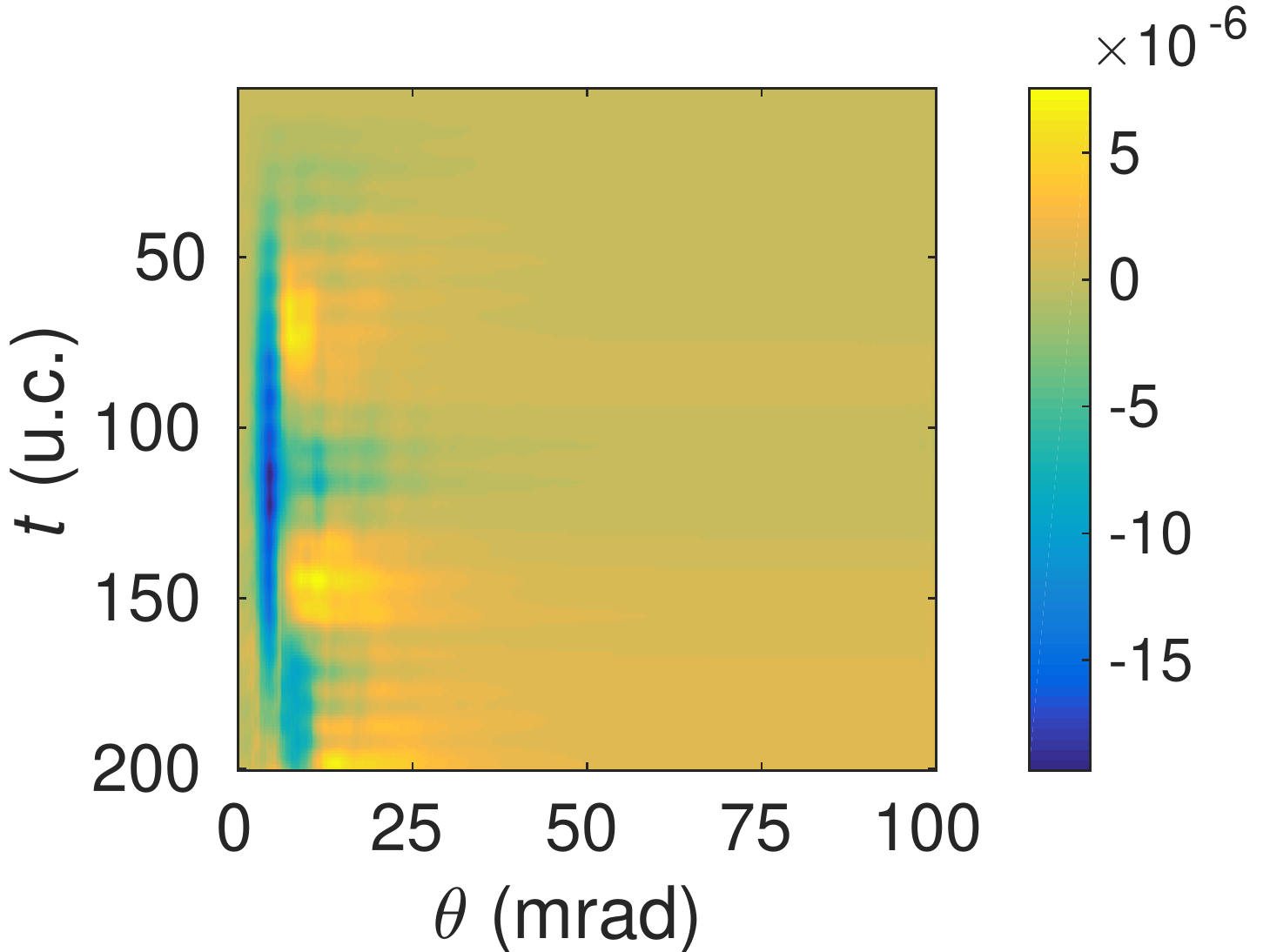}
        \caption{$V_\text{acc} = 300~\text{kV}$, $\alpha = 6~\text{mrad}$, $l=\pm1$.}
    \end{subfigure}
	\begin{subfigure}[b]{0.23\textwidth}
        \includegraphics[width=\textwidth]{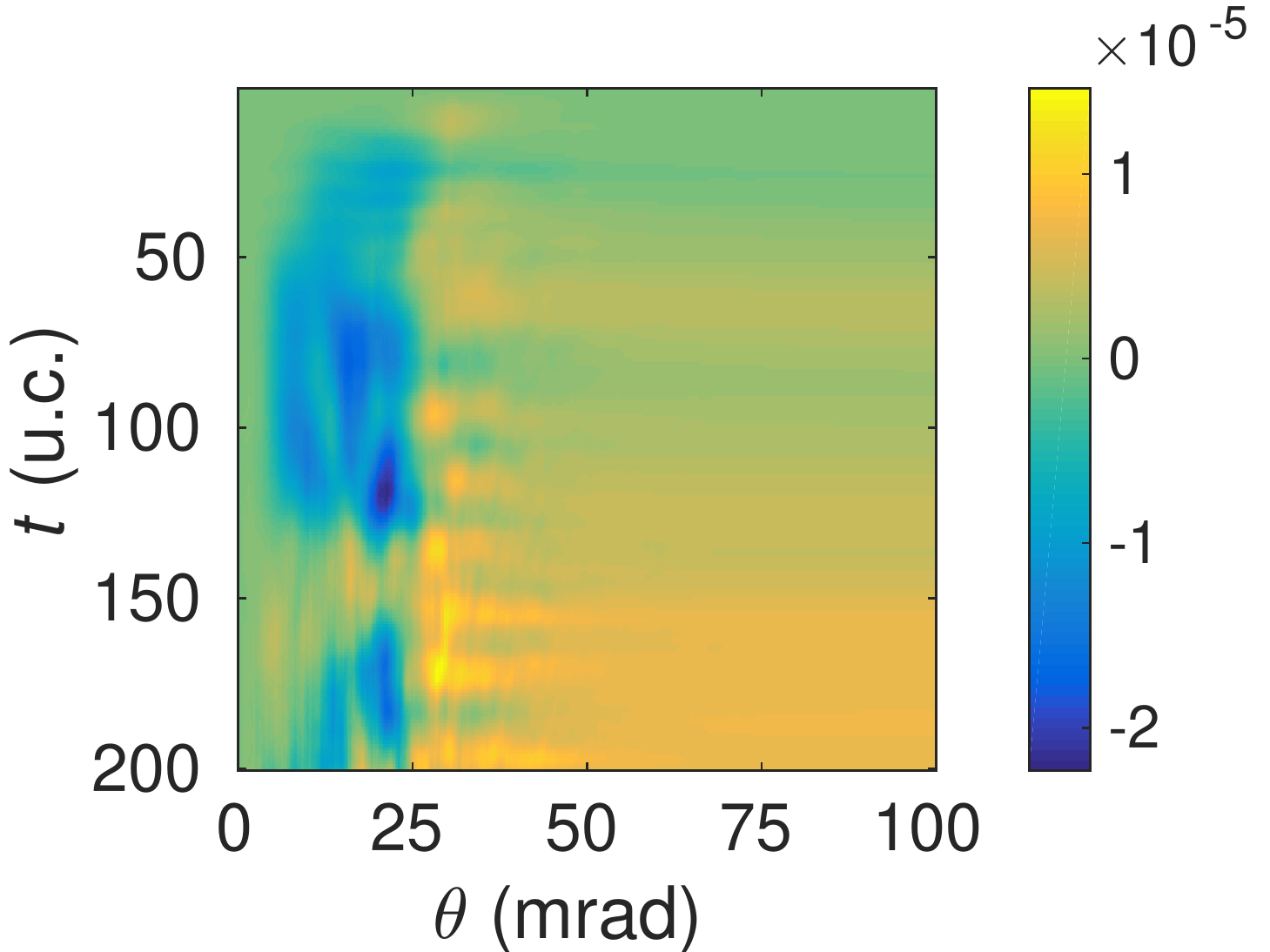}
        \caption{$V_\text{acc} = 100~\text{kV}$, $\alpha = 30~\text{mrad}$, $l=\pm1$.}
    \end{subfigure}
    \begin{subfigure}[b]{0.23\textwidth}
        \includegraphics[width=\textwidth]{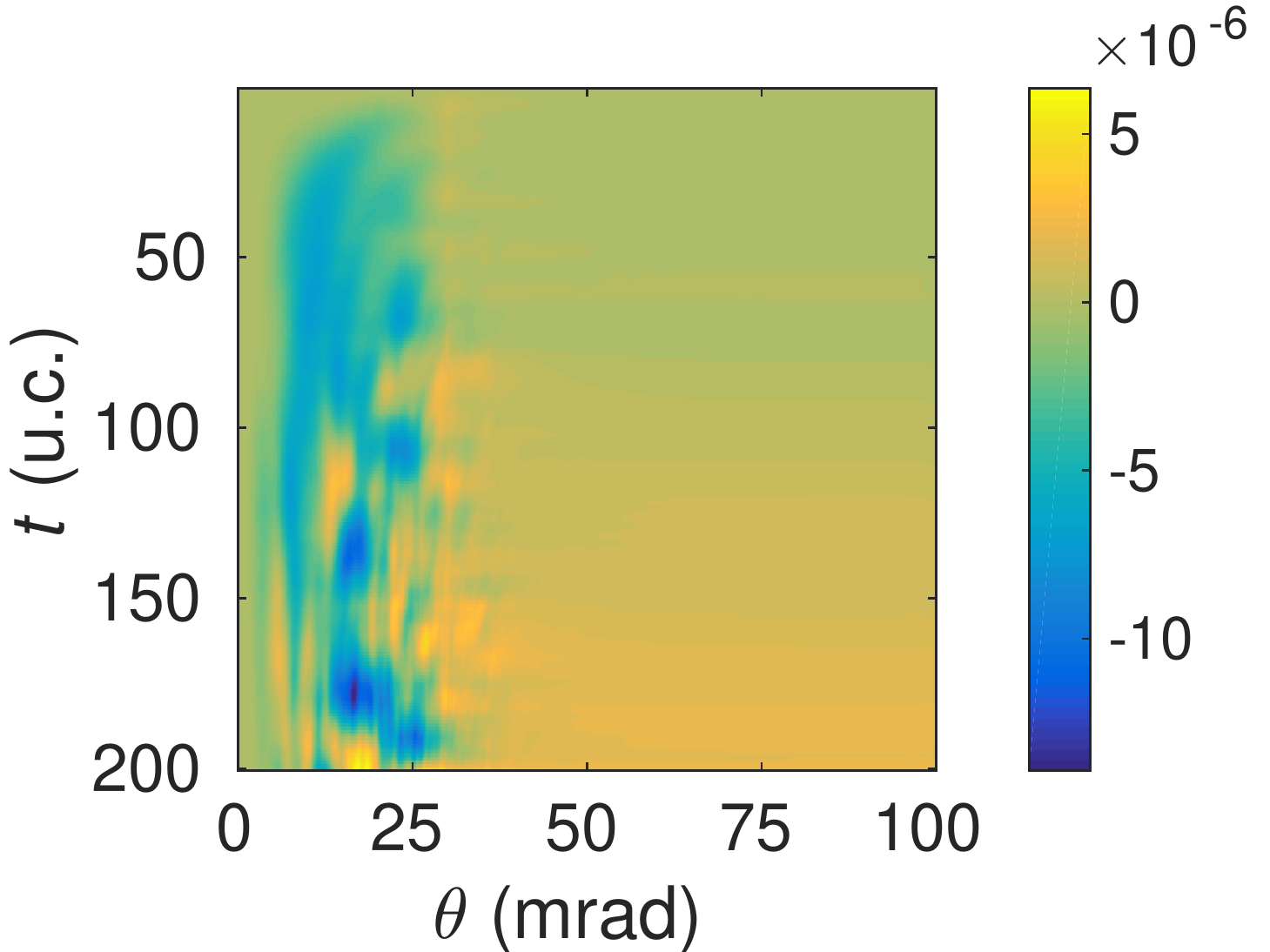}
        \caption{$V_\text{acc} = 300~\text{kV}$, $\alpha = 30~\text{mrad}$, $l=\pm1$.}
    \end{subfigure}  
	\begin{subfigure}[b]{0.23\textwidth}
        \includegraphics[width=\textwidth]{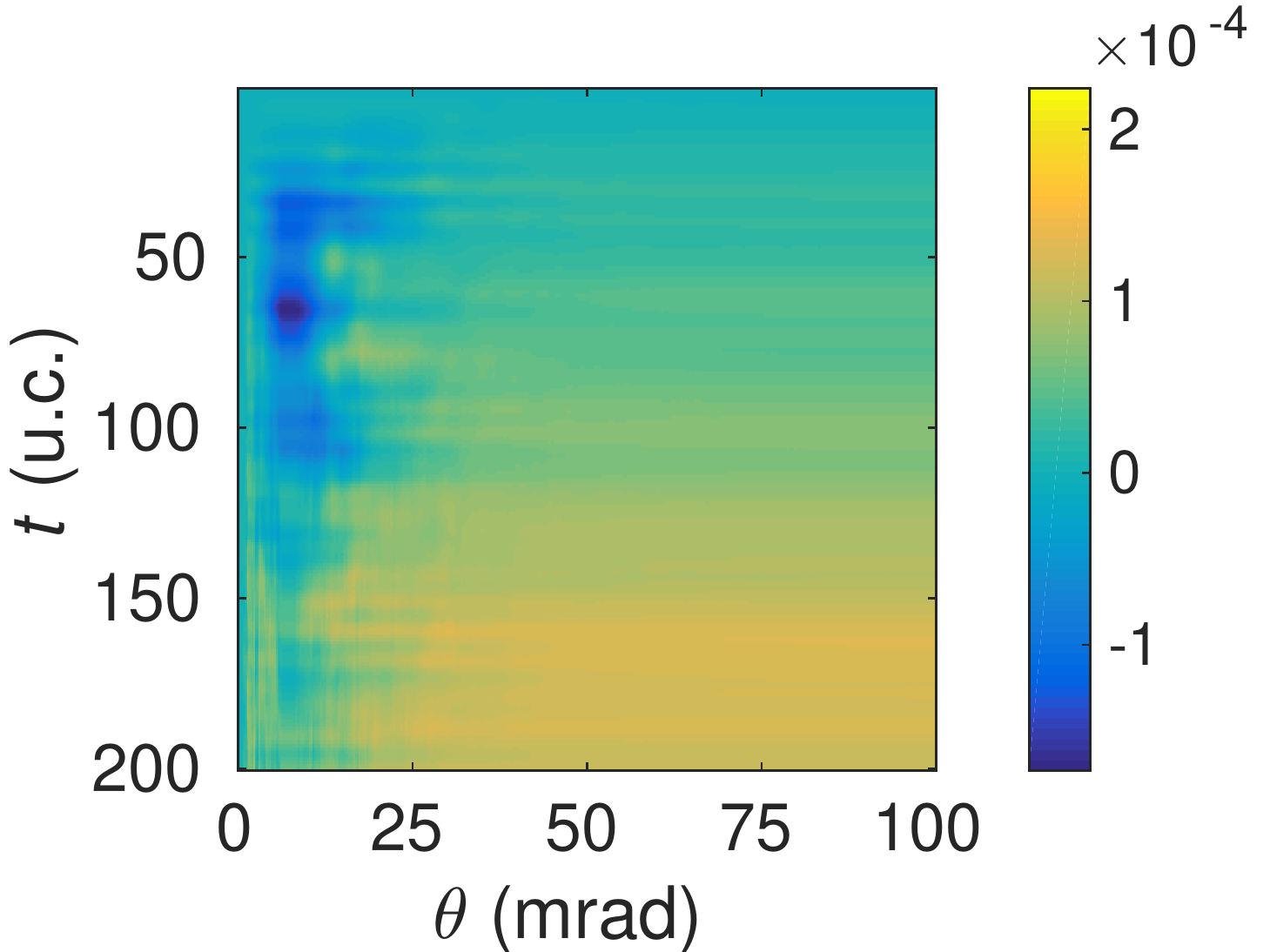}
        \caption{$V_\text{acc} = 100~\text{kV}$, $\alpha = 6~\text{mrad}$, $l=\pm10$.}
    \end{subfigure}
    \begin{subfigure}[b]{0.23\textwidth}
        \includegraphics[width=\textwidth]{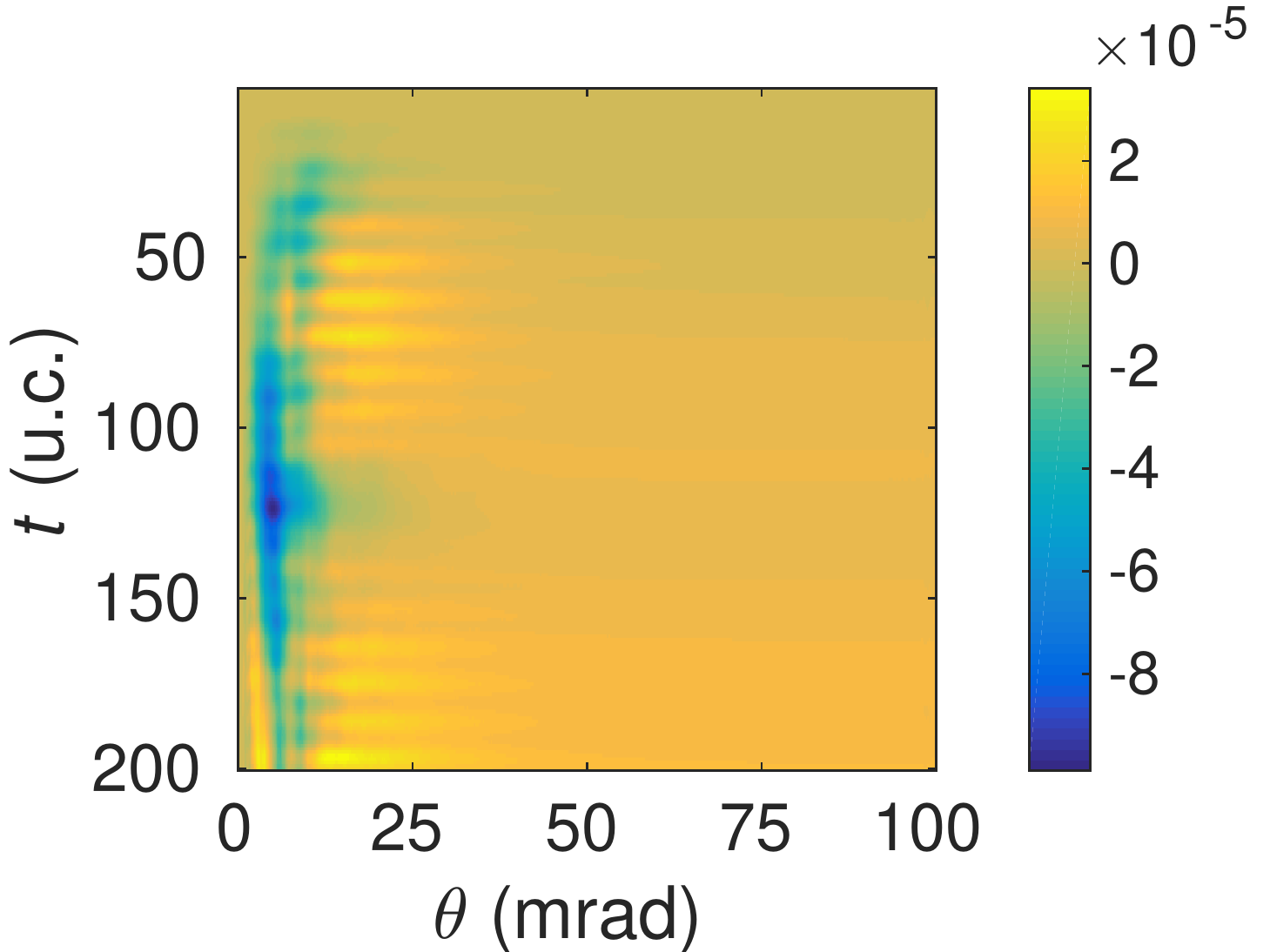}
        \caption{$V_\text{acc} = 300~\text{kV}$, $\alpha = 6~\text{mrad}$, $l=\pm10$.}
        \label{fig.V300alpha6l10}
    \end{subfigure}
	\begin{subfigure}[b]{0.23\textwidth}
        \includegraphics[width=\textwidth]{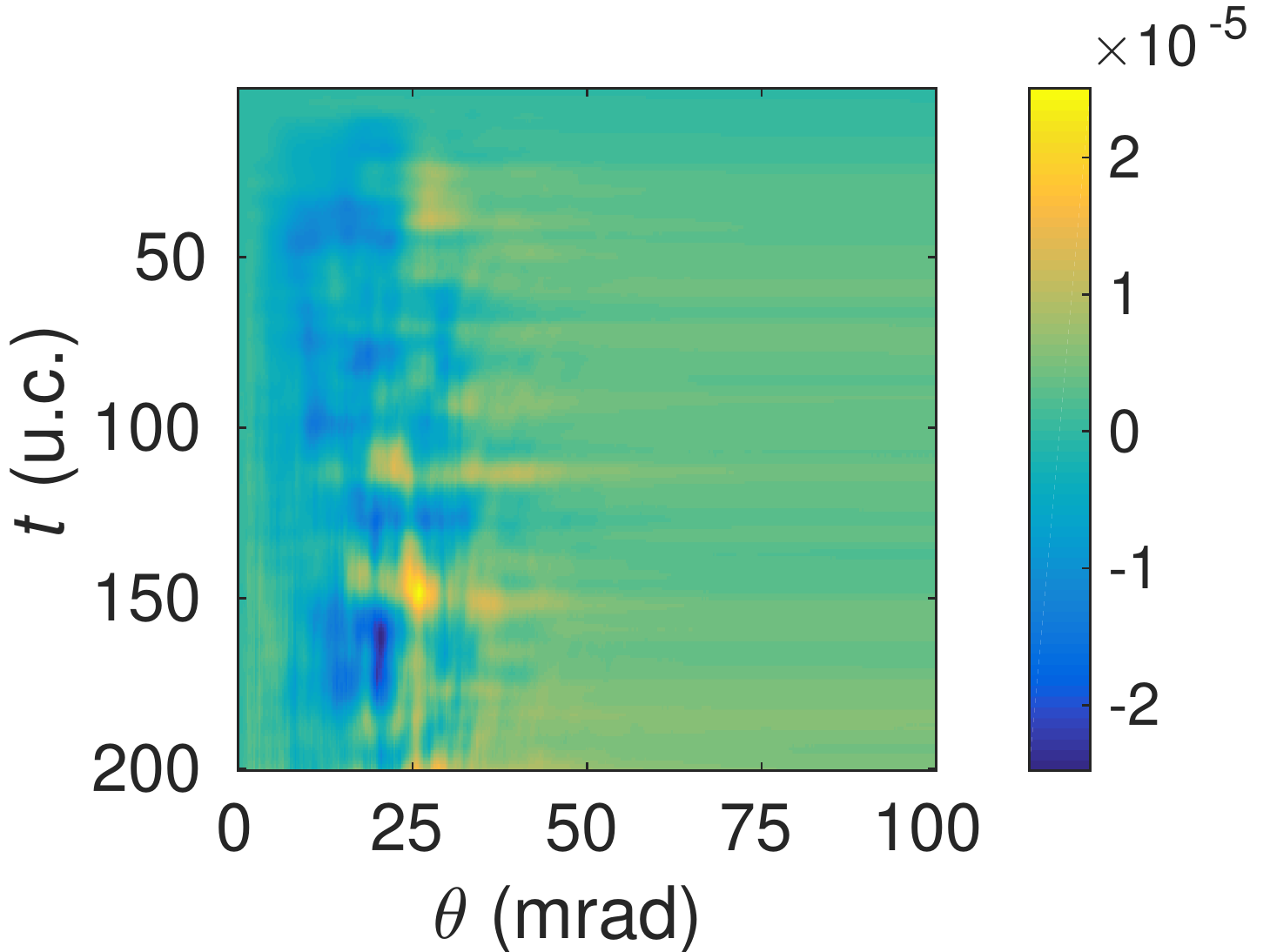}
        \caption{$V_\text{acc} = 100~\text{kV}$, $\alpha = 30~\text{mrad}$, $l=\pm10$.}
    \end{subfigure}
    \begin{subfigure}[b]{0.23\textwidth}
        \includegraphics[width=\textwidth]{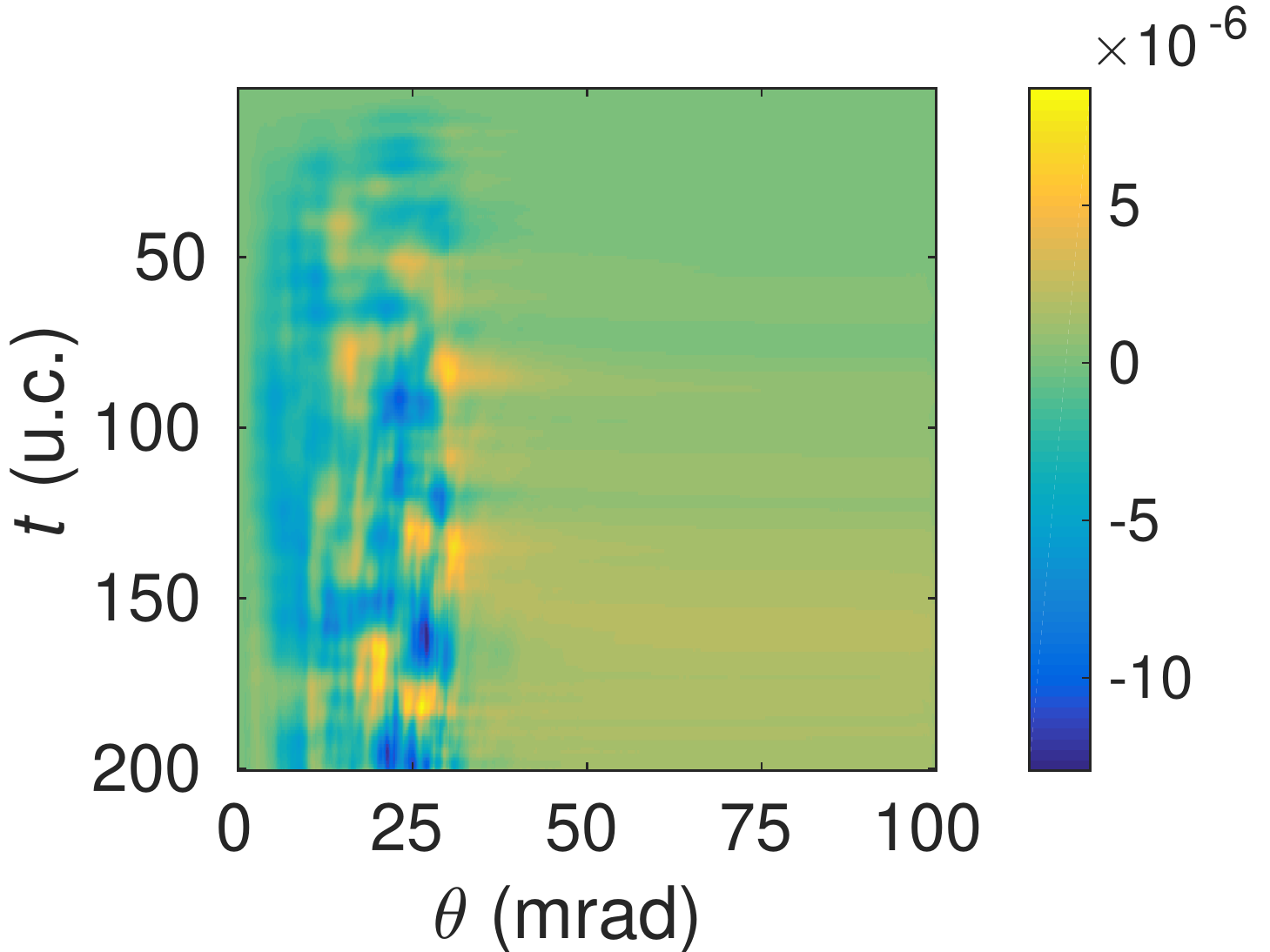}
        \caption{$V_\text{acc} = 300~\text{kV}$, $\alpha = 30~\text{mrad}$, $l=\pm10$.}
    \end{subfigure}        
	\begin{subfigure}[b]{0.23\textwidth}
        \includegraphics[width=\textwidth]{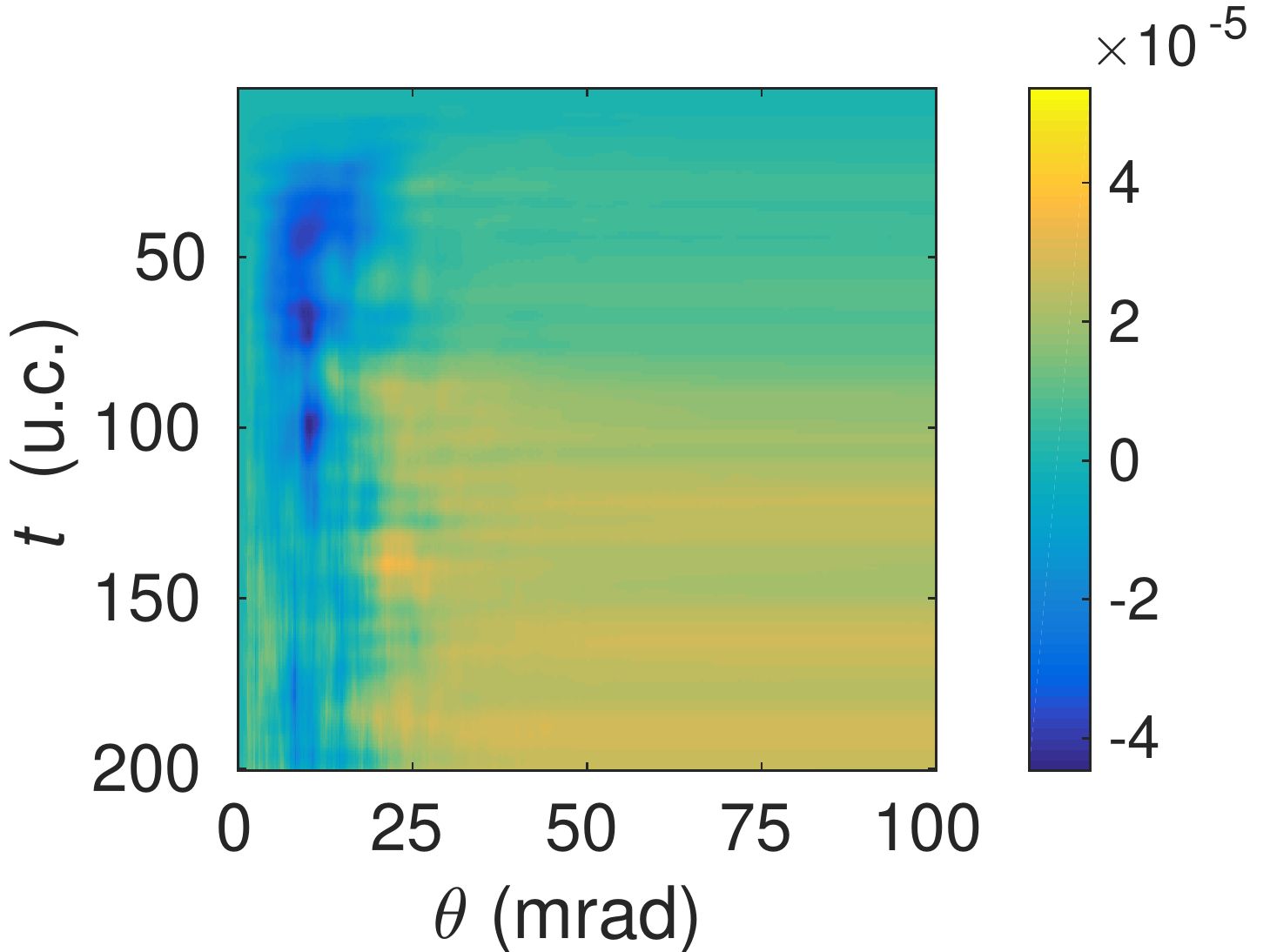}
        \caption{$V_\text{acc} = 100~\text{kV}$, $\alpha = 15~\text{mrad}$, $l=\pm10$.}
    \end{subfigure}
    \begin{subfigure}[b]{0.23\textwidth}
        \includegraphics[width=\textwidth]{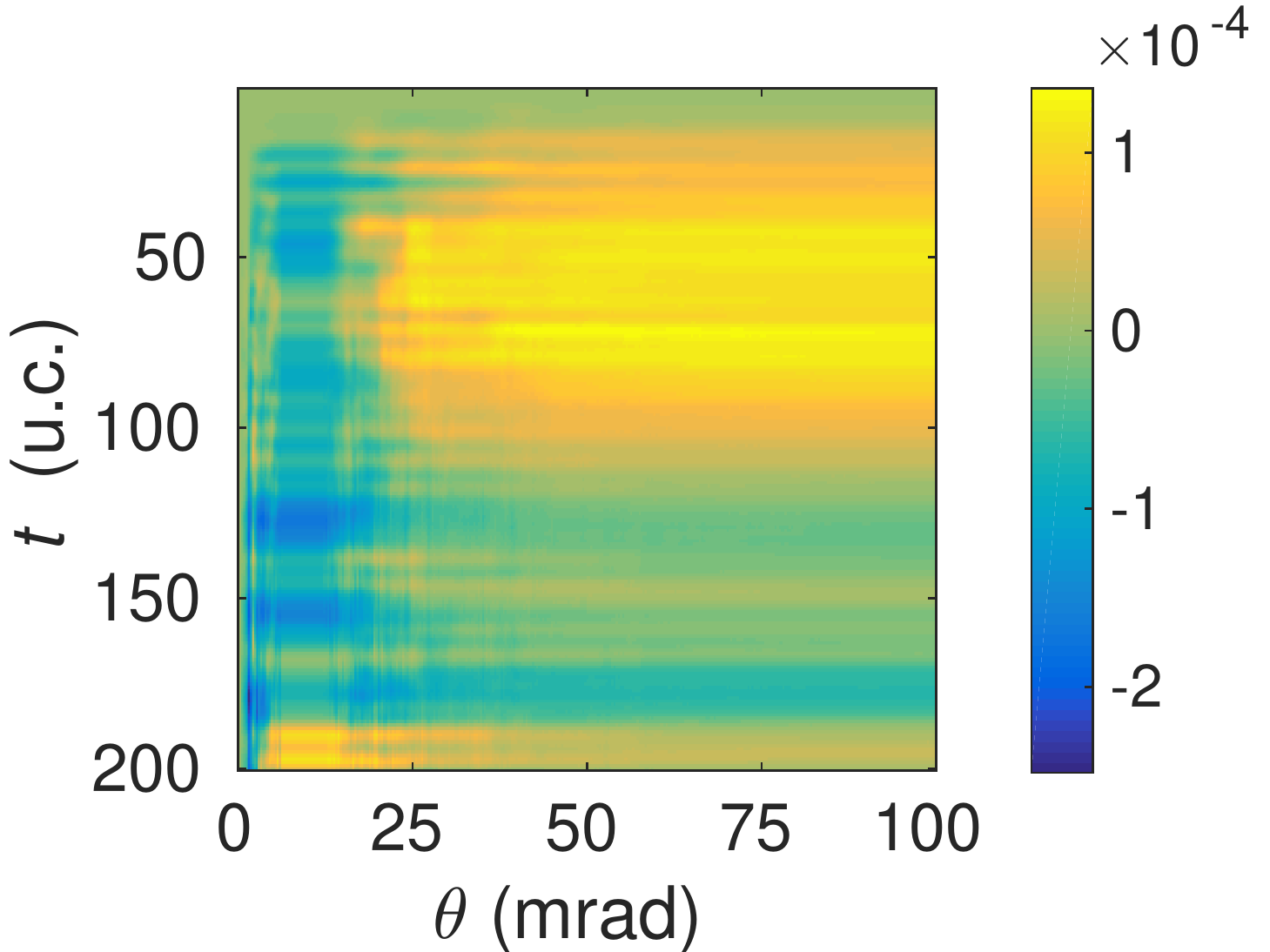}
        \caption{$V_\text{acc} = 60~\text{kV}$, $\alpha = 6~\text{mrad}$, $l=\pm10$.}
    \end{subfigure}     
	\caption{OAM magnetic signal, as functions of sample thickness $t$ and collection angle, $\theta$, for various OAM $l$, acceleration voltages $V_\text{acc}$ and convergence angles $\alpha$.}
	\label{fig.FePtsigoftandtheta}
\end{figure}

After every unit cell the radial intensity distribution, i.e. integral of the diffraction pattern over a disk shaped region up to some maximum collection angle $\theta$, was computed since the difference between such distributions for opposite values of OAM yields an OAM magnetic signal\cite{edstrom16}. Such a signal may be experimentally measured in the TEM by conventional annular detectors in a straight forward manner. The magnetic signal can then easily be studied as function of both collection angles and thickness for each parameter combination in Table~\ref{table1}. Plots of the OAM magnetic signal as function of the collection angle and sample thickness are shown for some example parameters, specified in the captions, in Figs.~\ref{fig.FePtsigoftandtheta}. In all cases one can see a magnetic signal reaching magnitudes of order $10^{-5}$ varying strongly with both thickness and collection angle. Some general traits that can be observed regarding the regions with most magnetic signal include that higher acceleration voltages moves these regions to larger sample thicknesses while larger convergence angles spreads out this region over larger collection angles. Both of these observations are expected because larger acceleration voltage, roughly speaking, will effectively make the sample appear thinner, while a larger convergence angle will cause the beam to spread out more. Still it will most likely be difficult to easily predict where the largest magnetic signals can be found for a given set of parameters. Therefore, computational studies such as that presented here will be an important help to experimental work attempting to detect such a magnetic signal. Another observation in several of the figures, e.g. Fig.~\ref{fig.V300alpha6l10}, is that there is a region with relatively large negative magnetic signal for small collection angles, just under $10~\text{mrad}$, while there is a smaller positive signal spread out over larger collection angles. This behavior results from the from the predominantly positive magnetic coupling term in case of the negative OAM beam reducing its lateral momentum with respect to the positive OAM beam possessing a negative magnetic coupling term. Thus the magnetic field has a tendency to localize the negative OAM beam and delocalize the positive OAM beam around the propagation axis. Note that at sufficiently large collection angles the magnetic signal is expected to go to zero since it is obtained as a difference between two intensities, which are both normalized and should approach one at large collection angles. However, in the plots shown here the collection angles are restricted to $100~\text{mrad}$ for two reasons; firstly for better visibility of the interesting region with strongest magnetic signal at rather small collection angles, typically around $25~\text{mrad}$ or smaller, and secondly because the computational methods used are less accurate for very large scattering angles, rendering such data less reliable\cite{PhysRevB.92.235114}.

\begin{figure}[hbt!]
	\centering
	\begin{subfigure}[b]{0.4\textwidth}
        \includegraphics[width=\textwidth]{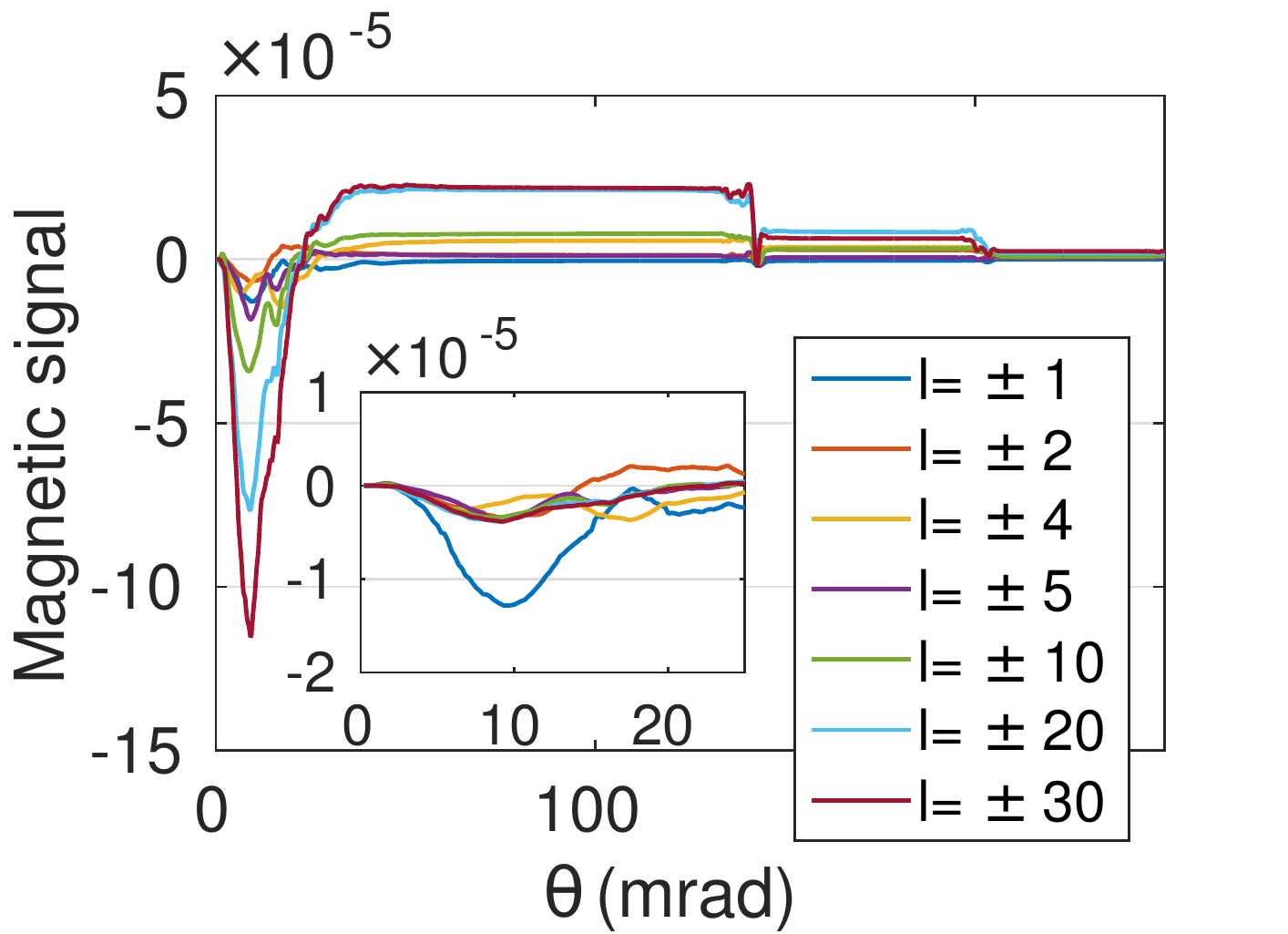}
        \caption{$V_\text{acc} = 100~\text{kV}$, $\alpha = 15~\text{mrad}$, varying $l$.}
        \label{fig.varyonepar_l}
    \end{subfigure}
	\begin{subfigure}[b]{0.4\textwidth}
        \includegraphics[width=\textwidth]{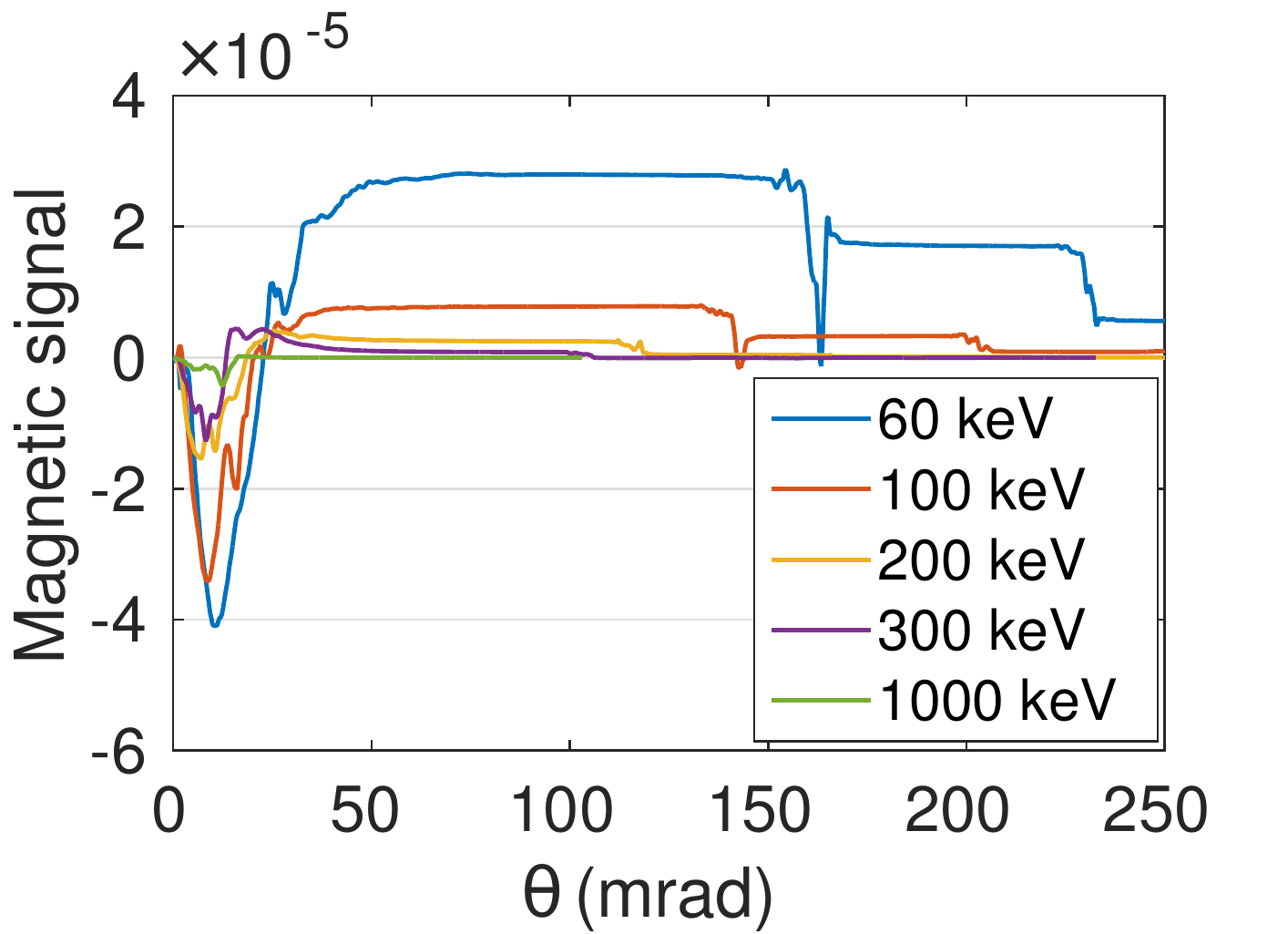}
        \caption{$\alpha = 16~\text{mrad}$, $l=\pm10$, varying $V_\text{acc}$.}
        \label{fig.varyonepar_V}
    \end{subfigure} 
    \begin{subfigure}[b]{0.4\textwidth}
        \includegraphics[width=\textwidth]{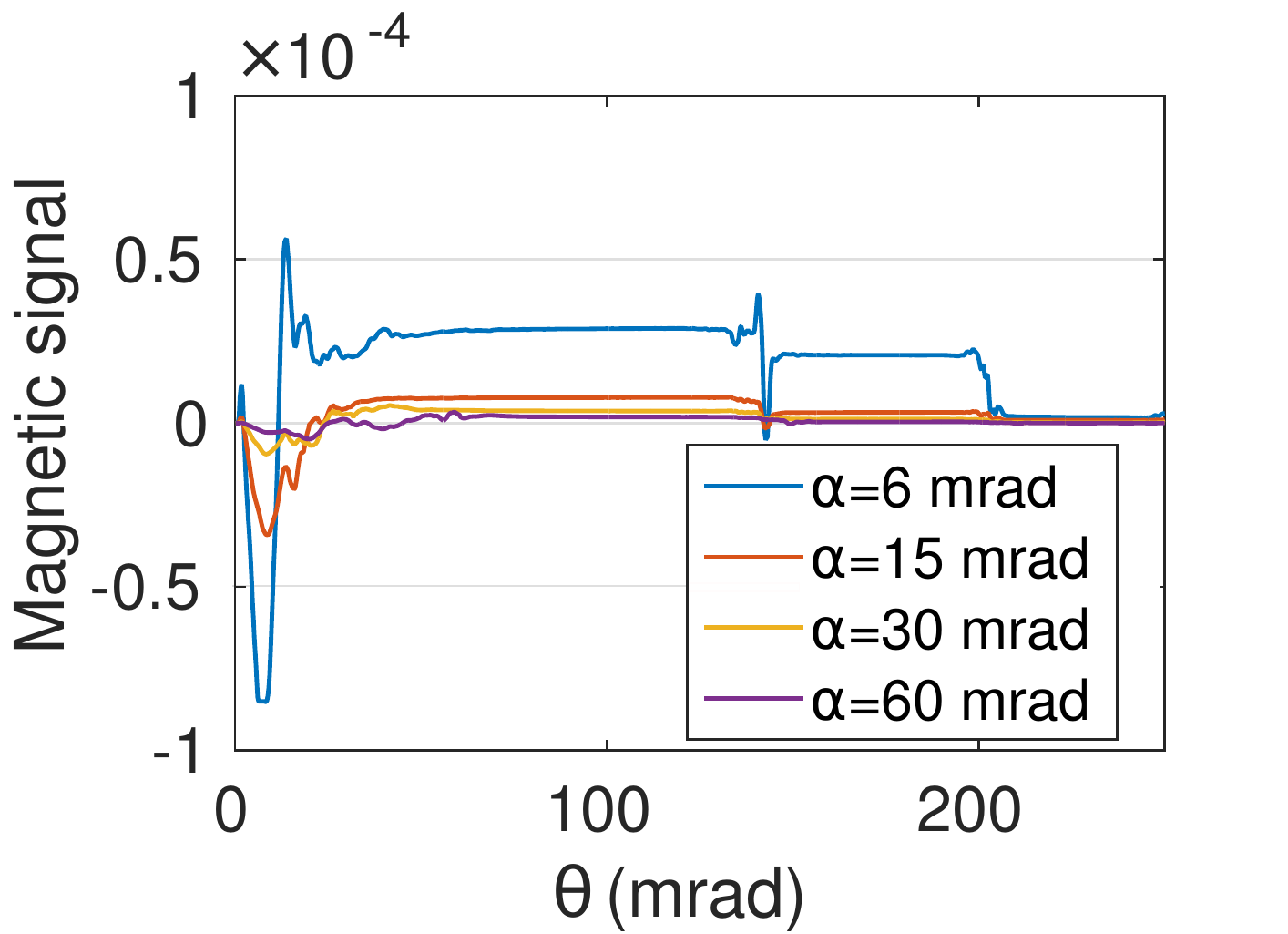}
        \caption{$V_\text{acc} = 100~\text{kV}$, $l=\pm10$, varying $\alpha$.}
        \label{fig.varyonepar_conv}
    \end{subfigure}
	\caption{OAM magnetic signal as function of collection angle after 50 unit cells of FePt for various beam parameters. The inset in a) shows magnetic signals normalized w.r.t. to the OAM, i.e. magnetic signal divided by $l$.}
	\label{fig.varyonepar}
\end{figure}

For better insight into how the magnetic signal varies with a given parameter while keeping others fixed, plots of the magnetic radial profiles at a fixed thickness of $t=50~\text{u.c.}=18.6~\text{nm}$ are shown in Fig.~\ref{fig.varyonepar}, varying either the initial OAM $l$ in Fig.~\ref{fig.varyonepar_l}, the acceleration voltage $V_\text{acc}$ in Fig.~\ref{fig.varyonepar_V} or the convergence angle $\alpha$ in Fig.~\ref{fig.varyonepar_conv}, while keeping the others fixed at values specified in the captions. 
Fig.~\ref{fig.varyonepar_l} shows an increasing magnetic signal with large OAM. As has been pointed out before\cite{edstrom16}, with large OAM and correspondingly large beam size there is a proportionality between magnetic signal and OAM as expected also from the discussions in Sec.~\ref{constantB}. This is clearly illustrated in the inset, which shows the magnetic signal divided by the magnitude of the OAM. For $l=5$ or larger the curves fall almost on top of each other while the deviation for smaller $l$ is due to the increasing interaction of the localized vortex probe with the periodic non-uniform part of the magnetic field rather than the uniform part corresponding to the saturation magnetization.. 

In Fig.~\ref{fig.varyonepar_V} one can observe a trend that a stronger magnetic signal is obtained with lower acceleration voltages. It was pointed out previously\cite{edstrom16} that one possible reason for this is a factor $\gamma^{-1}$ appearing in connection to the magnetic terms in Eq.~\ref{paraxialSeq} but not with the electrostatic potential. As $\gamma$ increases with $V_\text{acc}$, the magnetic effects are therefore expected to decrease with larger $V_\text{acc}$. However, multiplying each of the curves in Fig.~\ref{fig.varyonepar_V} with the corresponding $\gamma$ (not shown), will only result in a rather small change with most of the increase in magnetic signal at smaller $V_\text{acc}$ remaining, indicating that this is only part of the explanation. As larger acceleration voltages and faster electrons should effectively make the sample appear thinner, another part of the explanation for the increase in magnetic signal with smaller acceleration voltage seen here can also be due to the sample thickness being fixed and hence a thicker sample might be needed to accumulate the same magnitude magnetic signal for larger acceleration voltages. Furthermore, the acceleration voltage affects the beam width as a beam with higher energy will be of smaller spatial extent. It is possible that this also affects the magnetic signal and that a beam of larger spatial extent will interact more strongly with the magnetism in the sample. This idea is supported by looking at Fig.~\ref{fig.varyonepar_conv}, where the convergence angle $\alpha$ is varied. Here the trend appears to be that smaller convergence angles, i.e. larger beam sizes, yields a stronger magnetic signal. Unfortunately, this is inauspicious for atomic resolution imaging of magnetism via elastic scattering of electron vortex beams. 

\begin{figure}[hbt!]
	\centering
	\includegraphics[width=0.48\textwidth]{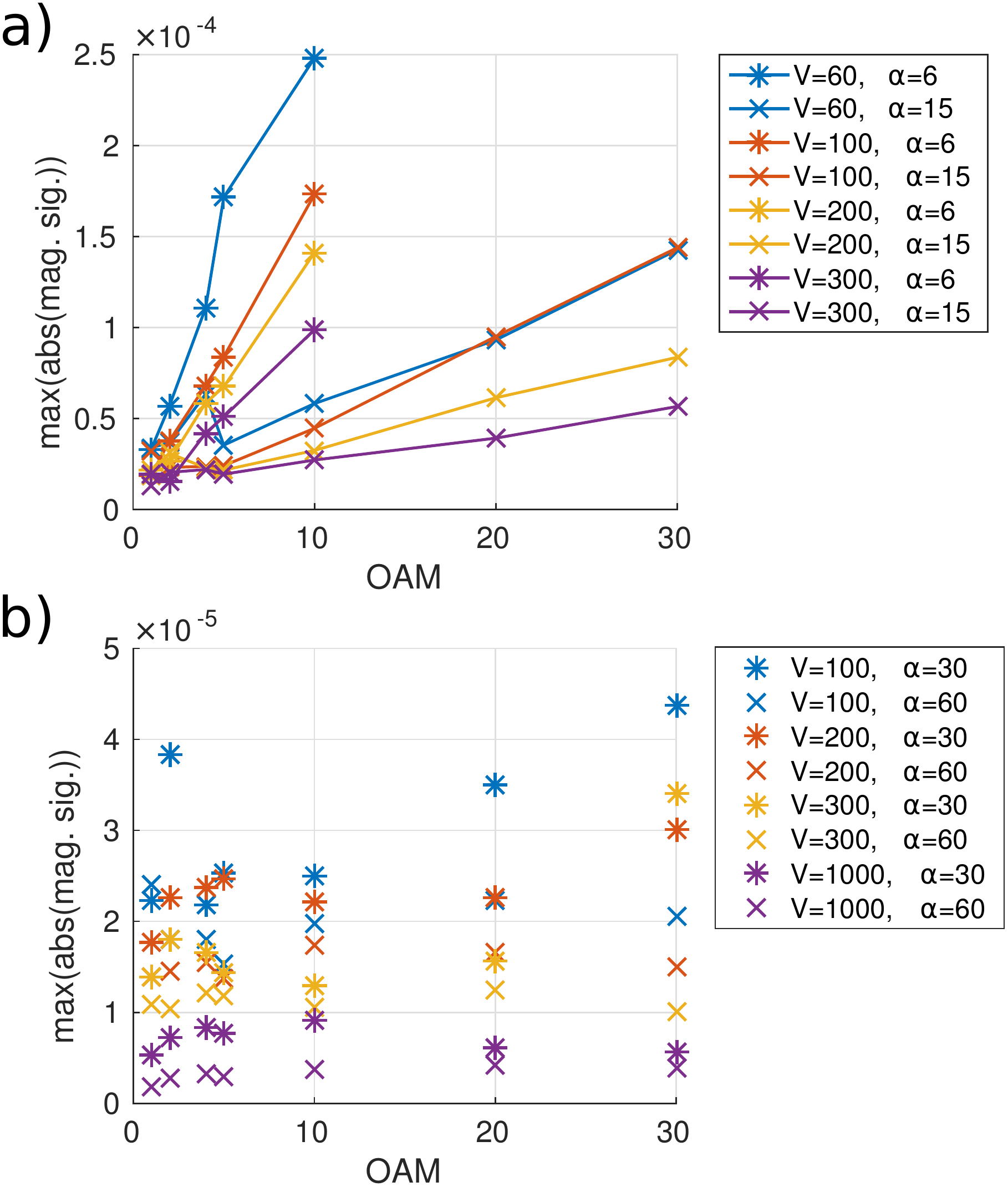}
	\caption{Maximum of absolute values of the OAM magnetic signals as shown in Fig.~\ref{fig.FePtsigoftandtheta} w.r.t. thickness and collection angle as function of $l$ for various acceleration voltages (in kV) and convergence angles (in mrad) specified in the legends. }
	\label{fig.MaxofAbsSig}
\end{figure}

To provide a more complete view of how the strength of the magnetic effects in elastic scattering of electron vortex beams depends on the various parameters considered here, the absolute value of the maximal magnetic signal w.r.t. thickness and collection angle (restricted to $100~\text{mrad}$) for the various acceleration voltages and convergence angles have been plotted against $l$. The data has been split up so that Fig.~\ref{fig.MaxofAbsSig}a) contains data with smaller convergence angles of $\alpha=6-15~\text{mrad}$, which yields larger sized beams, while Fig.~\ref{fig.MaxofAbsSig}b) contains data with larger convergence angles of $\alpha=30-60~\text{mrad}$, which yields smaller sized beams.
In Fig.~\ref{fig.MaxofAbsSig}a) it can once again be seen that larger beam sizes result in a close to linear increase in the magnetic signal strengths with $l$. Furthermore, it can be seen that for a given $l$, smaller convergence angles or smaller acceleration voltages, i.e. larger spatial beam sizes, result in a stronger magnetic signal, also in agreement with previous observations regarding Fig.~\ref{fig.varyonepar}. In Fig.~\ref{fig.MaxofAbsSig}b) it is observed that, for the smaller sized beams, there is a rather weak dependence in the magnetic signal as function of $l$. For acceleration voltage and convergence angle, on the other hand, it still appears that smaller values, corresponding to wider beams, yield stronger magnetic signals. As pointed out above, this presents a difficulty for experiments aiming at very high spatial resolution since the results presented here indicate that the signal weakens for the smaller beam sizes required for such experiments. 

\begin{figure}[hbt!]
	\centering
	\includegraphics[width=0.48\textwidth]{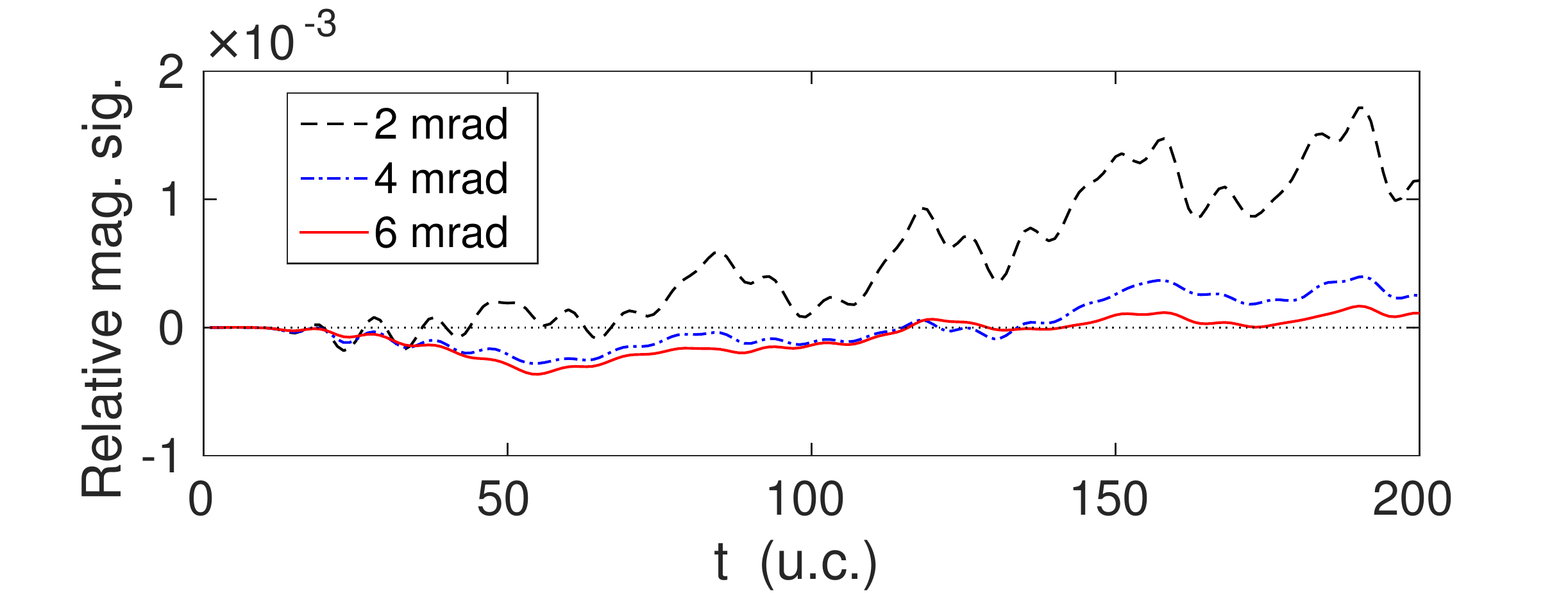}
	\caption{Thickness dependence of the relative OAM magnetic signal for a beam with $V_\text{acc} = 100~\text{kV}$, $\alpha = 6~\text{mrad}$ and $l=\pm10$ in FePt.}
	\label{FePt_V100_l10_conv6_sigoft}
\end{figure}

The largest OAM magnetic signals above $10^{-4}$ have been observed for large OAM, small convergence angles and low acceleration voltages. To further elaborate on the feasibility to experimentally measure the magnetic signals, the thickness dependence of the relative OAM magnetic signal, i.e. difference in intensity for opposite OAM divided by the sum, has been plotted for $V_\text{acc} = 100~\text{kV}$, $\alpha = 6~\text{mrad}$ and $l=\pm10$ in Fig.~\ref{FePt_V100_l10_conv6_sigoft}. The collection angles for the disk shaped regions in the diffraction plane are indicated in the legend.
For a small collection angle the largest relative magnetic signal is above $10^{-3}$, compared to previous values reported just below $10^{-3}$ with $l=\pm30$\cite{edstrom16}. With larger values of the OAM one can then expect relative magnetic signals in the order of a percent or larger. 

In order to look at the effect of magnetism in a solid on electrons with different spin, for the case of $l=0$, calculations were performed for each of the spin polarizations, up and down, with respect to the $z$-direction. Similar data as that above, which was presented for opposite values of $l$, can then be obtained for opposite spins $s$ to yield a spin magnetic signal. It has previously been shown\cite{edstrom16} that similar effects of the same order of magnitude are obtained for $s = \pm 1/2$ as for $l = \pm 1$, which is expected since the gyromagnetic factor for the electron is equal to two. 
\begin{figure}[hbt!]
	\centering
	\begin{subfigure}[b]{0.245\textwidth}
        \includegraphics[width=\textwidth]{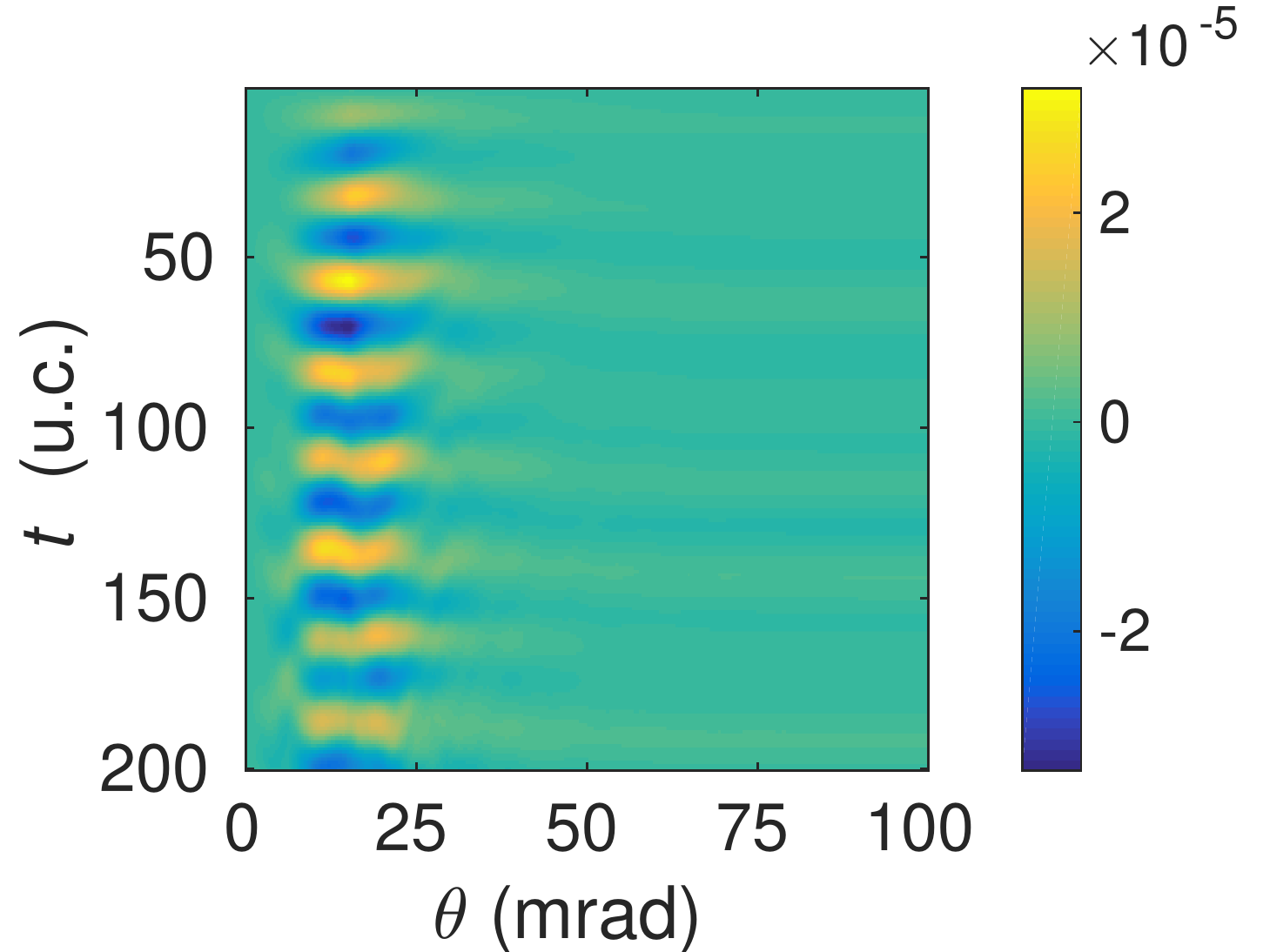}
        \caption{$V_\text{acc} = 100~\text{kV}$, $\alpha= 15~\text{mrad}$.}
        \label{FePt_V100_conv15_l0_spin}
    \end{subfigure}%
	\begin{subfigure}[b]{0.245\textwidth}
        \includegraphics[width=\textwidth]{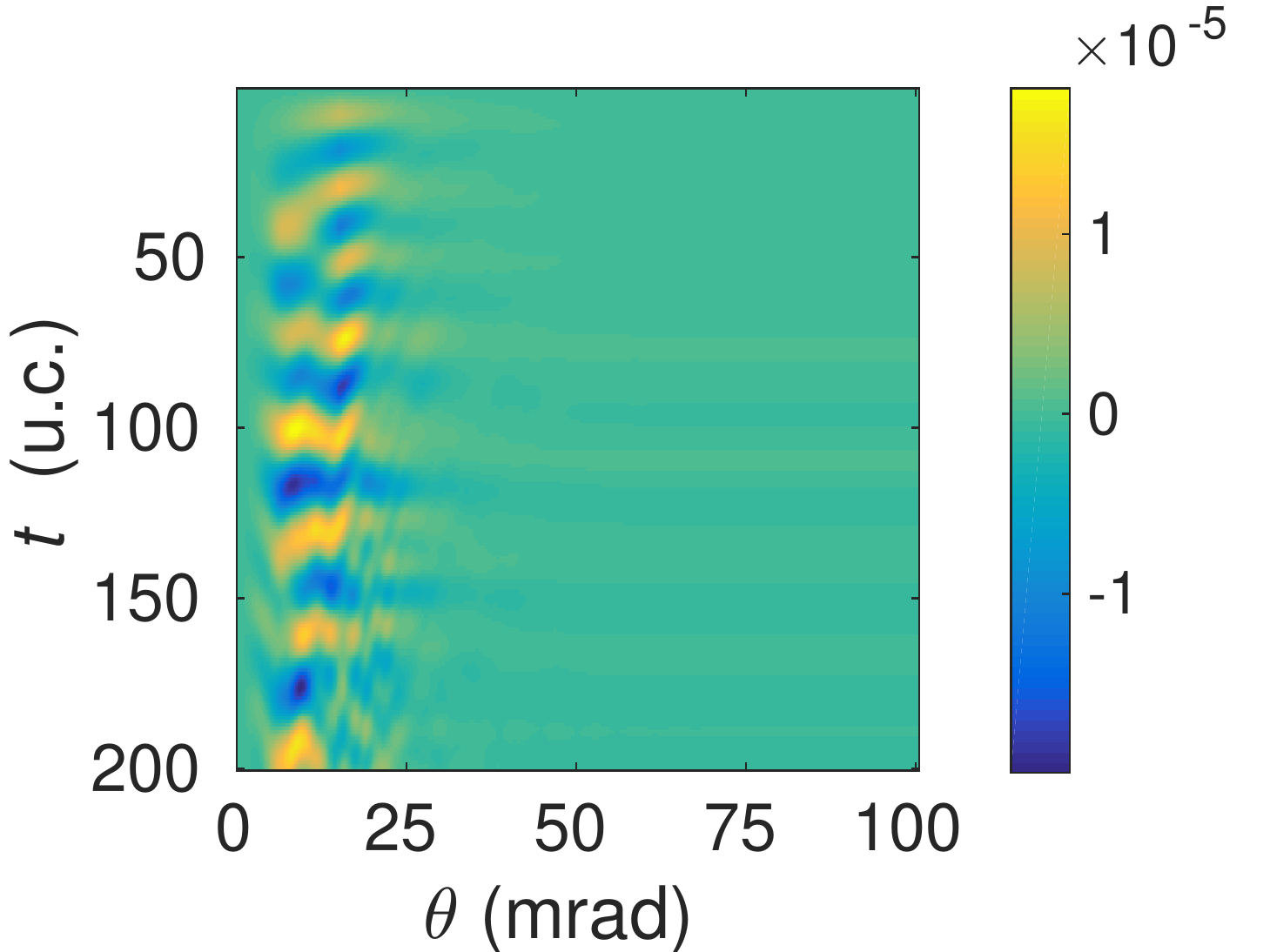}
        \caption{$V_\text{acc} = 200~\text{kV}$, $\alpha= 15~\text{mrad}$.}
        \label{FePt_V200_conv15_l0_spin}
    \end{subfigure}	   
	\begin{subfigure}[b]{0.48\textwidth}
        \includegraphics[width=\textwidth]{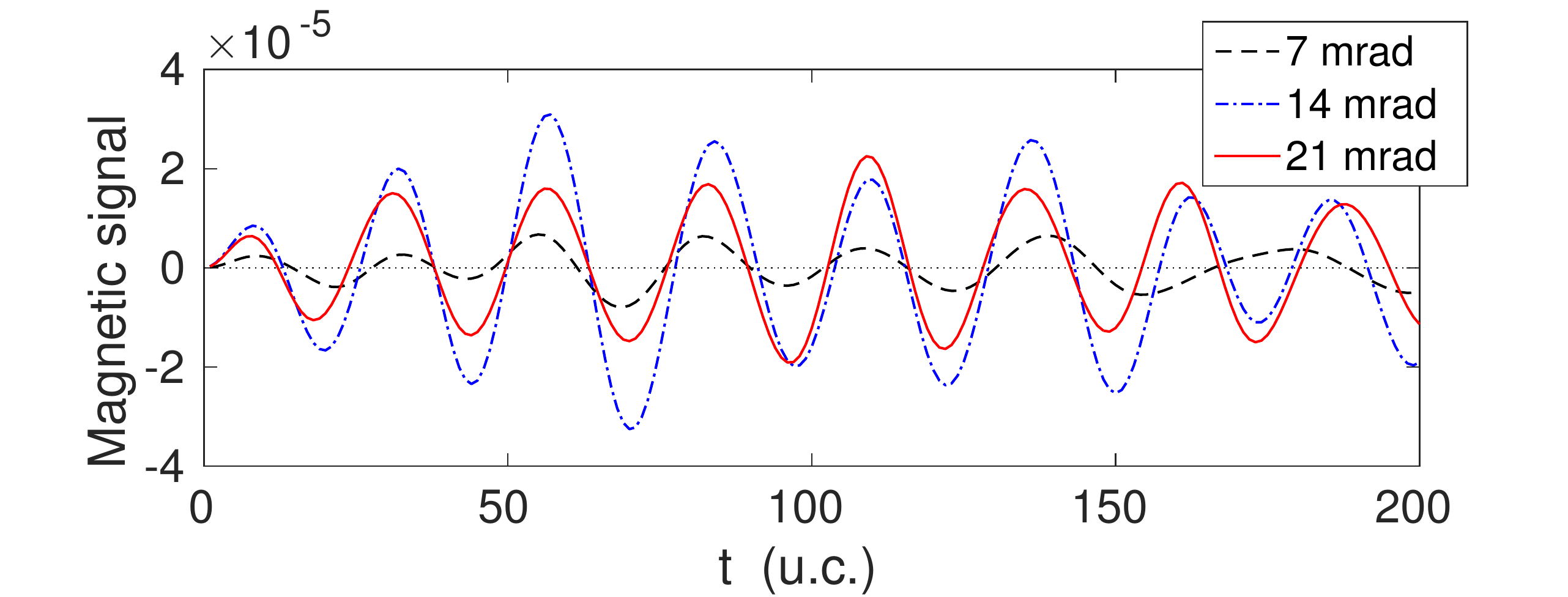}
        \caption{Magnetic signal as a function of thickness for different collection angles and $V_\text{acc} = 100~\text{kV}$, $\alpha= 15~\text{mrad}$.}
        \label{spin_sigoft_V100conv15l0}
    \end{subfigure}	         
	\begin{subfigure}[b]{0.4\textwidth}
        \includegraphics[width=\textwidth]{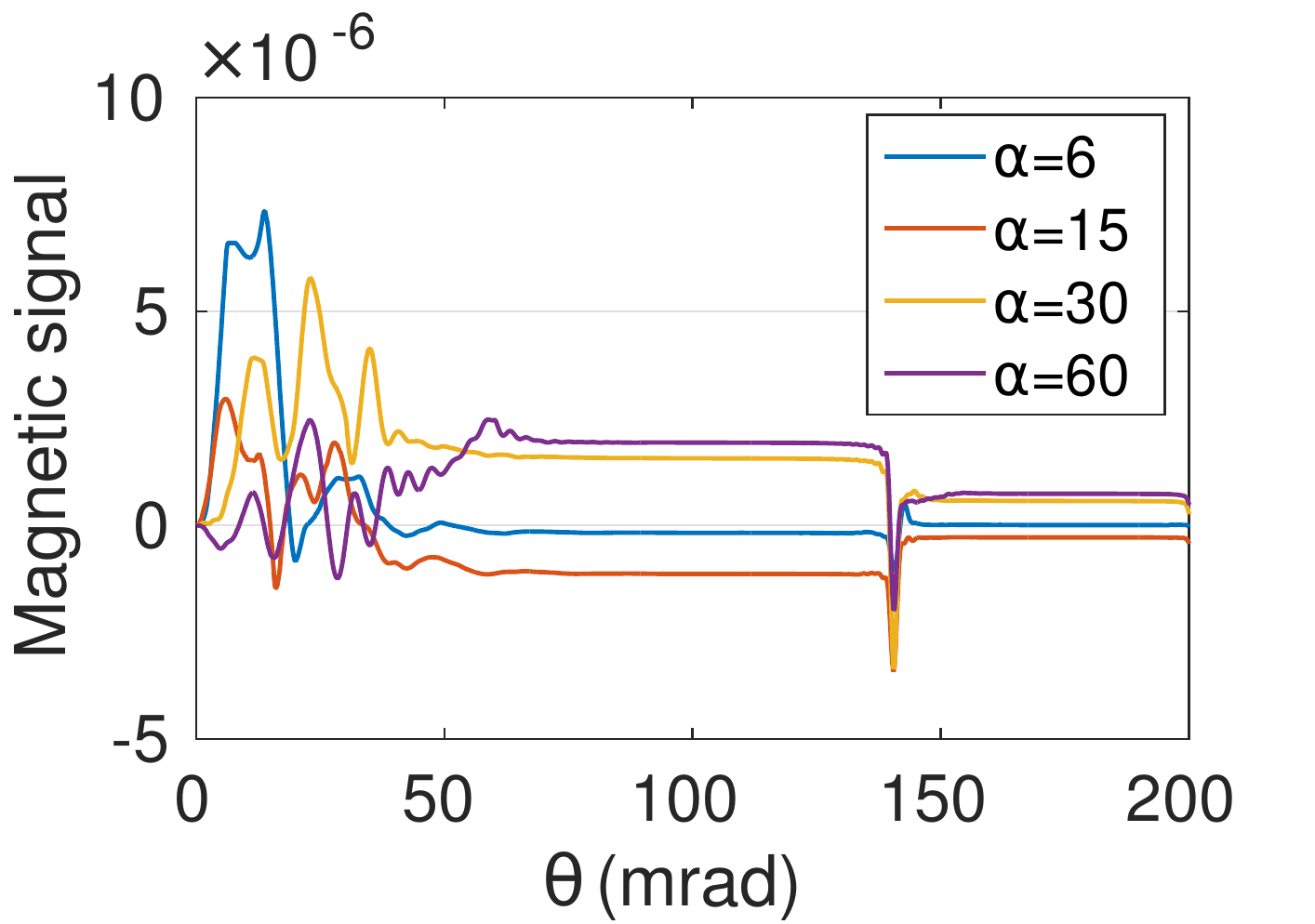}
        \caption{$V_\text{acc} = 100~\text{kV}$, varying $\alpha$.}
        \label{FePt_t50_l0_varyconv}
    \end{subfigure}
    \begin{subfigure}[b]{0.4\textwidth}
        \includegraphics[width=\textwidth]{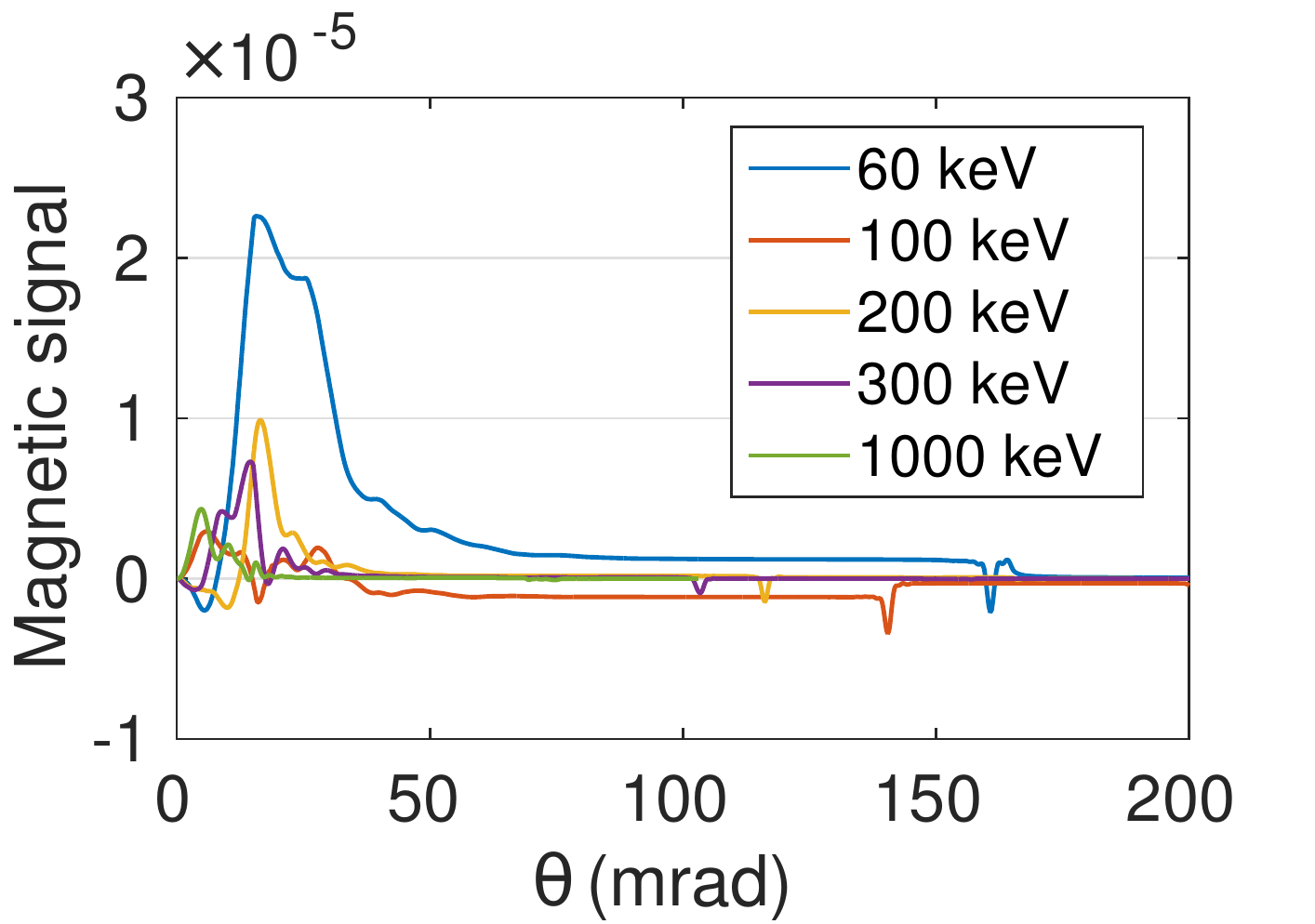}
        \caption{$\alpha = 15~\text{mrad}$, varying $V_\text{acc}$.}
        \label{FePt_t50_l0_varyV}
    \end{subfigure}
	\caption{Spin magnetic signal as function of thickness and/or collection angle in FePt for various beam parameters with $l=0$. }
	\label{fig.varyonepar_spin}
\end{figure}
Fig.~\ref{fig.varyonepar_spin} shows spin magnetic signals in FePt as function of both thickness and collection angle in a)-b), thickness dependence with fixed collection angles in c) and dependence on collection angle with fixed thickness of $t=50~\text{u.c.}$ in d)-e). Acceleration voltages $V_\text{acc}$ and convergence angles $\alpha$ are specified in captions and legends. Fig.~\ref{FePt_V100_conv15_l0_spin} displays the spin magnetic signal with $V_\text{acc} = 100~\text{kV}$ and $\alpha= 15~\text{mrad}$ as function of both thickness and collection angle for a disk shaped region in the diffraction plane and it reaches values of magnitude in the order of $10^{-5}$, which is comparable to the OAM magnetic signals seen with rather low OAM. As pointed out, this is due to the value of two for the gyromagnetic ratio of the electron. Fig.~\ref{FePt_V200_conv15_l0_spin} contains the same type of data but with a higher acceleration voltage of $V_\text{acc}=200~\text{kV}$ which results in a moderately weaker signal, similarly as was observed for the OAM magnetic signal.  Fig.~\ref{FePt_V100_conv15_l0_spin} also reveals an oscillatory thickness behaviour in the spin magnetic signal, which becomes somewhat more blurred out and distorted at larger thickness, presumably because of the deformation of the beam wave function as it propagates through the sample. A somewhat similar thickness behaviour is seen in Fig.~\ref{FePt_V200_conv15_l0_spin}, although more distorted. This distortion might be because the beam with larger acceleration voltage is more focused and thus scatters more strongly at an atomic column. The thickness dependence for $V_\text{acc} = 100~\text{kV}$ and $\alpha= 15~\text{mrad}$ is more clearly shown in Fig.~\ref{spin_sigoft_V100conv15l0}, where an oscillatory behaviour with periodicity of approximately 24 u.c., independent of collection angle, is distinctly seen. Comparison with the intensity for a given spin channel (not shown), indicates that this behaviour could be related to \emph{Pendell\"osung} oscillations.
   
Fig.~\ref{FePt_t50_l0_varyconv} - \ref{FePt_t50_l0_varyV} indicate similar results as those observed in the case of the OAM magnetic signal, namely that a stronger magnetic signal is obtained with low acceleration voltages and small convergence angles. An apparently contrary behavior is shown in Fig.~\ref{FePt_t50_l0_varyV}, where lower magnetic signals are observed for $V_\text{acc} = 100~\text{kV}$ than for $V_\text{acc} = 200~\text{kV}$. This can, however, be understood by looking at Fig.~\ref{FePt_V100_conv15_l0_spin} or Fig.~\ref{spin_sigoft_V100conv15l0}, where $t=50~\text{u.c.}$ happens to be a thickness where the spin magnetic signal is close to zero independent of collection angle for these parameters. 
  
\begin{figure*}[hbt!]
	\centering
	\includegraphics[width=\textwidth]{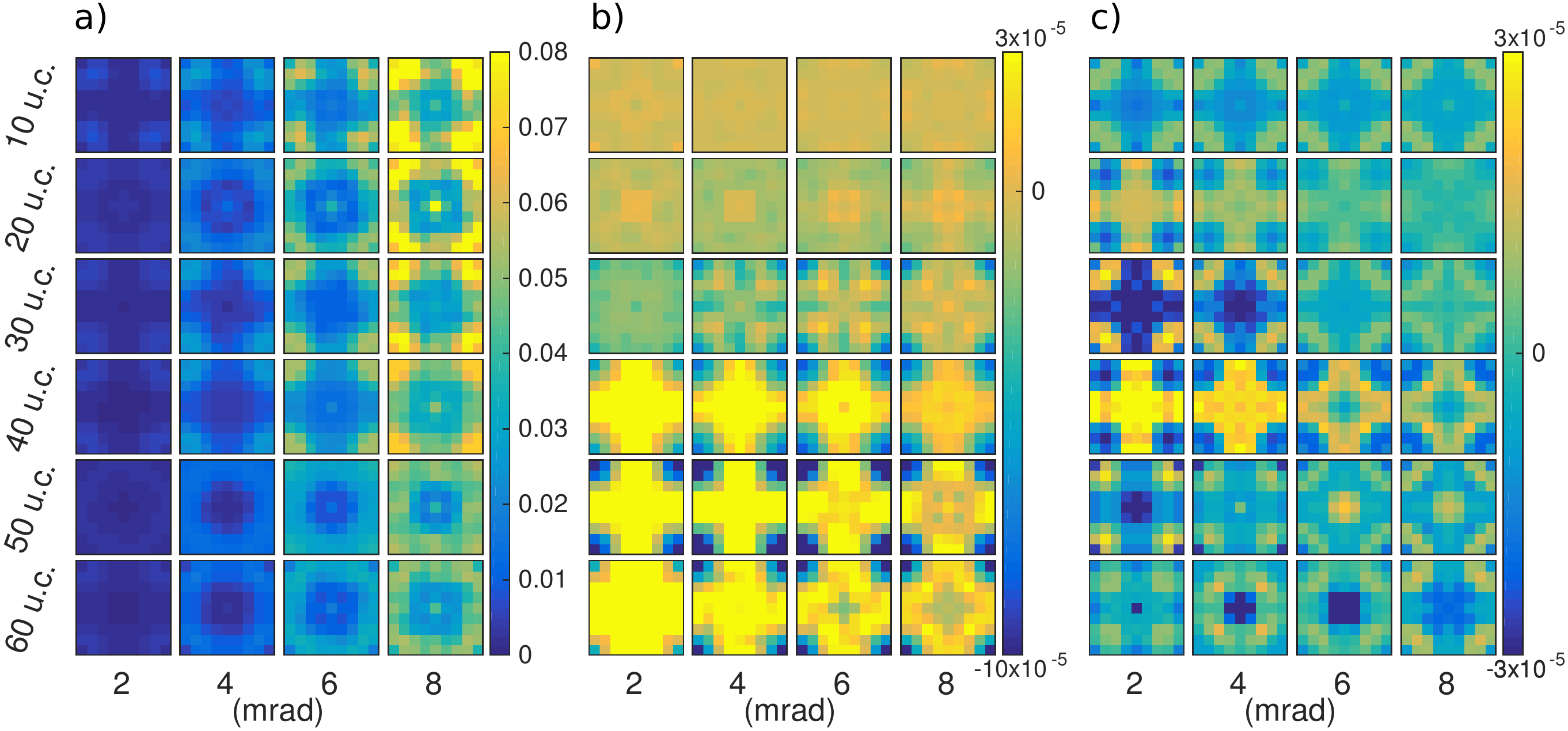}
	\caption{Simulated STEM images over one unit cell of FePt for an electron probe with $V_\text{acc}=100~\text{kV}$, $\alpha=30~\text{mrad}$ and $l=\pm1$. Different rows show different sample thicknesses while different columns show different maximum semi-angles for the disk shaped collection region. a) Total intensity for a $l+1$ beam averaged over spin channels b) magnetic signal obtained as difference between $l=+1$ and $l=-1$ probes after mirroring in the $y=x$ line c) magnetic signal as obtained for a fixed $l=+1$ and by taking a difference between spin channels. }
	\label{fig:FePt_STEM}
\end{figure*}

When exploiting magnetic effects in electron scattering for novel high resolution imaging techniques to probe magnetism in the electron microscope, it is highly desirable to achieve atomic spatial resolution. In previous calculations only very weak magnetic signals were visible in the atomic resolution STEM simulations\cite{edstrom16}. Hence, a further investigation of the parameter space, as performed here, to obtain better signals is of crucial importance. For atomic resolution STEM imaging, the parameters included in Fig.~\ref{fig.MaxofAbsSig}b), which yield small beam sizes, are the most relevant. Out of these parameters, those which yield the strongest magnetic effects appear to be $V_\text{acc}=100~\text{kV}$ and $\alpha=30~\text{mrad}$. Hence, using these parameters in combination with $l=\pm1$, a set of calculations was performed at beam positions $(x,y) = \frac{a}{8}(i,j)$ for $i,j = 0, 1, 2, 3, 4$, i.e., a $5\times5$ grid covering a quarter of the unit cell from which the remaining unit cell is reconstructed using the four-fold rotational symmetry. The results of these calculations are shown in Fig.~\ref{fig:FePt_STEM} for different sample thicknesses in the range 10-60 unit cells as indicated in the different rows and with varying collection angles as indicated in the different columns. Fig.~\ref{fig:FePt_STEM}a) shows the total intensity for an $l=+1$ beam averaged over spin channels (although the results for different spins differ negligibly). Due to the symmetry aspects of electron vortex beams\cite{Rusz2014}, if neglecting the magnetic interactions studied in this work, the images obtained with $l=-1$ would be identical to that obtained by taking the $l=+1$ image and applying a mirror symmetry operation of the crystal (visually the image obtained is identical to the mirror image of Fig.~\ref{fig:FePt_STEM}a) also with magnetism since the magnetic scattering is relatively weak), whereas this symmetry is broken by magnetism, which changes sign with mirror operations. As has previously been pointed out\cite{edstrom16}, this means that an atomic resolution magnetic signal can be obtained by taking the difference between the signal for a $+l$ OAM beam at $(x,y)$ and $-l$ OAM beam at $(y,x)$, if mirroring at the $y=x$ line belongs to the crystal symmetries, as it is for L1$_0$ FePt with the $c$-axis oriented along the propagation direction $z$. The magnetic image obtained from such a procedure is shown in Fig.~\ref{fig:FePt_STEM}b). As one might expect, the strongest magnetic signals are seen for beams located on top or near the Fe atomic columns (at the corners of the unit cell). After 50 unit cells a magnetic signal of magnitude $10^{-5}$ is obtained if using a collection angle of $8~\text{mrad}$. Albeit small, this is one order of magnitude larger than values previously reported\cite{edstrom16}. Nevertheless, atomic resolution imaging of magnetism with the considered scheme will remain challenging unless further improvements can be made. 

With spin polarized beams, a magnetic signal can be obtained by taking the difference between spin up and spin down beams at the same beam position, without considering mirror operations. Such images, obtained by taking a difference over spin channels for fixed $l=+1$, are shown in Fig.~\ref{fig:FePt_STEM}. As for the case of OAM magnetic signal, the strongest magnetic signals have a tendency to be localized around the Fe columns. At a thickness of 50 or 60 unit cells and with a collection angle of 6 or 8 mrad, a reasonable signal is also seen for beams exactly on the Pt column. This might be related to the very localized spin density and corresponding $B_z$ observed at $c/2$ in Fig.~\ref{fig.FePt-fields}.

\subsection{FePt with Magnetization Perpendicular to the Propagation Direction}\label{FePtxmag}

In the calculations presented thus far as well as in previous work\cite{edstrom16}, the magnetization of the sample, the beam propagation direction as well as the OAM and spin quantization axis were parallel. Such a setup is expected to increase magnetic effects as it makes $\mathbf{L}\cdot\mathbf{B}$ or $\mathbf{S}\cdot\mathbf{B}$ large. However, other magnetic effects such as spin flip processes might be enhanced in a setup where the sample magnetization is perpendicular to the spin quantization axis, as has been suggested before\cite{GrilloKarimi} and can also be seen in Eq.~\ref{paraxialPeq} by noting that the only off-diagonal terms are proportional to the $x$- and $y$-components of the magnetic field. In this section we explore this effect by performing simulations for an FePt sample which has been rotated, while keeping the spin quantization axis parallel to the propagation direction. Since FePt is a very hard magnet with easy magnetization axis along the $c$-direction, both the crystal axis and the magnetization direction were rotated, resulting in an $a \times c \times a$ unit cell, which was, however, still discretized on a $64\times 64 \times 32$ grid. 

\begin{figure}[hbt!]
	\centering
	\begin{subfigure}[b]{0.46\textwidth}
        \includegraphics[width=\textwidth]{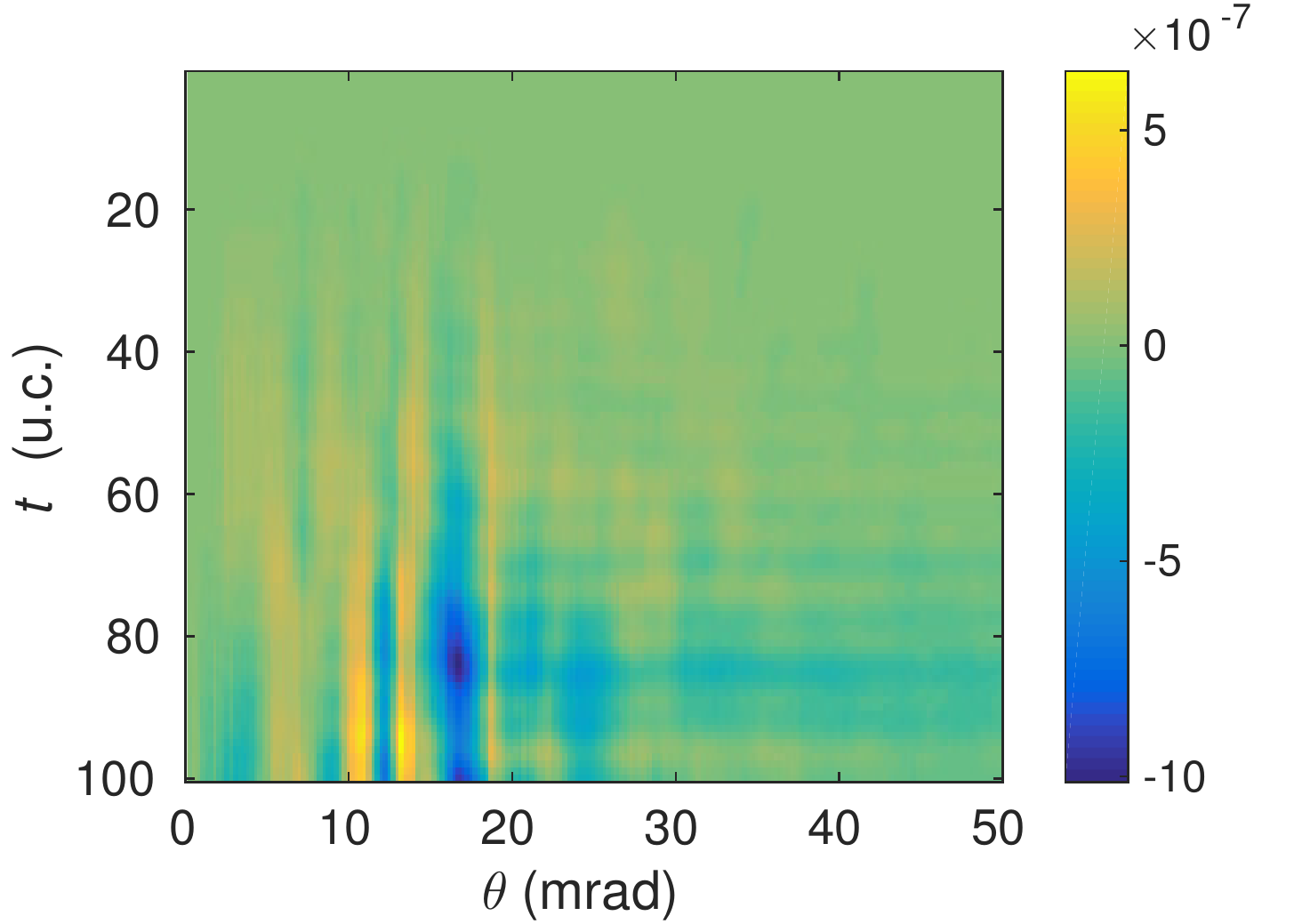}
        \caption{OAM magnetic signal as function of thickness and collection angle.}
    \end{subfigure} \\%
    \begin{subfigure}[b]{0.45\textwidth}
        \includegraphics[width=\textwidth]{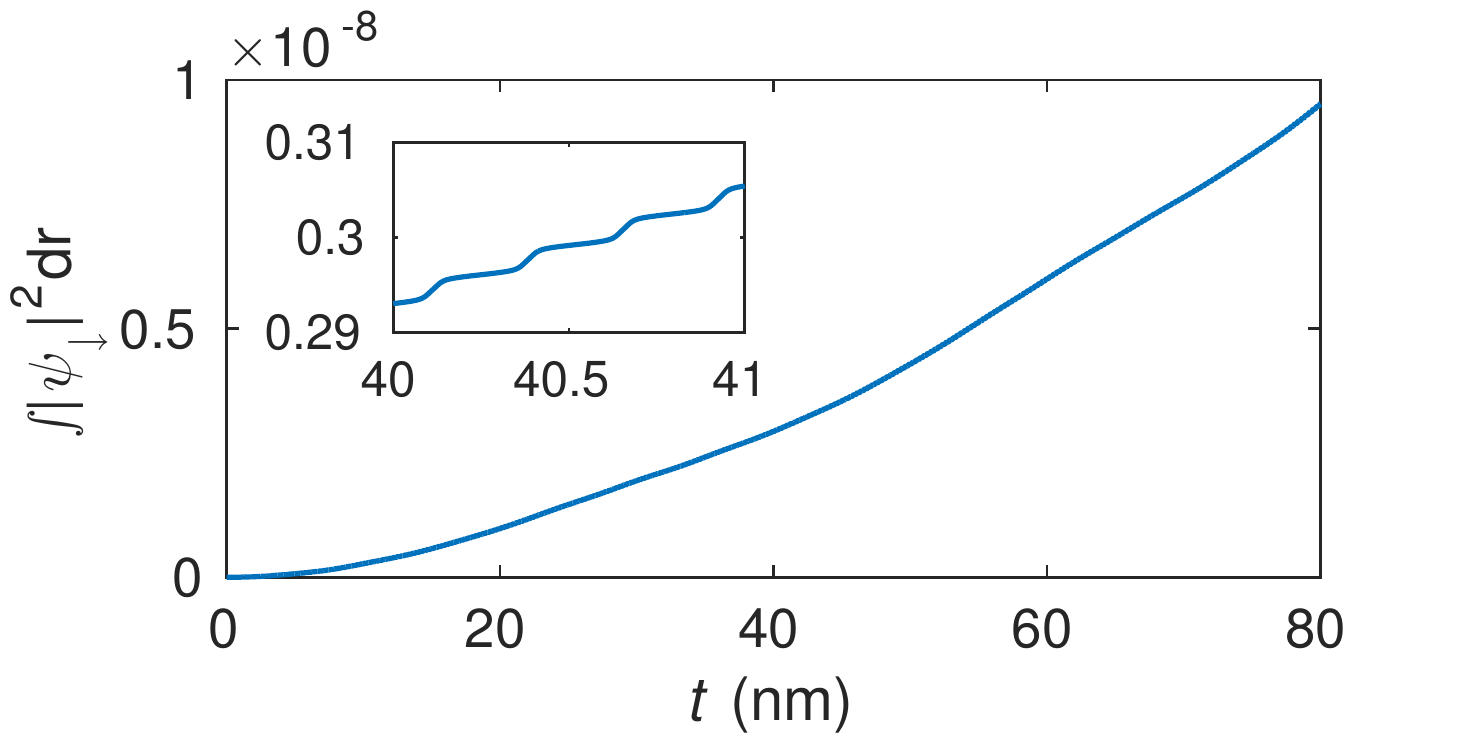}
        \caption{Fraction of spin down electrons from an initially spin up polarized beam as function of thickness with $l=+10$.}
    \end{subfigure}
    	\begin{subfigure}[b]{0.45\textwidth}
        \includegraphics[width=\textwidth]{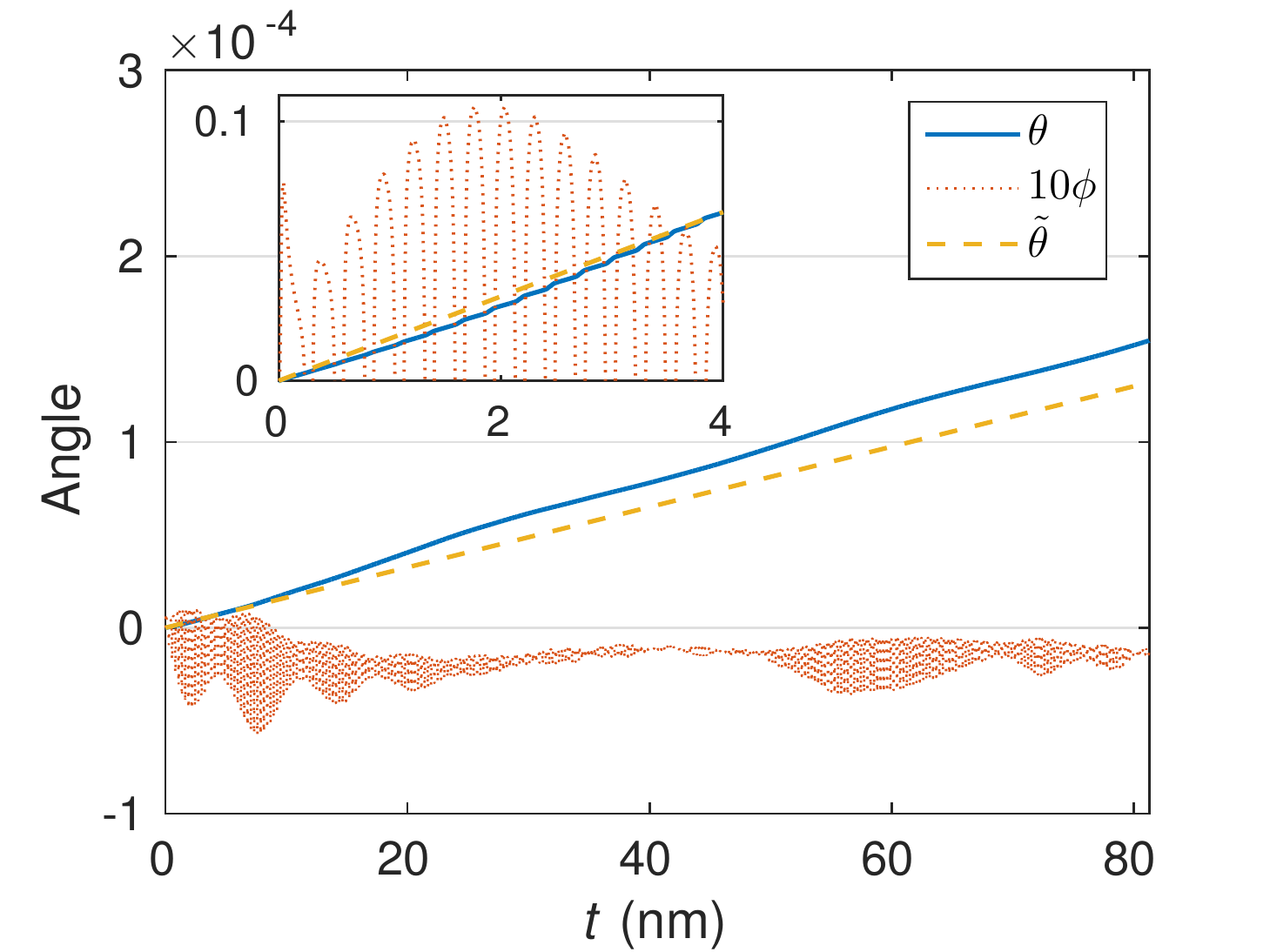}
        \caption{Rotation of the expectation value of the spin operator w.r.t. the propagation direction in terms of polar and azimuthal angles $\theta$ and $\phi$. The angle  $\tilde{\theta}$, is that expected for a spin in a constant magnetic field corresponding to the saturation magnetization of FePt.}
    \end{subfigure}%
	\caption{Simulations of $l=\pm 10$ vortex beams with 100 keV, 15~mrad convergence angle and spin polarization in the propagation direction for FePt with magnetization perpendicular to the propagation direction. }
	\label{fig.FePt_rotmag}
\end{figure}

Calculations have been performed with $V_\text{acc} = 100~\text{kV}$, convergence angle $15~\text{mrad}$ and $l=\pm 10$ and the results are summarized in Fig.~\ref{fig.FePt_rotmag}. Fig.~\ref{fig.FePt_rotmag}a) shows the OAM magnetic signal as function of thickness and collection angle. This signal is significantly weaker than that observed in the previous section as one would expect when the magnetic field is mainly perpendicular to the OAM, but nevertheless reaches magnitudes of order $10^{-6}$ for certain thicknesses and collection angles. Fig.~\ref{fig.FePt_rotmag}b) shows the proportion of spin down electrons in the initially completely spin up polarized beam, as a function of sample thickness. Within the thicknesses studied this is a monotonically increasing function and it reaches approximately $10^{-8}$ after $80~\text{nm} \approx 300~\text{u.c.}$ Albeit being a very small fraction of the electrons, this is several orders of magnitude more than was seen for magnetizations parallel to spin polarization where less than $10^{-12}$ of the majority spin states scattered into minority spin states, in agreement with the expectations mentioned above. The inset shows a close up of the region around $40~\text{nm}$ and allows one to see a regular step-like behaviour with periodicity somewhat less than one third of a nanometer, i.e. corresponding to one unit cell with thickness $0.271~\text{nm}$.

Fig.~\ref{fig.FePt_rotmag}c) illustrates the rotation of the spin of the beam as function of thickness in terms of the polar and azimuthal angles $\theta$ and $\phi$ describing the direction of the spin expectation value, $\mean{\mathbf{S}}$, relative to the sample and propagation direction. The polar angle $\theta$ reaches values in the order of $10^{-4}$ within the thicknesses studied while the azimuthal angle $\phi$ is of similar size for very small thicknesses but remains at least one or two orders of magnitude smaller than $\theta$ for larger thicknesses. The inset shows a close up for small thicknesses and displays clear oscillations in $\phi$ with periodicity of the lattice, which can be understood as the variation of $\phi$ should be mainly due to interactions of the beam with the non-uniform part of the magnetic field ($\mathbf{B}_\text{p}$). For comparison, the angle 
\begin{equation}
\tilde{\theta} = \frac{2 m \mu_0 M_\text{s} \mu_\text{B}}{\hbar^2 k}t ,
\end{equation}
expected for a spin in a homogeneous magnetic field according to Eq.~\ref{spininField}, is shown. $t$ denotes thickness, i.e., the $z$ coordinate. The angles $\theta$ and $\tilde{\theta}$ are similar, especially for small thicknesses up to about $2~\text{nm}$, after which $\theta$ is somewhat enhanced compared to $\tilde{\theta}$. 

\subsection{AFM LaMnAsO}\label{LaMnAsOresults}

\begin{figure}[hbt!]
	\centering
	\includegraphics[width=0.45\textwidth]{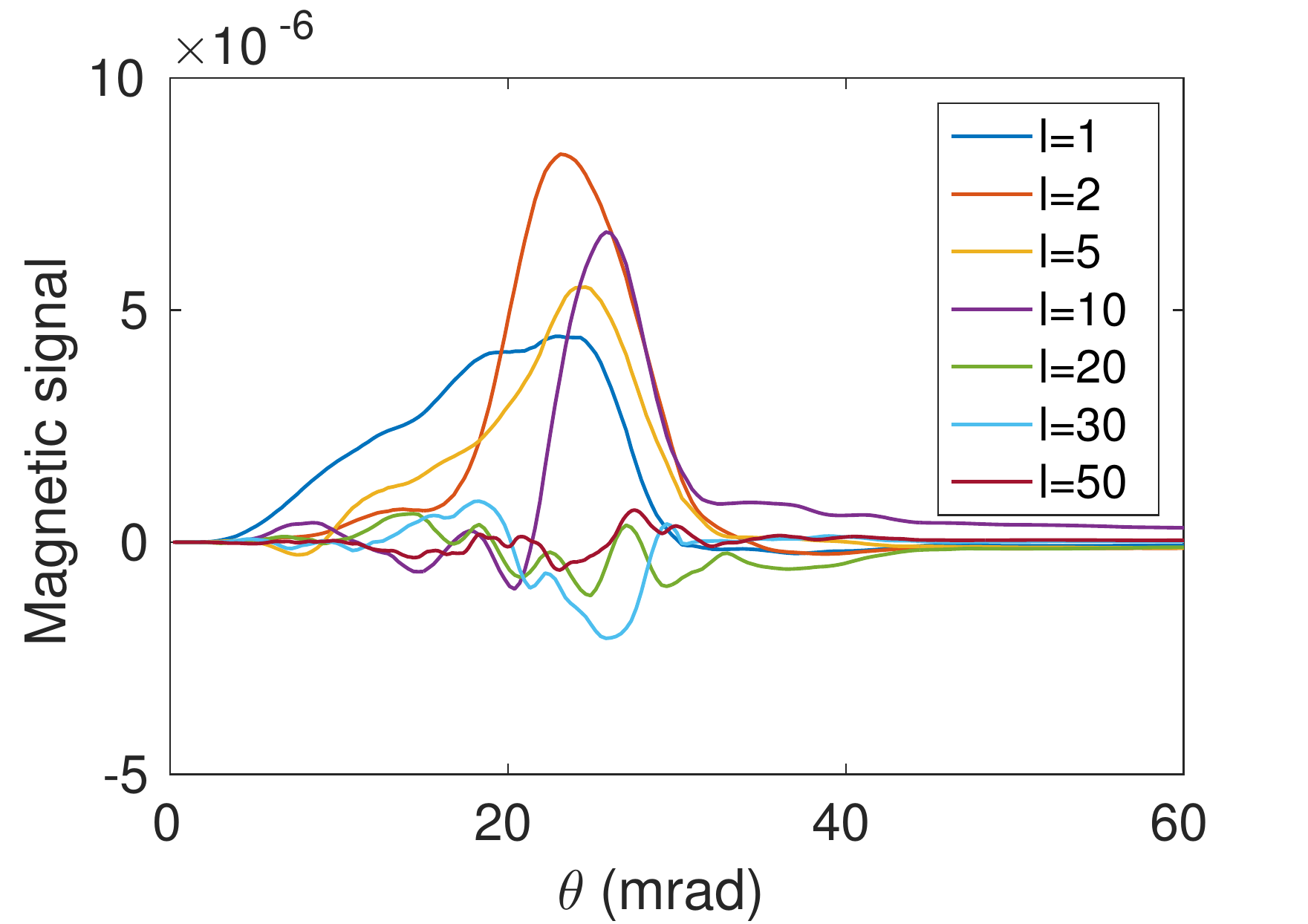}
	\caption{Magnetic signal as a function of collection angle for $\alpha=30~\text{mrad}$, $V_\text{acc} = 100~\text{kV}$ and various values of $l$ after the beam passes through 20 u.c. of LaMnAsO.}
	\label{fig.sigofOAM}
\end{figure}

In an antiferromagnetic (AFM) compound such as LaMnAsO, the saturation magnetization is zero and consequently the volume average of the magnetic field is also zero, $\mathbf{B}_\text{avg}=\mu_0 \mathbf{M}=0$. In this case there is no reason to expect any proportionality between the magnetic signal and OAM, for nanometer sized beams. Instead one can expect that a large OAM and correspondingly large spatial extent of the beam, for a given convergence angle, results in a vanishingly small magnetic signal as the beam size grows significantly beyond the dimensions of one unit cell. This is confirmed in Fig.~\ref{fig.sigofOAM}, which illustrates the magnetic signal for various values of the initial OAM $l$, with convergence angle $\alpha=30~\text{mrad}$ and acceleration voltage $V_\text{acc} = 100~\text{kV}$, after the beam passes through 20 u.c. of LaMnAsO with each unit cell discretized on a $64\times64\times64$ grid. These beams have initial widths (calculated as the diameter of the ring with maximum intensity) increasing with $l$ and ranging from 0.6 \AA, a fraction of an atomic distance, to 22 \AA, well above atomic distances. The beam with $l=10$ has a diameter of 5 \AA, i.e. slightly more than the in-plane lattice parameter. Reasonably large magnetic signals can be observed with $l \leq 10$, while it becomes notably reduced for larger values, when the beam widths increase significantly beyond the size of one unit cell. 

\begin{figure*}[hbt!]
	\centering
	\includegraphics[width=\textwidth]{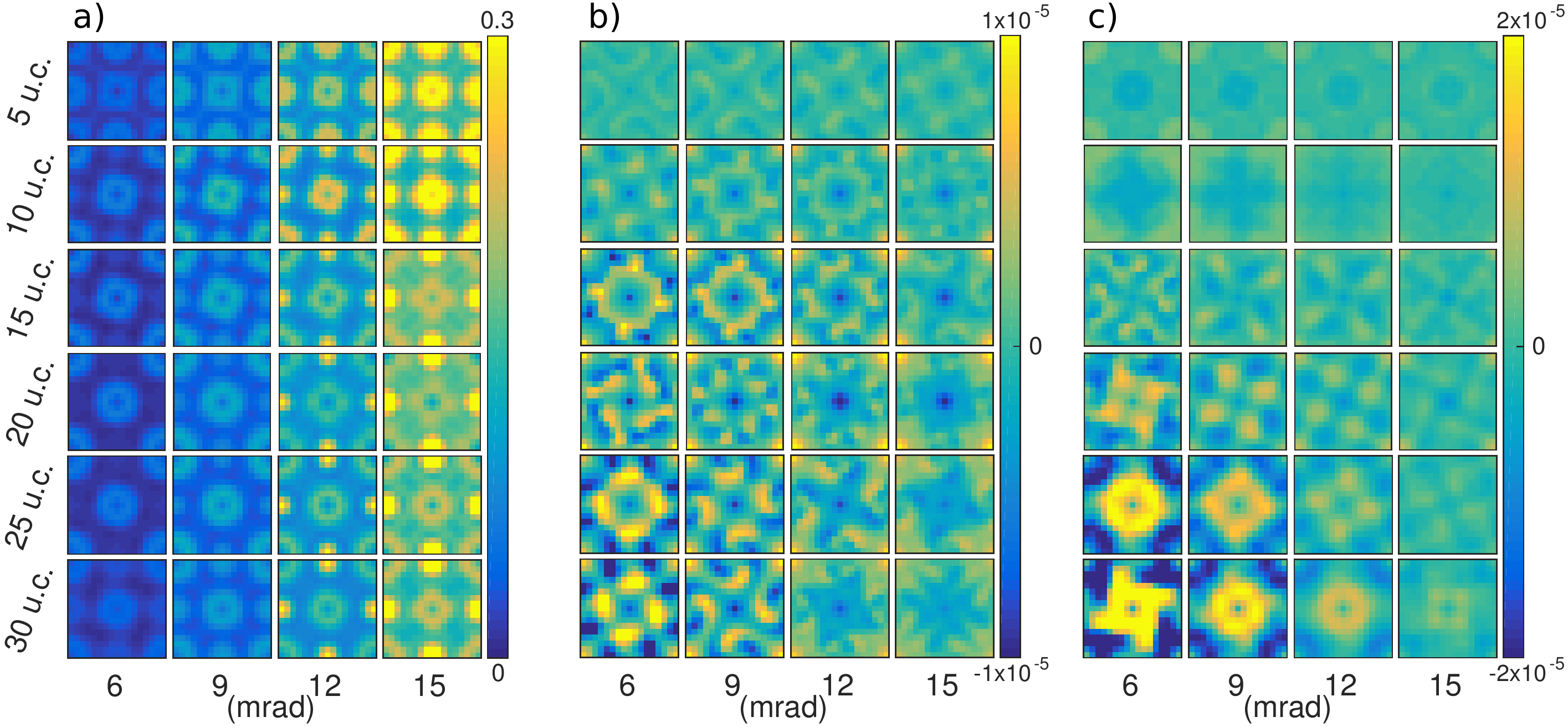}
	\caption{Simulated STEM images over a unit cell of FePt for an electron probe with $V_\text{acc}=100~\text{kV}$, $\alpha=30~\text{mrad}$ and $l=\pm1$. Different rows show different sample thicknesses while different columns show different maximum semi-angles for the disk shaped collection region. a) Total intensity for a $l+1$ beam averaged over spin channels b) magnetic signal obtained as difference between $l=+1$ and $l=-1$ probes after mirroring in the $y=0$ line c) magnetic signal as obtained for a fixed $l=+1$ and by taking a difference between spin channels.}
	\label{fig:LMAO_STEM}
\end{figure*}

Since an AFM has a vanishing saturation magnetization and the magnetic scattering of electron vortex beams cannot be enhanced by using large OAM beams, these materials should mainly be of interest if atomic resolution imaging can be performed. Hence, an atomic resolution STEM simulation has been done with the same parameters as was used in the FePt case, i.e. $V_\text{acc}=100~\text{kV}$, $\alpha=30~\text{mrad}$ and $l=\pm1$. One unit cell was discretized on a $16 \times 16$ grid and calculations were performed for beam positions over one half of the unit cell.  For FePt the mirror operation in the $y=x$ line was a symmetry operation of the crystal and this was used to obtain a magnetic signal by taking differences in intensity between opposite OAM beams for mirror points. For the LaMnAsO crystal structure illustrated in Fig.~\ref{fig.crystalstructs}b), a mirroring in $y=x$ is not a symmetry operation of the crystal, whereas mirroring in $y=0$ (as well as $x=0$) is, whereby such mirror points were used to create a magnetic signal from opposite OAM beams, with the result shown in Fig.~\ref{fig:LMAO_STEM}b). The strength of this magnetic signal is of similar size as that observed in the case of FePt, although slightly stronger maximum signals were observed for FePt. As one would expect there is signal of different sign at the corner Mn atom with spin up and the central spin down Mn atom. Furthermore, it can be noted that essentially no magnetic signal is seen at the other atomic columns containing atoms around which the spin density is two orders of magnitude smaller (Fig.~\ref{fig.LMAO-fields}). 
Fig.~\ref{fig:LMAO_STEM}c) shows the spin magnetic signal, obtained as difference between opposite spin channels for fixed OAM of $l=+1$, at a given beam position. Again this is of similar order of magnitude as the magnetic signal seen in Fig.\ref{fig:LMAO_STEM}b), indicating that it might be easier to use spin polarization than electron vortex beams for STEM imaging of magnetism in the atomic resolution, if spin polarized beams and detectors can be made available, as it does not require comparison between measurements at different beam positions. For some of the larger thicknesses and smaller collection angles, "doughnut" shapes characteristic to vortex beams are seen around the Mn columns, similarly to the shapes observed also in Fig.\ref{fig:LMAO_STEM}a), most likely related to the radial shape of the electron vortex beams. 

\subsection{Phase aberrated beams}\label{aberrations}

It was recently suggested\cite{emcdc34} that electron vortex beams are merely a special case of a wider class of beams with non-trivial phase distributions and that magnetism can be observed in EMCD experiments using phase aberrated electron beams. Such beams are readily available in modern aberration corrected electron microscopes and their creation was recently discussed\cite{PhysRevB.93.104420} and experimentally demonstrated\cite{Idrobo2016}. Motivated by this, the magnetic interaction between a phase aberrated beam and a magnetic sample in the elastic scattering regime is investigated in this section, in order to see whether such beams are potentially useful in imaging magnetic materials also in elastic scattering experiments. Phase aberrated electron beams have a phase distribution described by an aberration function\cite{Krivanek19991, PhysRevB.93.104420}
\begin{multline}
\chi(k_\perp,\phi_k) = \frac{2\pi}{\lambda} \sum_{n,m} \frac{\theta^{n+1}}{n+1} \left[ C_{n,m}^a\cos(m\phi_k) + \right. \\ \left. C_{n,m}^b\sin(m\phi_k) \right],
\label{eq.aberrationfunction}
\end{multline}
where $n$ is a non-negative integer, which denotes the order of the aberration and the non-negative integer $m=n+1, n-1, n-3, ...$ denotes the order of the rotational symmetry of the aberration. $\theta=\arctan(k_\perp/kz)$ is the axial angle, while $\phi_k=\arctan(k_y/k_x)$ is the azimuthal angle, as in Eq.~\ref{eq.vortex}. The idea behind using aberrated electron beams to observe magnetism is that a magnetic signal should be obtained by comparing a beam with $+C_{n,m}^{i}$ aberration, where $i=a$ or $b$, to one with $-C_{n,m}^{i}$ aberration, e.g. by looking at the intensity difference in a disk shaped region in the diffraction plane, similarly as done with electron vortex or spin polarized beams in other sections of this work. For this to yield a magnetic signal requires that the mirror symmetry operations of the crystal map the $+C_{n,m}^{i}$ term onto $-C_{n,m}^{i}$. Changing the sign of the aberration is then equivalent to a mirror operation of the crystal, which leaves everything invariant except magnetism, which changes sign. It is also important that the rotational  symmetry operations of the crystal do not map $+C_{n,m}^{i}$ to $-C_{n,m}^{i}$ since the magnetic signal is then expected to be zero. It can also be noted that the expectation value of the OAM operator w.r.t. a beam with a phase distribution such as that described by Eq.~\ref{eq.aberrationfunction} is zero. Based on Eq.~\ref{Lplus2S} one would therefore not expect a magnetic signal from large beams interacting mainly with the uniform part of the magnetic field, but mainly from atomic resolution electron beams where the phase distribution can locally couple to the microscopic magnetic field in the sample.

\begin{figure}[hbt!]
	\centering
	\includegraphics[width=0.45\textwidth]{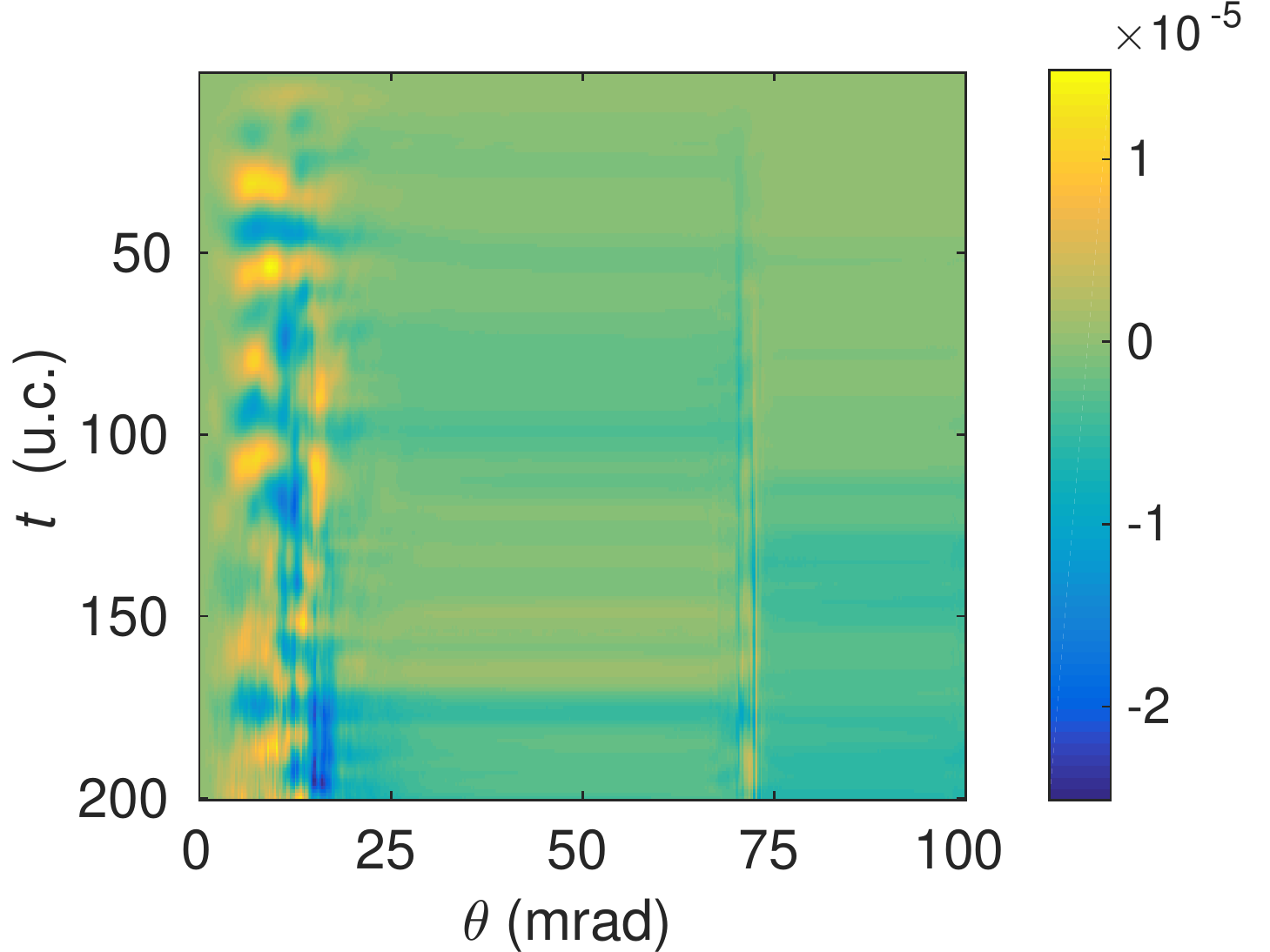}
	\caption{Magnetic signal obtained as intensity difference for opposite aberration values of an electron probe with aberration of $\pm C_{3,4b}$. The results shown are for a beam with $V_\text{acc}=100~\text{kV}$ and $\alpha = 30~\text{mrad}$.}
	\label{aberrationfig}
\end{figure}

Calculations with aberrated electron probes have been performed for the ferromagnetic FePt crystal for which $C_{4v}$ is a subgroup of the crystallographic point group. Similarly as in the case of EMCD\cite{PhysRevB.93.104420}, the lowest order aberration expected to be useful in observing magnetism is then $C_{34}^b$ (i.e. four-fold astigmatism). For a beam with $100~\text{keV}$, convergence angle $\alpha = 30~\text{mrad}$ and the lattice parameters of FePt, it has been shown that $C_{3,4}^b = \pm 14~\mu\text{m}$ is useful for observing magnetism in EMCD\cite{PhysRevB.93.104420}, whereby the same value is attempted here. The magnetic signal obtained from the intensity difference of opposite value $C_{3,4}^b$-aberrations as function of thickness and collection angle is illustrated in Fig.~\ref{aberrationfig}. The maximum values of the magnetic signal is of order $10^{-5}$, which is similar to what was found both in the spin magnetic signal and in the OAM magnetic signal with low values of $l$. This indicates that aberrated electron beams are potentially useful in observing magnetism in elastic scattering experiments in a similar way that spin polarized or vortex beams can be. The possible advantage of aberrated beams is that aberration modification technology is more readily available than spin polarization or vortex beam generation technology. Moreover, there is no implicit intensity loss due to Fresnel zone plates or magnetic needles required for producing vortex beam, when employing aberrated beams.

\subsection{Noise and error analysis}\label{noise}

In the preceding sections it has been shown that magnetic signals result from the elastic scattering of electrons with angular momentum or phase aberrations as they scatter through magnetic matter. The signals discussed have, however, been found to be weak. In order to estimate the feasibility of experimental observations of the discussed phenomena, it is therefore crucial to consider the effect of statistical noise as well as systematic errors, as is done in this section. 

Firstly, regardless the precision of experimental equipment, statistical noise in the form of Poisson (i.e. shot) noise must be taken into account. To provide an idea of the acquisition times required, a beam current of 100 pA together with either a large OAM beam with $l=\pm10$, convergence angle of 6~mrad and 100~keV (see Fig.~\ref{FePt_V100_l10_conv6_sigoft}) or a small OAM atomic resolution case of $l=\pm1$, convergence angle of 30~mrad and 100 keV (see Fig.~\ref{fig:FePt_STEM}) is considered. In the large OAM case, considering a 6~mrad collection angle and sample thickness of 54 u.c., the magnitude of the magnetic signal is $10^{-4}$, whereby the acquisition time required to obtain a signal-to-noise ratio (SNR) of 3 is $144~\mu\text{s}$. For the small OAM case, the largest magnitude magnetic signal, out of the conditions shown in Fig.~\ref{fig:FePt_STEM}, is obtained with a collection angle of 8~mrad and sample thickness of 50 u.c., where it is approximately $10^{-5}$. The acquisition time needed for a SNR of 3 is then $1.44~\text{ms}$. 

Any experimental setup will suffer from some degree of mechanical noise due to for example vibrations, drift or tilt of the sample and the effect of this will now be addressed for the same large and small OAM cases as above. Together with the required acquisition times stated above an idea of the experimental feasibility can then be obtained. Firstly, drift is considered by performing calculations for beam positions ranging from zero to five steps, in the $x$-direction, in units of the smallest grid spacing, i.e. $\frac{a}{64} = 4.23~\text{pm}$ so the biggest shift is $21~\text{pm}$ from an Fe atomic column, which corresponds to realistic scanning distortions in STEM experiments.
\begin{figure}[hbt!]
	\centering
    \begin{subfigure}[b]{0.4\textwidth}
        \includegraphics[width=\textwidth]{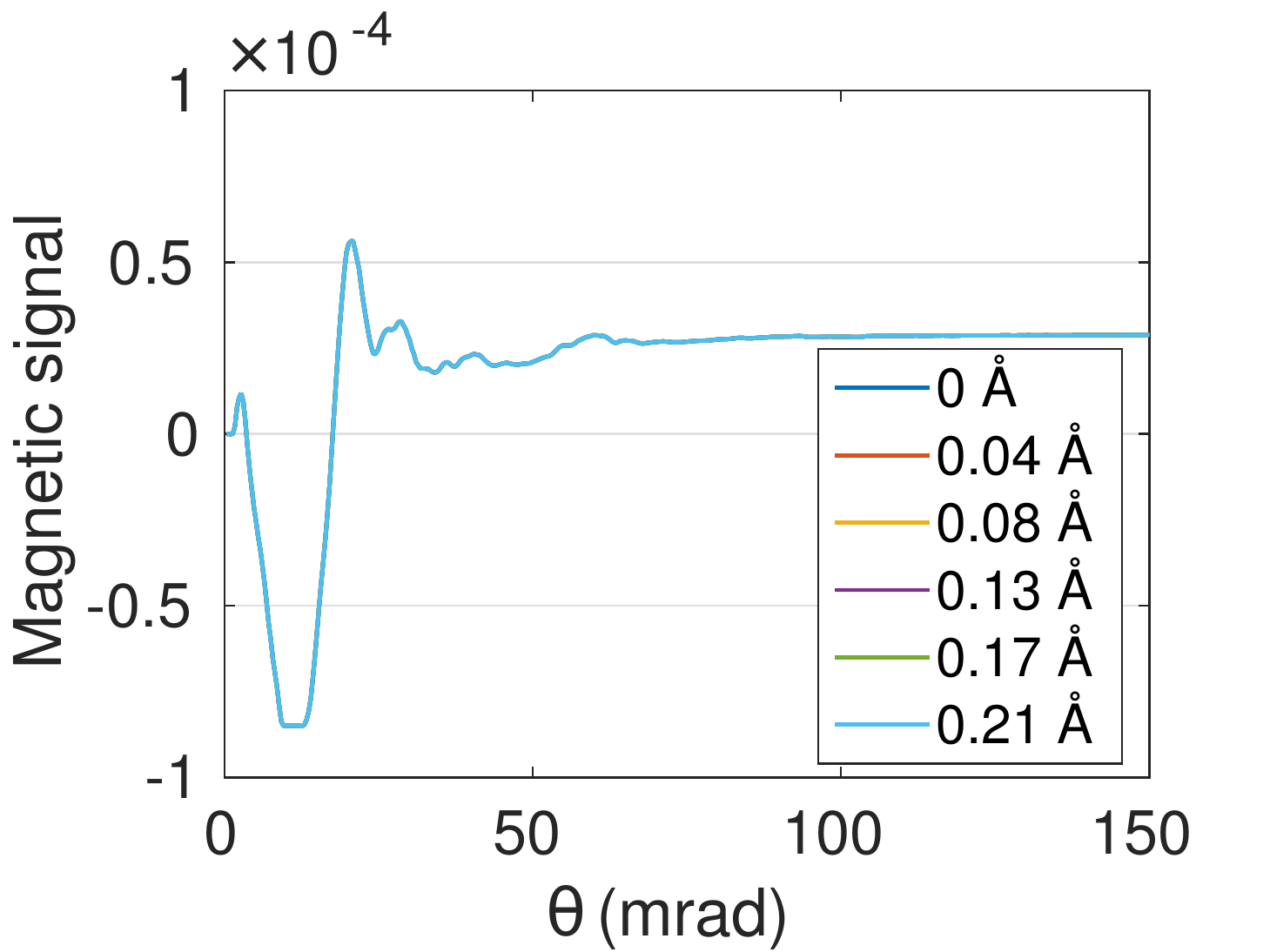}
        \caption{$l=\pm10$, 6~mrad and 100~keV.}
    \end{subfigure} \\%
    \begin{subfigure}[b]{0.4\textwidth}
        \includegraphics[width=\textwidth]{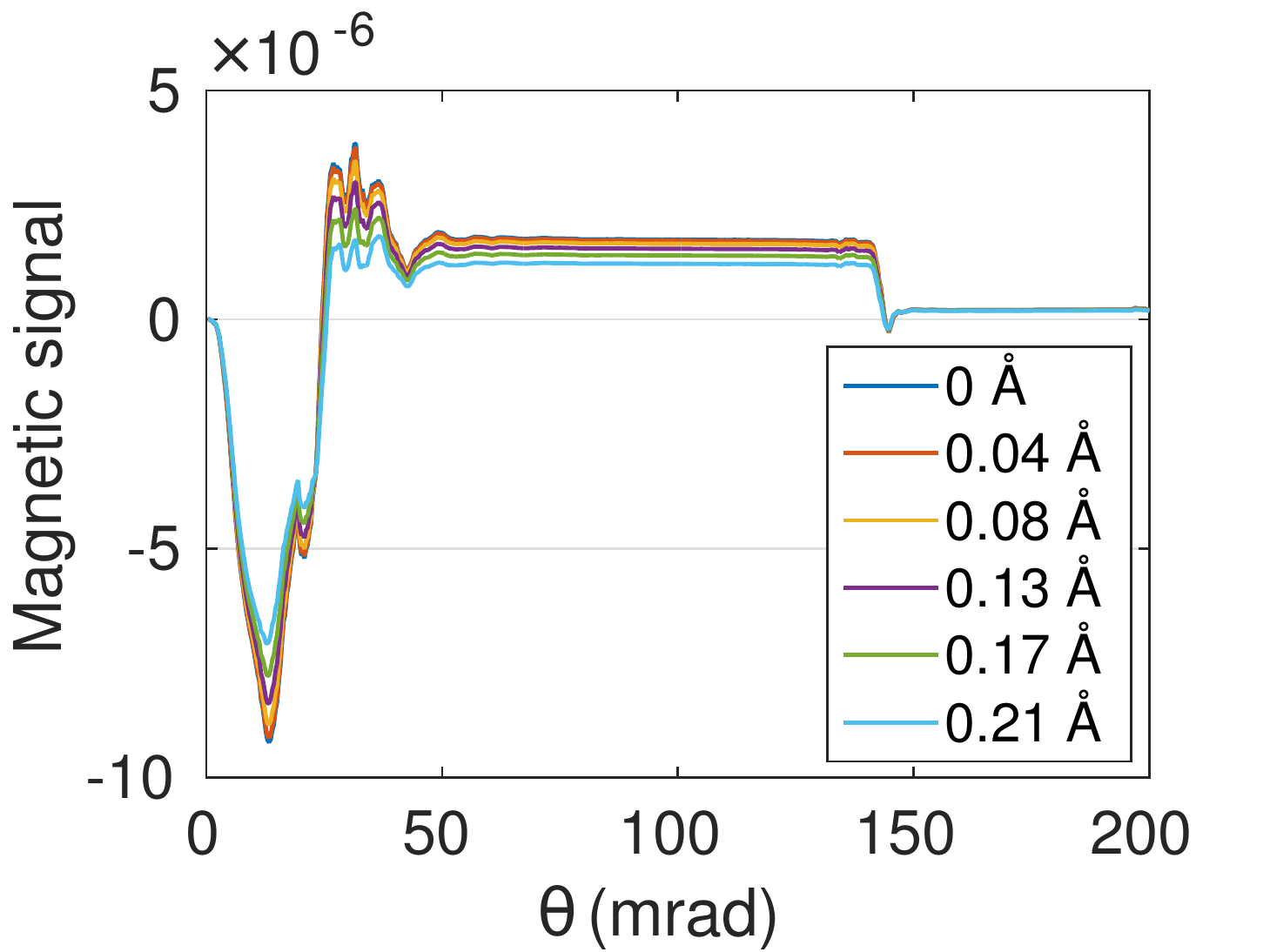}
        \caption{$l=\pm1$, 30~mrad and 100~keV.}
    \end{subfigure} \\%
	\caption{OAM magnetic signals at a given beam position for two sets of beam parameters and various small beam shifts.}
	\label{fig.driftnoise}
\end{figure}
The result of such calculations is shown in Fig.~\ref{fig.driftnoise} with OAM magnetic signal as a function of collection angle for the two sets of beam parameters in a) and b) respectively and various beam shifts indicated in the legend, after 50 unit cells of FePt. Clearly the considered shifts are of no relevance for the large OAM case whereas a small quantitative change is observed for the small OAM case. This change should, however, only yield a small quantitative error and mechanical drift in the considered range should not present a significant problem, neither in the large nor small OAM case. 

\begin{figure}[hbt!]
	\centering
     \includegraphics[width=0.4\textwidth]{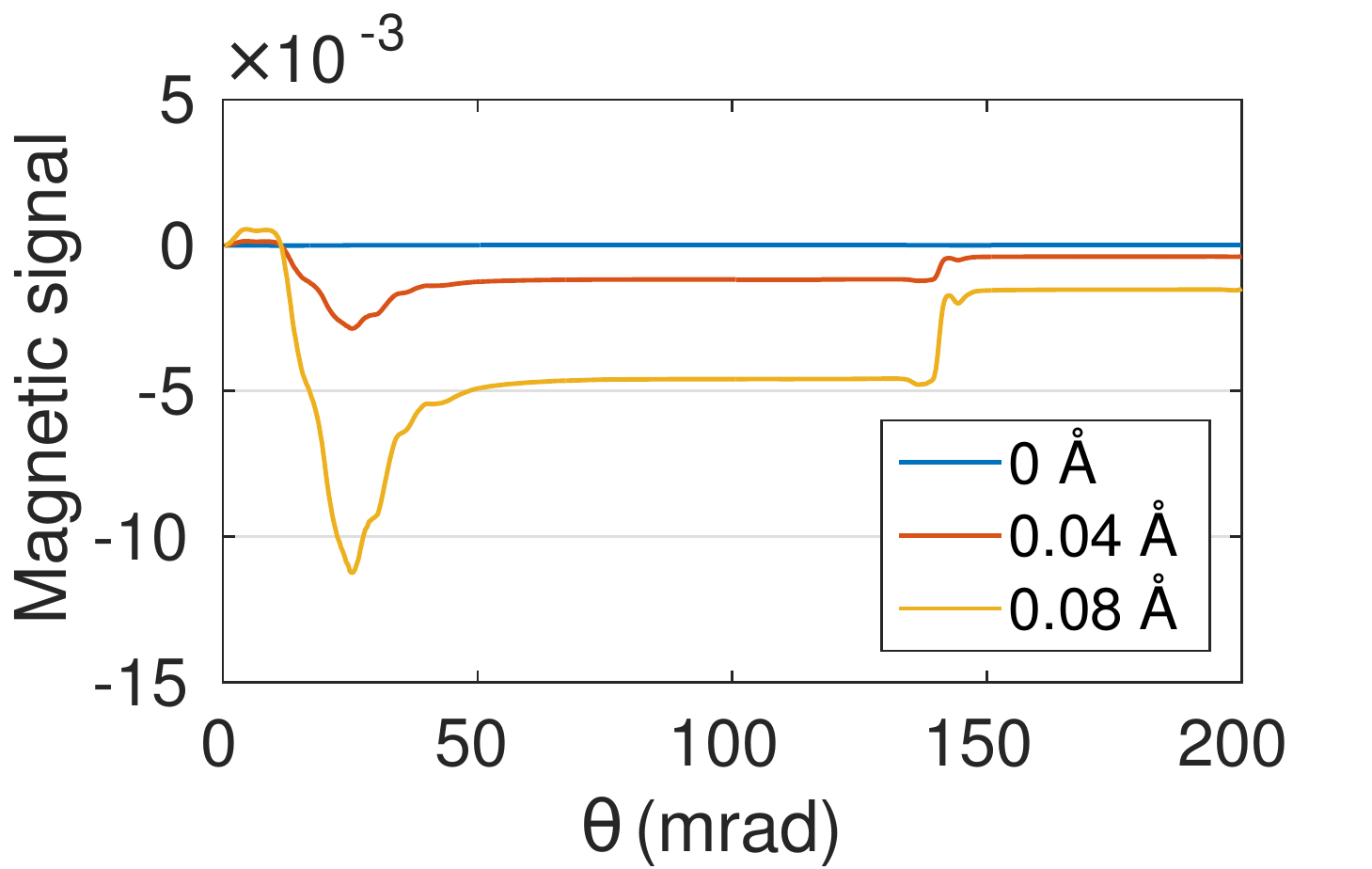}
	\caption{OAM magnetic signal from a positive OAM that has been shifted by a given amount and a negative OAM beam at the Fe atomic column, with $l=\pm1$, 30~mrad and 100~keV.}
	\label{fig.driftnoise2}
\end{figure}

The data presented in Fig.~\ref{fig.driftnoise} assumed that the signals for positive and negative OAM beams were collected for identical beam positions, which might be difficult to achieve in experiments. Hence, in Fig.~\ref{fig.driftnoise2} similar data is presented for the small OAM case except that the magnetic signal is obtained as the difference in intensity for a shifted positive OAM beam and a non-shifted negative OAM beam. Here it can be seen that a shift of {0.04 \AA} is already enough for a difference that is much greater than the magnetic signal to appear. In the large OAM case (not shown), however, the data is visually identical to that in Fig.~\ref{fig.driftnoise}a). Hence, drift noise should not present a problem for large beams with spatial dimensions well above the interatomic distances while it might present a problem for atomic resolution measurements if positive and negative OAM beams are measured at slightly different beam positions.

\begin{figure}[hbt!]
	\centering
    \begin{subfigure}[b]{0.4\textwidth}
      \includegraphics[width=\textwidth]{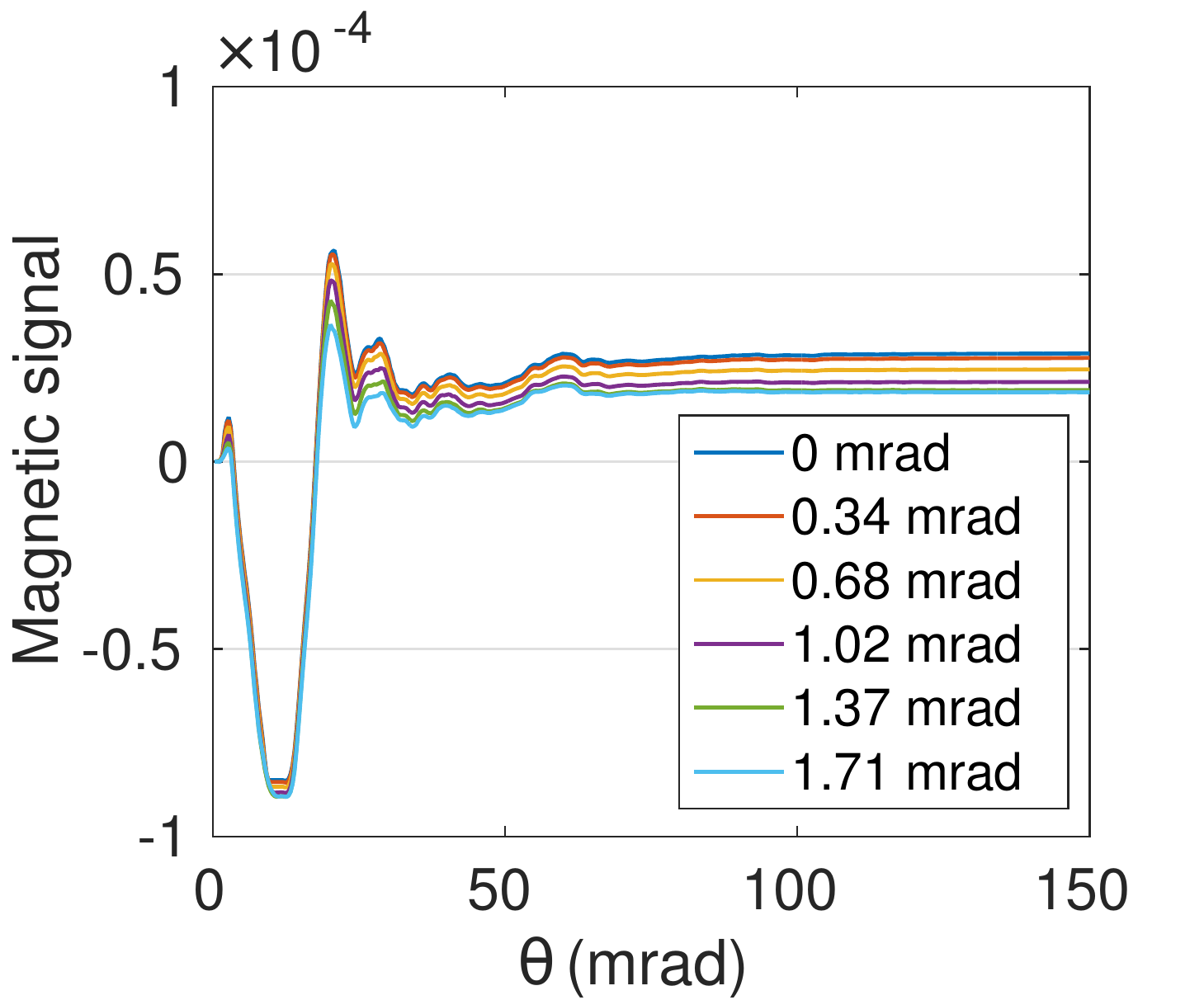}
        \caption{Same tilt for the positive and negative OAM beams.}
    \end{subfigure} \\%
     \begin{subfigure}[b]{0.4\textwidth}
        \includegraphics[width=\textwidth]{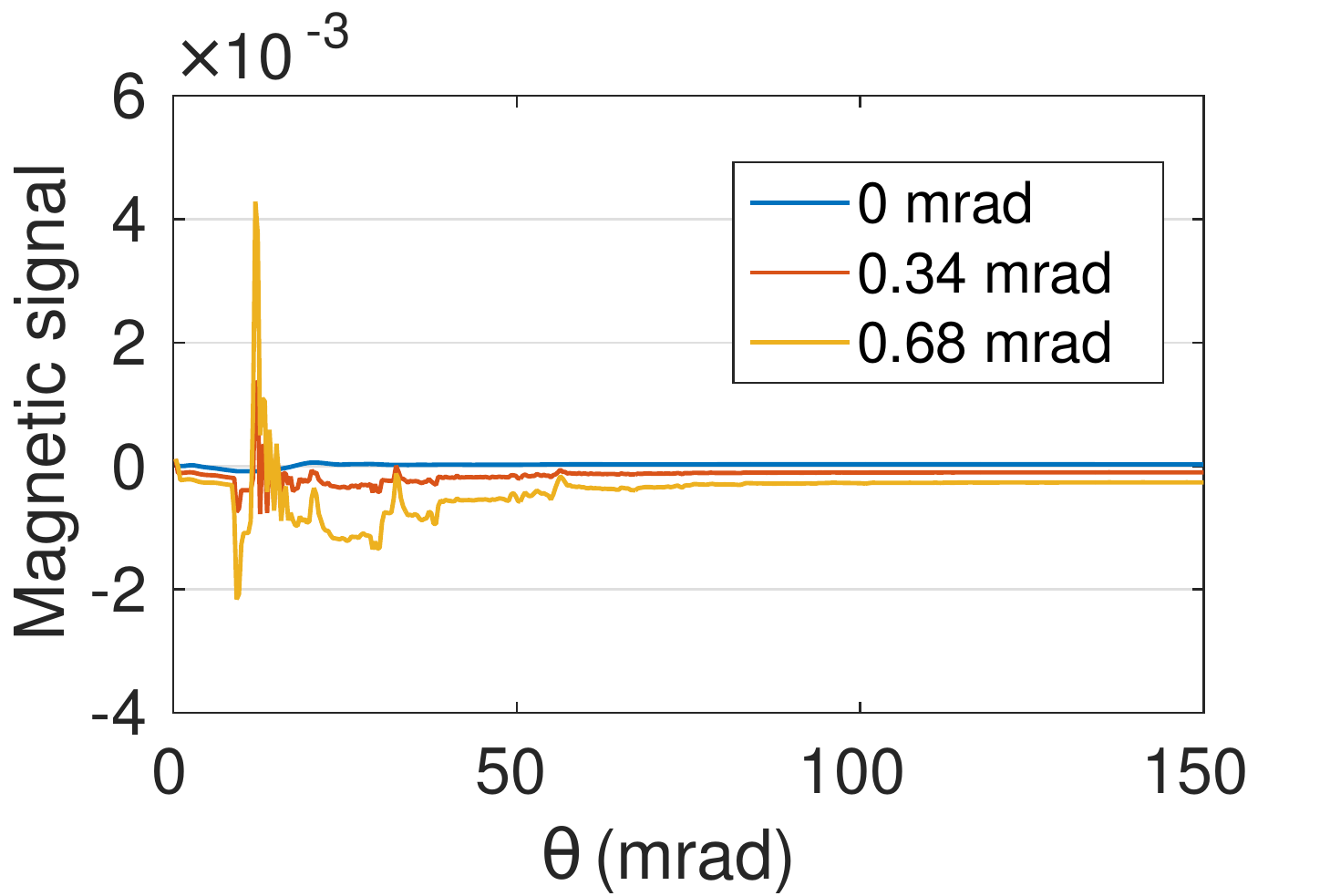}
        \caption{Tilted positive OAM beam and non-tilted negative OAM beam.}
    \end{subfigure} \\%
	\caption{Effect of tilt on the OAM magnetic signal as function of collection angle for a beam with $l=\pm10$, 6~mrad and 100~keV after passing through 50 unit cells of FePt.}
	\label{fig.tilt}
\end{figure}

Next, tilt is considered by shifting the origin of the Fourier transform of the initial beam by $(\Delta k_x , \Delta k_y)$, which yields a tilt of approximately $\Delta k_x / k$ and $\Delta k_y / k$ radians around the $y$ and $x$-axes respectively. The smallest tilt possible for a 100 keV beam and a system size of $60\times60$ unit cells in the $xy$-plane is then 0.34~mrad, corresponding to a realistic misalignment in STEM experiments.. The result of calculations for the large OAM beam after passing through 50 unit cells of FePt is contained in Fig.~\ref{fig.tilt}, with the OAM magnetic signal obtained either from the difference of positive and negative OAM beams with the same tilt in a) or with a positive OAM beam with tilt and a negative OAM beam with no tilt in b). If the tilt is the same for both beams, it does not appear to have a significant effect on the magnetic signal in this case, whereas if the tilt is different for the two beams, 0.34~mrad is enough to cause a difference significantly greater than the magnetic signal. Hence, sample tilt is expected to be problematic for large OAM beams only if the diffraction pattern for opposite OAM beams are acquired for different tilt angles. In the case of the small OAM beam (not shown), a large error is caused by the 0.34~mrad tilt also if measurements for both beams are done at the same tilt. Atomic resolution measurements would hence require extremely high precision mechanics in the experimental setup. 

\section{Summary and Conclusions}

A comprehensive computational study has been presented regarding magnetic effects in the paraxial regime of elastic electron scattering in magnetic crystals. This has been done using recently implemented methods\cite{edstrom16} to provide a realistic description of the coupling between a highly energetic electron beam, such as that used in transmission electron microscopy, with magnetism in a solid. Such effects are especially interesting in the context of high resolution TEM imaging of magnetism. In particular three ways of obtaining a magnetic signal have been considered: electron vortex beams, spin polarized beams or beams with phase aberrations. In all three cases a magnetic signal is obtainable by taking a difference in radial intensity distributions for beams with opposite sign angular momentum or aberration coefficients. This potentially allows for three different novel methods of imaging magnetism in a transmission electron microscope in a relatively simple experimental setup compared to EELS experiments such as EMCD. Two of the suggested methods were already discussed in recent work\cite{edstrom16}, where large OAM vortex beams was pointed out as the most feasible method, which unfortunately restricts the spatial resolution, while atomic resolution measurements with small OAM or spin polarized beams were deemed technically very challenging if not impossible. The more comprehensive computational study presented here indicates significantly stronger magnetic signals, increased by one or two orders of magnitude both for large and small OAM beams, for certain beam parameters, compared the previous work. Specifically it appears promising to use relatively low acceleration voltages and convergence angles, unfortunately again setting restrictions to spatial resolutions. In addition to vortex beams and spin polarization, the new effect of a magnetic signal from elastic scattering of phase aberrated electron beams has been demonstrated. This effect is of similar order of magnitude as that obtained with low OAM vortex beams and thus deemed possible but challenging to detect experimentally. Considering continuous technological improvements related to electron vortex beams\cite{Uchida2010, Verbeeck2010, McMorran2011, VerbeeckAPL2011, Saitoh2012, Pohl2015, Grillo2015}, spin polarization technology\cite{tanaka} and aberration correctors\cite{Krivanek19991, Krivanek2008179}, the present work will hopefully stimulate various experimental efforts to detect magnetism based on the suggested effects. An analysis of errors due to, e.g., sample drift and tilt suggests again that it will be very challenging to perform atomic resolution measurements while nanometer resolution measurements are expected to be more feasible. 

We acknowledge Swedish Research Council and G\"oran Gustafsson's Foundation for financial support. A.L. acknowledges support from DIP, the German-Israeli Project Cooperation. Computational work was performed using resources from the Swedish National Infrastructure for Computing (SNIC).

\appendix
\section{Magnetic field and vector potential of a periodic system}\label{AppA}

The magnetic field $\mathbf{B}$ should fulfill Maxwell's equations
\begin{equation}
\nabla \cdot \mathbf{B}(\mathbf{r}) = 0, \quad \nabla \times \mathbf{B}(\mathbf{r}) = \mu_0 \mathbf{J}(\mathbf{r}),
\label{Max}
\end{equation}
where $\mathbf{J}$ is the current density, together with physical boundary conditions.  In the case of periodic boundary conditions one more constraint, e.g. specifying $\mathbf{B}$ at a given point or its volume average, is necessary and sufficient for a unique solution to exist\cite{Jackson}. Similarly, any periodic $\mathbf{B}$ can be decomposed into a periodic part with volume average zero $\mathbf{B}_\text{p}$ and a uniform part $\mathbf{B}_\text{avg}$ corresponding to the volume average of $\mathbf{B}$, as discussed also in Sec.~\ref{MagFields} of this paper. Such a periodic $\mathbf{B}$ is suitable to Fourier transform according to
\begin{equation}
\mathbf{B}(\mathbf{r}) = \int_\mathbf{\text{BZ}} \mathbf{b}(\mathbf{k}) \ee^{\img \mathbf{k}\cdot\mathbf{r}}\dd \mathbf{k},
\end{equation}
with integration over the Brillouin zone (BZ), so that Eqs.~\ref{Max} read 
\begin{equation}
\mathbf{k}\cdot \mathbf{b}(\mathbf{k})=0, \quad \img\mathbf{k}\times\mathbf{b}(\mathbf{k})=\mu_0 \mathbf{j}(\mathbf{k}),
\label{Max_k}
\end{equation}
where $\mathbf{b}(0)$ corresponds to the volume average of $\mathbf{B}(\mathbf{r})$, $\mathbf{j}(\mathbf{k})$ is the Fourier transform of $\mathbf{J}(\mathbf{r})$,  and it was assumed that $\mathbf{J}(\mathbf{r})$ was also periodic with a zero volume average, i.e.  $\mathbf{j}(0)=0$ (otherwise Eqs.~\ref{Max_k} cannot be fulfilled). 

The vector potential $\mathbf{A}(\mathbf{r})$ should fulfill the defining equation $\mathbf{B} = \nabla \times \mathbf{A}$ together with some gauge choice, rather than physical boundary conditions such as periodicity. If, nevertheless, one assumes that $\mathbf{A}(\mathbf{r})$ is periodic, its Fourier transform $\mathbf{a}(\mathbf{k})$ must fulfill $\img\mathbf{k}\times\mathbf{a}(\mathbf{k}) = \mathbf{b}(\mathbf{k})$, which clearly can only be fulfilled in the case that $\mathbf{b}(0)=0$, i.e. if the volume average of $\mathbf{B}(\mathbf{r})$ is zero. As has been pointed out before\cite{Brown1969313}, this leads to the conclusion that a periodic vector potential is only possible to construct in the case that the volume average of the magnetic field is zero. With respect to the division of the magnetic field according to $\mathbf{B} = \mathbf{B}_\text{p} + \mathbf{B}_\text{avg}$, this corresponds to $\mathbf{B}_\text{avg}=0$. However, by doing a corresponding decomposition of the vector potential into a periodic part $\mathbf{A}_\text{p}$ and a non-periodic part $\mathbf{A}_\text{np}$, as done in this paper, and relating 
\begin{equation}
\mathbf{B}_\text{p} = \nabla \times \mathbf{A}_\text{p} , \quad \mathbf{B}_\text{avg} = \nabla \times \mathbf{A}_\text{np},
\end{equation}
it is clear from the above arguments that $\mathbf{A}_\text{p}$ can be chosen to be periodic whereas $\mathbf{A}_\text{np}$ necessarily is non-periodic for non-zero $\mathbf{B}_\text{avg}$. In Coulomb gauge one obtains $\mathbf{A}_\text{np} = \frac{1}{2} \mathbf{B}_\text{avg} \times \mathbf{r}$, possibly with the addition of an arbitrary constant, as mentioned in Sec.~\ref{MagFields}.  

The impossibility of relating a periodic vector potential to a non-zero, uniform magnetic field can also be seen by considering the line integral of such a periodic $\mathbf{A}_\text{p}$ along a closed path $\gamma$ around a parallelogram of lattice vectors\cite{Brown1969313}. Such integral is zero due to contributions of equal magnitude but opposite sign from opposing sides of the parallelogram, whereby 
\begin{equation}
\oint_\gamma \mathbf{A}_\text{p}\cdot\dd \mathbf{r} = \int_S \mathbf{B}\cdot \dd \mathbf{S} = 0
\end{equation}
for any surface $S$ with boundary $\gamma$. The only uniform $\mathbf{B}$ which can fulfill this is $\mathbf{B}=0$.

\bibliographystyle{apsrev4-1}
\bibliography{literature}{}

\end{document}